\documentclass[aps,prb,twocolumn,groupedaddress,amsmath,amssymb]{revtex4-1}
\usepackage{graphicx}
\usepackage{amscd}
\usepackage{tensor}
\usepackage{float}

\usepackage{xy}
\xyoption{matrix}
\xyoption{frame}
\xyoption{arrow}
\xyoption{arc}
\xyoption{curve}
\xyoption{all}
\xyoption{knot}
\xyoption{pdf}
\usepackage{mathdots}

\newdir{|>}{%
!/4.5pt/@{|}*:(1,-.2)@^{>}*:(1,+.2)@_{>}}

\newcommand{\smalllabels}{
	\def\objectstyle{\scriptstyle}
	\def\labelstyle{\scriptstyle}
}

\newcommand{\unknot}[1]{
	\xygraph{
		!{0;/r2.0pc/:}
		!{\vcap-}
		!{\vcap}
	}
}

\newcommand{\qdim}[2]{
	\smalllabels
	\xy
		(-4,0)="L"+(-2,0)*{#1};
		(4,0)="R";
		(0,4)="T";
		(0,-4)="B";
		"L";"T"*\crv{"L"+(0,2)&"T"+(-2,0)};
		"T";"R"*\crv{"T"+(2,0)&"R"+(0,2)};
		"R";"B"*\crv{"R"+(0,-2)&"B"+(2,0)};
		"B";"L"*\crv{"B"+(-2,0)&"L"+(0,-2)};
		"L"+(.05,1)*\dir{#2};
	\endxy
}

\newcommand{\anyon}[1]{
	\smalllabels
	\xy
		(0,6)*{}="T";
		(0,-6)*{}="B";
		"B"*{};"T"**\dir{-}?(.6)*\dir{>} + (2,-3)*{#1};
	\endxy
}

\newcommand{\longanyon}[1]{
	\smalllabels
	\xy
		(0,8)*{}="T";
		(0,-8)*{}="B"+(-2,3)*{#1};
		"B"*{};"T"**\dir{-}?(.6)*\dir{>}; 
	\endxy
}

\newcommand{\conjanyon}[1]{
	\smalllabels
	\xy
		(0,6)*{}="T";
		(0,-6)*{}="B";
		"B"*{};"T"**\dir{-}?(.4)*\dir{<} + (2,-1)*{#1};
	\endxy
}

\newcommand{\vac}[1]{
	\smalllabels
	\xy
		(0,6)*{}="T";
		(0,-6)*{}="B";
		"B"*{};"T"**\dir{.}?(.6)*\dir{} + (2,-3)*{#1};
	\endxy
}

\newcommand{\vertexket}[4]{
	\smalllabels
	\xy
		(-4.24,4.24)*{}="a";
		(4.24,4.24)*{}="b";
		(0,-6)*{}="c";
		(0,0)*{}="mid" +(2,-1)*{#4};
		"mid"; "a" **\dir{-}?(.75)*\dir{>} + (-2,0)*{#1};
		"mid"; "b" **\dir{-}?(.75)*\dir{>} + (2,0)*{#2};
		"c"; "mid"**\dir{-}?(.7)*\dir{>}+(2,-3)*{#3};
	\endxy 
}

\newcommand{\vacvertexket}[3]{
	\smalllabels
	\xy
		(-4.24,4.24)*{}="a";
		(4.24,4.24)*{}="b";
		(0,-6)*{}="c";
		(0,0)*{}="mid";
		"mid"; "a" **\dir{.}?(.75)*\dir{} + (-2,0)*{#1};
		"mid"; "b" **\dir{.}?(.75)*\dir{} + (2,0)*{#2};
		"c"; "mid"**\dir{.}?(.7)*\dir{}+(2,-3)*{#3};
	\endxy 
}

\newcommand{\vacLket}[3]{
	\smalllabels
	\xy
		(-4.24,4.24)*{}="a";
		(4.24,4.24)*{}="b";
		(0,-6)*{}="c";
		(0,0)*{}="mid" +(2,-1)*{};
		"mid"; "a" **\dir{.}?(.75)*\dir{} + (-2,0)*{#1};
		"mid"; "b" **\dir{-}?(.75)*\dir{>} + (2,0)*{#2};
		"c"; "mid"**\dir{-}?(.7)*\dir{>}+(2,-3)*{#3};
	\endxy 
}

\newcommand{\vacRket}[3]{
	\smalllabels
	\xy
		(-4.24,4.24)*{}="a";
		(4.24,4.24)*{}="b";
		(0,-6)*{}="c";
		(0,0)*{}="mid" +(2,-1)*{};
		"mid"; "a" **\dir{-}?(.75)*\dir{>} + (-2,0)*{#1};
		"mid"; "b" **\dir{.}?(.75)*\dir{} + (2,0)*{#2};
		"c"; "mid"**\dir{-}?(.7)*\dir{>}+(2,-3)*{#3};
	\endxy 
}

\newcommand{\newvacRket}[3]{
	\smalllabels
	\xy
		(-4.24,4.24)*{}="a";
		(4.24,4.24)*{}="b";
		(0,-6)*{}="c";
		(0,0)*{}="mid" +(2,-1)*{};
		"mid"; "a" **\dir{-}?(.75)*\dir{>} + (-2,0)*{#1};
		"mid"; "b" **\dir{--}?(.75)*\dir{} + (2,0)*{#2};
		"c"; "mid"**\dir{-}?(.7)*\dir{>}+(2,-3)*{#3};
	\endxy 
}

\newcommand{\newvacBket}[3]{
	\smalllabels
	\xy
		(-4.24,4.24)*{}="a";
		(4.24,4.24)*{}="b";
		(0,-6)*{}="c";
		(0,0)*{}="mid" +(2,-1)*{};
		"mid"; "a" **\dir{-}?(.75)*\dir{>} + (-2,0)*{#1};
		"mid"; "b" **\dir{-}?(.75)*\dir{>} + (2,0)*{#2};
		"c"; "mid"**\dir{--}?(.7)*\dir{}+(2,-3)*{#3};
	\endxy 
}

\newcommand{\vacBket}[3]{
	\smalllabels
	\xy
		(-4.24,4.24)*{}="a";
		(4.24,4.24)*{}="b";
		(0,-6)*{}="c";
		(0,0)*{}="mid" +(2,-1)*{};
		"mid"; "a" **\dir{-}?(.75)*\dir{>} + (-2,0)*{#1};
		"mid"; "b" **\dir{-}?(.75)*\dir{>} + (2,0)*{#2};
		"c"; "mid"**\dir{.}?(.7)*\dir{}+(2,-3)*{#3};
	\endxy 
}

\newcommand{\vertexbra}[4]{
	\smalllabels
	\xy
		(-4.24,-4.24)*{}="a";
		(4.24,-4.24)*{}="b";
		(0,6)*{}="c";
		(0,0)*{}="mid" + (2,1)*{#4};
		"a"; "mid"**\dir{-}?(.7)*\dir{>} + (-4,-1.5)*{#1};
		"b"; "mid"**\dir{-}?(.7)*\dir{>} + (4,-1.5)*{#2};
		"mid"; "c"**\dir{-}?(.7)*\dir{>}+(2,0)*{#3};
	\endxy 
}

\newcommand{\condensatevertexbraanyon}[1]{
	\smalllabels
	\xy
		(-4.24,-4.24)*{}="a";
		(4.24,-4.24)*{}="b";
		(0,6)*{}="c";
		(0,0)*{}="mid" + (2,1)*{};
		"a"; "mid"**\dir{--}?(.7)*\dir{} + (-4,-1.5)*{};
		"b"; "mid"**\dir{-}?(.7)*\dir{>} + (4,-1.5)*{#1};
		"mid"; "c"**\dir{-}?(.7)*\dir{>}+(2,0)*{};
	\endxy 
}

\newcommand{\vertexbraket}[6]{
	\smalllabels
	\xy
		(0,6)*{}="c" +(-2,-1)*{#1};
		(-3,0)*{}="a"+(-1,0)*{#2};
		(3,0)*{}="b"+(1,0)*{#3};
		(0,-6)*{}="cc" +(-2,1)*{#4};
		(0,3)*{}="abc"+(2,0)*{#5};
		(0,-3)="abcc"+(2,0)*{#6};
		"abc";"c"**\dir{-}?(.7)*\dir{>};	
		"abcc";"abc"**\crv{"a"}?(.6)*\dir{>};
		"abcc";"abc"**\crv{"b"}?(.6)*\dir{>};		
		"cc";"abcc"**\dir{-}?(.7)*\dir{>};	
	\endxy
}

\newcommand{\vertexbraketbubbleL}[4]{
	\smalllabels
	\xy
		(-3,6)*{}="a"+(3,0)*{#2};
		(0.5,4)*{}="a1";
		(-3,3)*{}="a2";
		(4,6)*{}="b"+(2,0)*{#1};
		(-3,-6)*{}="c"+(-1,1)*{};
		(0.5,-4)*{}="c2";
		(-3,-3)*{}="c1";
		(4,-6)*{}="d"+(1.5,0)*{#4};
		(0,0)*{}="e"+(2,0)*{};
		(4,2)*{}="e2"+(2,0)*{};
		(4,-2)*{}="e1"+(1.5,3)*{#3};
		"e2";"a1"**\dir{-}?(1)*\dir{>} ;
		"a2";"c1"**\dir{-}?(.7)*\dir{} ;
		"e2";"b"**\dir{-}?(.7)*\dir{>} ;
		"c2";"e1"**\dir{-}?(.7)*\dir{} ;
		"d";"e1"**\dir{-}?(.7)*\dir{>} ;
		"e1";"e2"**\dir{-}?(.7)*\dir{>};
		"a1";"a2"**\crv{"a"};
		"c1";"c2"**\crv{"c"};
	\endxy
}

\newcommand{\vertexbraketbubbleR}[4]{
	\smalllabels
	\xy
		(3,6)*{}="a"+(-3,0)*{#3};
		(-0.5,4)*{}="a1";
		(3,3)*{}="a2";
		(-4,6)*{}="b"+(-2,0)*{#1};
		(3,-6)*{}="c"+(1,1)*{};
		(-0.5,-4)*{}="c2";
		(3,-3)*{}="c1";
		(-4,-6)*{}="d"+(-1.5,0)*{#4};
		(0,0)*{}="e"+(-2,0)*{};
		(-4,2)*{}="e2"+(-2,0)*{};
		(-4,-2)*{}="e1"+(-1.5,3)*{#2};
		"e2";"a1"**\dir{-}?(1)*\dir{>} ;
		"a2";"c1"**\dir{-}?(.7)*\dir{} ;
		"e2";"b"**\dir{-}?(.7)*\dir{>} ;
		"c2";"e1"**\dir{-}?(.7)*\dir{} ;
		"d";"e1"**\dir{-}?(.7)*\dir{>} ;
		"e1";"e2"**\dir{-}?(.7)*\dir{>};
		"a1";"a2"**\crv{"a"};
		"c1";"c2"**\crv{"c"};
	\endxy
}

\newcommand{\onetothreeL}[7]{
	\smalllabels
	\xy
		(-12,6)*{}="a" + (-1,-1)*{#1};
		(-4,6)*{}="b"+ (-3.5,-1)*{#2};
		(4,6)*{}="c"+ (-3.5,-1)*{#3};
		(-4,-8)*{}="d"+(-2,1)*{#4};
		(-6,0)*{}="e"+(-1.3,-1.3)*{#5};
		(-8,2)*{}="ab"+(-1.4,-1.2)*{#6};
		(-4,-2)*{}="cd"+(-1.4,-1.2)*{#7};
		"ab";"a"**\dir{-}?(.7)*\dir{>};	
		"ab";"b"**\dir{-}?(.7)*\dir{>};
		"cd";"c"**\dir{-}?(.75)*\dir{>};
		"d";"cd"**\dir{-}?(.7)*\dir{>};
		"cd";"ab"**\dir{-}?(.7)*\dir{>};
	\endxy
}

\newcommand{\onetothreeR}[7]{
	\smalllabels
	\xy
		(-12,6)*{}="a"+(-1,-1)*{#1};;
		(-4,6)*{}="b"+(-1,-1)*{#2};;
		(4,6)*{}="c"+(-3.5,-1)*{#3};;
		(-4,-8)*{}="d"+(-2,1)*{#4};
		(-2,0)*{}="f"+(2,-2)*{#5};
		(0,2)*{}="bc"+(1.4,-1.2)*{#6};
		(-4,-2)*{}="ad"+(1.4,-1.2)*{#7};
		"ad";"a"**\dir{-}?(.75)*\dir{>};			
		"bc";"b"**\dir{-}?(.7)*\dir{>};
		"bc";"c"**\dir{-}?(.7)*\dir{>};
		"d";"ad"**\dir{-}?(.7)*\dir{>};
		"ad";"bc"**\dir{-}?(.7)*\dir{>};
	\endxy
}

\newcommand{\twoanyons}[2]{
	\smalllabels
	\xy
		(-2,6)*{}="a2";
		(-2,-6)*{}="a1"+(-1,3)*{#1};
		(2,6)*{}="b2";
		(2,-6)*{}="b1"+(1,3)*{#2};
		"a1"*{};"a2"**\dir{-}?(.58)*\dir{>};
		"b1"*{};"b2"**\dir{-}?(.58)*\dir{>};
	\endxy 
}

\newcommand{\twototwoS}[7]{
	\smalllabels
	\xy
		(-2,6)*{}="a" + (-2,-1)*{#1};
		(2,6)*{}="b"+ (2,-1)*{#2};
		(-2,-6)*{}="c"+ (-2,1)*{#3};
		(2,-6)*{}="d"+ (2,1)*{#4};
		(0,0)="e"+ (0,2)*{#5};
		(-2,2)*{}="e2"+(-1.5,-.5)*{#6};
		(2,-2)*{}="e1"+(1.5,.5)*{#7};
		"c";"a"**\dir{-};
		"c"+(0,3)*\dir{>}+(0,8)*\dir{>};
		"d";"b"**\dir{-};
		"d"+(0,3)*\dir{>}+(0,8)*\dir{>};
		"e1";"e2"**\dir{-}?(.7)*\dir{>} ;		
	\endxy
}

\newcommand{\twototwoM}[7]{
	\smalllabels
	\xy
		(-4,6)*{}="a"+(-1,-1)*{#1};
		(4,6)*{}="b"+(1,-1)*{#2};
		(-4,-6)*{}="c"+(-1,1)*{#3};
		(4,-6)*{}="d"+(1,1)*{#4};
		(0,0)*{}="e"+(-2,0)*{#5};
		(0,2)*{}="e2"+(2,0)*{#6};
		(0,-2)*{}="e1"+(2,0)*{#7};
		"e2";"a"**\dir{-}?(.7)*\dir{>} ;
		"e2";"b"**\dir{-}?(.7)*\dir{>} ;
		"c";"e1"**\dir{-}?(.7)*\dir{>} ;
		"d";"e1"**\dir{-}?(.7)*\dir{>} ;
		"e1";"e2"**\dir{-}?(.7)*\dir{>};
	\endxy
}

\newcommand{\twototwoSvacL}[7]{
	\smalllabels
	\xy
		(-2,6)*{}="a" + (-2,-1)*{#1};
		(2,6)*{}="b"+ (2,-1)*{#2};
		(-2,-6)*{}="c"+ (-2,1)*{#3};
		(2,-6)*{}="d"+ (2,1)*{#4};
		(0,0)="e"+ (0,2)*{#5};
		(-2,2)*{}="e2"+(-1.5,-.5)*{#6};
		(2,-2)*{}="e1"+(1.5,.5)*{#7};
		"c";"e2"**\dir{-};
		"e2";"a"**\dir{.};
		"c"+(0,3)*\dir{>}+(0,8)*\dir{};
		"d";"b"**\dir{-};
		"d"+(0,3)*\dir{>}+(0,8)*\dir{>};
		"e1";"e2"**\dir{-}?(.7)*\dir{>};		
	\endxy
}	

\newcommand{\twototwoSvacR}[7]{
	\smalllabels
	\xy
		(-2,6)*{}="a" + (-2,-1)*{#1};
		(2,6)*{}="b"+ (2,-1)*{#2};
		(-2,-6)*{}="c"+ (-2,1)*{#3};
		(2,-6)*{}="d"+ (2,1)*{#4};
		(0,0)="e"+ (0,2)*{#5};
		(-2,2)*{}="e2"+(-1.5,-.5)*{#6};
		(2,-2)*{}="e1"+(1.5,.5)*{#7};
		"c";"a"**\dir{-};
		"c"+(0,3)*\dir{>}+(0,8)*\dir{>};
		"d";"e1"**\dir{.};
		"e1";"b"**\dir{-};
		"d"+(0,3)*\dir{}+(0,8)*\dir{>};
		"e1";"e2"**\dir{-}?(.7)*\dir{>};		
	\endxy
}	

\newcommand{\twototwoMvacM}[4]{
	\smalllabels
	\xy
		(-4,6)*{}="a"+(-1,-1)*{#1};
		(4,6)*{}="b"+(1,-1)*{#2};
		(-4,-6)*{}="c"+(-1,1)*{#3};
		(4,-6)*{}="d"+(1,1)*{#4};
		(0,0)*{}="e"+(-2,0)*{};
		(0,2)*{}="e2"+(2,0)*{};
		(0,-2)*{}="e1"+(2,0)*{};
		"e2";"a"**\dir{-}?(.7)*\dir{>} ;
		"e2";"b"**\dir{-}?(.7)*\dir{>} ;
		"c";"e1"**\dir{-}?(.7)*\dir{>} ;
		"d";"e1"**\dir{-}?(.7)*\dir{>} ;
		"e1";"e2"**\dir{.}?(.7)*\dir{};
	\endxy
}

\newcommand{\twototwoSvacLvacR}[2]{
	\smalllabels
	\xy
		(-2,6)*{}="a" + (-2,-1)*{};
		(2,6)*{}="b"+ (2,-1)*{#1};
		(-2,-6)*{}="c"+ (-2,1)*{#1};
		(2,-6)*{}="d"+ (2,1)*{};
		(0,0)="e"+ (0,2)*{#2};
		(-2,2)*{}="e2"+(-1.5,-.5)*{};
		(2,-2)*{}="e1"+(1.5,.5)*{};
		"c";"e2"**\dir{-};
		"e2";"a"**\dir{.};
		"c"+(0,3)*\dir{>}+(0,8)*\dir{};
		"d";"e1"**\dir{.};
		"e1";"b"**\dir{-};
		"d"+(0,3)*\dir{}+(0,8)*\dir{>};
		"e1";"e2"**\dir{-}?(.7)*\dir{>};		
	\endxy
}	

\newcommand{\twototwoSvacLvacRvar}[1]{
	\smalllabels
	\xy
		(-2,6)*{}="a" + (-2,-1)*{};
		(2,6)*{}="b"+ (2,-1)*{};
		(-2,-6)*{}="c"+ (-2,1)*{#1};
		(2,-6)*{}="d"+ (2,1)*{};
		(0,0)="e"+ (0,2)*{};
		(-2,2)*{}="e2"+(-1.5,-.5)*{};
		(2,-2)*{}="e1"+(1.5,.5)*{};
		"c";"e2"**\dir{-};
		"e2";"a"**\dir{.};
		"c"+(0,3)*\dir{>}+(0,8)*\dir{};
		"d";"e1"**\dir{.};
		"e1";"b"**\dir{-};
		"d"+(0,3)*\dir{}+(0,8)*\dir{>};
		"e1";"e2"**\dir{-}?(.7)*\dir{};		
	\endxy
}

\newcommand{\operator}[5]{
	\smalllabels
	\xy
		(0,0)*{#1}="X";
		(-6,4)="L";
		(-6,-4)="LL";
		(6,4)="R";
		(6,-4)="RR";
		(-4,4)="a1";(-4,8)="A1"+(0,3)*{#2};
		(4,4)="a2";(4,8)="A2"+(0,3)*{#3};
		(-4,-4)="aa1";(-4,-8)="AA1"+(0,-2)*{#4};		
		(4,-4)="aa2";(4,-8)="AA2"+(0,-2)*{#5};
		"L";"R"**\dir{-};"RR"**\dir{-};"LL"**\dir{-};"L"**\dir{-};
		"a1";"A1"**\dir{-}?(.7)*\dir{>};
		"a2";"A2"**\dir{-}?(.7)*\dir{>};
		"AA1";"aa1"**\dir{-}?(.7)*\dir{>};
		"AA2";"aa2"**\dir{-}?(.7)*\dir{>};
		"X"+(0,6)*{\dots};
		"X"+(0,-6)*{\dots};
	\endxy
}

\newcommand{\tensoroperator}[1]{
	\smalllabels
	\xy
		(0,0)*{#1}="X";
		(-12,4)="L";
		(-12,-4)="LL";
		(12,4)="R";
		(12,-4)="RR";
		(-10,4)="a1";"a1"+(0,4)="A1";
		(-1.5,4)="a2";"a2"+(0,4)="A2";
		(-10,-4)="aa1";	"aa1"+(0,-4)="AA1";
		(-1.5,-4)="aa2";"aa2"+(0,-4)="AA2";
		(10,4)="b1";"b1"+(0,4)="B1";
		(1.5,4)="b2";"b2"+(0,4)="B2";
		(10,-4)="bb1";	"bb1"+(0,-4)="BB1";
		(1.5,-4)="bb2";"bb2"+(0,-4)="BB2";
		"L";"R"**\dir{-};"RR"**\dir{-};"LL"**\dir{-};"L"**\dir{-};
		"a1";"A1"**\dir{-}?(.7)*\dir{>};
		"a2";"A2"**\dir{-}?(.7)*\dir{>};
		"AA1";"aa1"**\dir{-}?(.7)*\dir{>};
		"AA2";"aa2"**\dir{-}?(.7)*\dir{>};
		"b1";"B1"**\dir{-}?(.7)*\dir{>};
		"b2";"B2"**\dir{-}?(.7)*\dir{>};
		"BB1";"bb1"**\dir{-}?(.7)*\dir{>};
		"BB2";"bb2"**\dir{-}?(.7)*\dir{>};
		"X"+(6,6)*{\dots};
		"X"+(6,-6)*{\dots};
		"X"+(-6,6)*{\dots};
		"X"+(-6,-6)*{\dots};
	\endxy
}

\newcommand{\quantumtrace}[3]{
	\smalllabels
	\xy
		(0,0)*{#1}="X";
		(-6,4)="L";
		(-6,-4)="LL";
		(6,4)="R";
		(6,-4)="RR";
		(-4,4)="a1";(-4,8)="A1"+(-2,0)*{#2};
		(4,4)="a2";(4,8)="A2"+(-2,0)*{#3};
		(-4,-4)="aa1";(-4,-8)="AA1"+(0,-2)*{};		
		(4,-4)="aa2";(4,-8)="AA2"+(0,-2)*{};
		"L";"R"**\dir{-};"RR"**\dir{-};"LL"**\dir{-};"L"**\dir{-};
		"a1";"A1"**\dir{-}?(.7)*\dir{>};
		"a2";"A2"**\dir{-}?(.7)*\dir{>};
		"AA1";"aa1"**\dir{-}?(.7)*\dir{>};
		"AA2";"aa2"**\dir{-}?(.7)*\dir{>};
		"X"+(0,6)*{\dots};
		"X"+(0,-6)*{\dots};
		"A1"+(4,4)="B1";
		"B1"+(18,0)="C1";
		"C1"+(4,-4)="D1";
		"A2"+(2,2)="B2";
		"B2"+(6,0)="C2";
		"C2"+(2,-2)="D2";
		"AA1"+(4,-4)="BB1";
		"BB1"+(18,0)="CC1";
		"CC1"+(4,4)="DD1";
		"AA2"+(2,-2)="BB2";
		"BB2"+(6,0)="CC2";
		"CC2"+(2,2)="DD2";
		"B1";"C1"**\dir{-};
		"B2";"C2"**\dir{-};	
		"BB1";"CC1"**\dir{-};
		"BB2";"CC2"**\dir{-};
		"D1";"DD1"**\dir{-};
		"D2";"DD2"**\dir{-};
		"A1";"B1"**\crv{"A1"+(0,2)&"B1"+(-2,0)};
		"A2";"B2"**\crv{"A2"+(0,1)&"B2"+(-1,0)};
		"D1";"C1"**\crv{"D1"+(0,2)&"C1"+(2,0)};
		"D2";"C2"**\crv{"D2"+(0,1)&"C2"+(1,0)};
		"AA1";"BB1"**\crv{"AA1"+(0,-2)&"BB1"+(-2,0)};
		"AA2";"BB2"**\crv{"AA2"+(0,-1)&"BB2"+(-1,0)};
		"DD1";"CC1"**\crv{"DD1"+(0,-2)&"CC1"+(2,0)};
		"DD2";"CC2"**\crv{"DD2"+(0,-1)&"CC2"+(1,0)};
		"X"+(18,6)*{\dots};
		"X"+(18,-6)*{\dots};
	\endxy
}

\newcommand{\fouranyonstate}[7]{
	\smalllabels
	\xy
		(-12,6)*{}="a" + (-1,-1)*{#1};
		(-4,6)*{}="b"+ (-3.5,-1)*{#2};
		(4,6)*{}="c"+ (-3.5,-1)*{#3};
		(12,6)="d" +(-3.5,-1)*{#4};
		(0,-6)*{}="bottom";
		(-6,0)*{}="e"+(-1.3,-1.3)*{#5};
		(-8,2)*{}="ab"+(-1.3,-1.3)*{#6};
		(-4,-2)*{}="cd"+(-1.3,-1.3)*{#7};
		"ab";"a"**\dir{-}?(.7)*\dir{>};	
		"ab";"b"**\dir{-}?(.7)*\dir{>};
		"cd";"c"**\dir{-}?(.75)*\dir{>};
		"bottom";"d"**\dir{-}?(.8)*\dir{>};
		"bottom";"cd"**\dir{-};
		"cd";"ab"**\dir{-}?(.7)*\dir{>};
	\endxy
}

\newcommand{\fouranyonstatecond}{
	\smalllabels
	\xy
		(-12,6)*{}="a" + (-1,-1)*{};
		(-4,6)*{}="b"+ (-3.5,-1)*{};
		(4,6)*{}="c"+ (-3.5,-1)*{};
		(12,6)="d" +(-3.5,-1)*{};
		(0,-6)*{}="bottom";
		(-6,0)*{}="e"+(-1.3,-1.3)*{};
		(-8,2)*{}="ab"+(-1.3,-1.3)*{};
		(-4,-2)*{}="cd"+(-1.3,-1.3)*{};
		"ab";"a"**\dir{--}?(.7)*\dir{};	
		"ab";"b"**\dir{--}?(.7)*\dir{};
		"cd";"c"**\dir{--}?(.75)*\dir{};
		"bottom";"d"**\dir{--}?(.8)*\dir{};
		"bottom";"cd"**\dir{--};
		"cd";"ab"**\dir{--}?(.7)*\dir{};
	\endxy
}

\newcommand{\braidket}[4]{
	\smalllabels
	\xy
		(-4.24,4.24)*{}="b" +(-2,-1)*{#2};
		(4.24,4.24)*{}="a"+(2,-1)*{#1};
		(0,1)="cross";
		(0,-7)*{}="c";
		(0,-2.5)*{}="mid" +(2,-1)*{#4};
		"c"; "mid"**\dir{-}?(.7)*\dir{>}+(-2,-3)*{#3};
		"mid";"cross"**\crv{"mid"+(-2,2)};
		"cross";"a"**\dir{-}?(.9)*\dir{>};
		"mid";"cross"+(.5,-.5)**\crv{"mid"+(2,2)};
		"cross"+(-.5,.5);"b"**\dir{-}?(.9)*\dir{>};
	\endxy 
}

\newcommand{\condensateRbraidket}[4]{
	\smalllabels
	\xy
		(-4.24,4.24)*{}="b" +(-2,-1)*{#2};
		(4.24,4.24)*{}="a"+(2,-1)*{#1};
		(0,1)="cross";
		(0,-7)*{}="c";
		(0,-2.5)*{}="mid" +(2,-1)*{#4};
		"c"; "mid"**\dir{-}?(.7)*\dir{>}+(-2,-3)*{#3};
		"mid";"cross"**\crv{"mid"+(-2,2)};
		"cross";"a"**\dir{--}?(.9)*\dir{};
		"mid";"cross"+(.5,-.5)**\crv{"mid"+(2,2)};
		"cross"+(-.5,.5);"b"**\dir{-}?(.9)*\dir{>};
	\endxy 
}

\newcommand{\braidbra}[4]{
	\smalllabels
	\xy
		(-4.24,-4.24)*{}="b" +(-2,1)*{#2};
		(4.24,-4.24)*{}="a"+(2,1)*{#1};
		(0,-1)="cross";
		(0,7)*{}="c"+(-3,-2)*{#3};
		(0,2.5)*{}="mid" +(2,1)*{#4};
		"c"; "mid"**\dir{-}?(.2)*\dir{<};
		"mid";"cross"**\crv{"mid"+(-2,-2)};
		"cross";"a"**\dir{-}?(.6)*\dir{<};
		"mid";"cross"+(.5,.5)**\crv{"mid"+(2,-2)};
		"cross"+(-.5,-.5);"b"**\dir{-}?(.6)*\dir{<};
	\endxy 
}

\newcommand{\braid}[2]{
	\smalllabels
	\xy
		(-4,-5)*{}="a1"+(-1,2)*{#1};
		(4,5)*{}="a2";
		(4,-5)*{}="b1"+(1,2)*{#2};
		(-4,5)*{}="b2";
		(0,0)="mid";
		"a1";"mid"**\dir{-};
		"mid";"a2"**\dir{-}?(.7)*\dir{>};
		"b1";"mid"+(.5,-.5)**\dir{-};
		"mid"+(-.5,.5);"b2"**\dir{-}?(.7)*\dir{>};
	\endxy 
}

\newcommand{\antibraid}[2]{
	\smalllabels
	\xy
		(-4,-5)*{}="a1"+(-1,2)*{#1};
		(4,5)*{}="a2";
		(4,-5)*{}="b1"+(1,2)*{#2};
		(-4,5)*{}="b2";
		(0,0)="mid";
		"a1";"mid"+(-.5,-.5)**\dir{-};
		"mid"+(.5,.5);"a2"**\dir{-}?(.7)*\dir{>};
		"b1";"mid"**\dir{-};
		"mid";"b2"**\dir{-}?(.7)*\dir{>};
	\endxy 
}

\newcommand{\doublebraidket}[4]{
	\smalllabels
	\xy
		(0,-12)="B"+(-2,1)*{#3};
		(0,-6)="M1"+(2,-1)*{#4};
		(-4,-3)="L1";
		(4,-3)="R1";
		(0,0)="M2";
		(-4,3)="L2";
		(4,3)="R2";
		(0,6)="M3";
		(-4,9)="L3"+(-2,-1)*{#1};
		(4,9)="R3"+(2,-1)*{#2};
		"B";"M1"**\dir{-}?(.7)*\dir{>};
		"M1";"L1"**\dir{-};
		"M1";"R1"**\dir{-};
		"L1";"R2"**\dir{-}?(.5)*{\hole}="hole1";
		"R1";"hole1"**\dir{-};"L2"**\dir{-};
		"L2";"R3"**\dir{-}?(.8)*\dir{>}?(.5)*{\hole}="hole2";
		"R2";"hole2"**\dir{-};"L3"**\dir{-}?(.8)*\dir{>};
	\endxy 
}
\newcommand{\doublebraid}[2]{
	\smalllabels
	\xy
		(-4,-8)="L1"+(-2,-1)*{#1};
		(4,-8)="R1"+(2,-1)*{#2};
		"L1"+(0,6)="L2";
		"R1"+(0,6)="R2";
		"L2"+(0,6)="L3";
		"R2"+(0,6)="R3";
		"L1";"R2"**\crv{"L1"+(3,4)&"R2"+(-3,-4)}?(.5)*+{\hole}="hole1";
		"L2";"R3"**\crv{"L2"+(3,4)&"R3"+(-3,-4)}?(.5)*+{\hole}="hole2";
		"R1";"hole1"**\crv{"R1"+(-1,2)};
		"L2"**\crv{"L2"+(0,-3)};
		"R2";"hole2"**\crv{"R2"+(-1,2)};
		"L3"**\crv{"L3"+(0,-3)};
		"L3";"L3"+(0,2)**\dir{-}?(.6)*\dir{>};
		"R3";"R3"+(0,2)**\dir{-}?(.6)*\dir{>};
		"L1"+(0,-2);"L1"**\dir{-};
		"R1"+(0,-2);"R1"**\dir{-};
	\endxy 
}

\newcommand{\twistright}[1]{
	\smalllabels
	\xy 
	(0,-8)*{}="T";
	(0,8)*{}="B";
	(0,-4)*{}="T'"+(-2,-2)*{#1};
	(0,6)*{}="B'";
	(3,0)*{}="MB";
	(7,0)*{}="LB";
	"T'";"T" **\dir{-};
	"B'";"B" **\dir{-}?(.4)*\dir{>};
	"T'";"LB" **\crv{(1,5) & (5,5)}; \POS?(.25)*{\hole}="2z";
	"LB"; "2z" **\crv{(8,-4) & (2,-4)};
	"2z"; "B'" **\crv{(0,2)};
\endxy 
}

\newcommand{\twistleft}[1]{
	\smalllabels
	\xy 
	(0,8)*{}="T";
	(0,-8)*{}="B";
	(0,5)*{}="T'";
	(0,-5)*{}="B'"+(-2,0)*{#1};
	(-3,0)*{}="MB";
	(-7,0)*{}="LB";
	"T'";"T" **\dir{-}?(.3)*\dir{>};
	"B";"B'" **\dir{-};
	"T'";"LB" **\crv{(-1,-3) & (-5,-4)}; \POS?(.25)*{\hole}="2z";
	"LB"; "2z" **\crv{(-8,6) & (-2,6)};
	"2z"; "B'" **\crv{(0,-3)};
\endxy 
}

\newcommand{\antitwistright}[1]{
	\smalllabels
	\xy 
	(0,8)*{}="T";
	(0,-8)*{}="B";
	(0,5)*{}="T'";
	(0,-5)*{}="B'"+(-2,0)*{#1};
	(3,0)*{}="MB";
	(7,0)*{}="LB";
	"T'";"T" **\dir{-}?(.3)*\dir{>};
	"B";"B'" **\dir{-};
	"T'";"LB" **\crv{(1,-3) & (5,-4)}; \POS?(.25)*{\hole}="2z";
	"LB"; "2z" **\crv{(8,6) & (2,6)};
	"2z"; "B'" **\crv{(0,-3)};
\endxy 
}

\newcommand{\antitwistleft}[1]{
	\smalllabels
	\xy 
	(0,-8)*{}="T";
	(0,8)*{}="B";
	(0,-4)*{}="T'"+(-2,-2)*{#1};
	(0,6)*{}="B'";
	(-3,0)*{}="MB";
	(-7,0)*{}="LB";
	"T'";"T" **\dir{-};
	"B'";"B" **\dir{-}?(.4)*\dir{>};
	"T'";"LB" **\crv{(-1,5) & (-5,5)}; \POS?(.25)*{\hole}="2z";
	"LB"; "2z" **\crv{(-8,-4) & (-2,-4)};
	"2z"; "B'" **\crv{(0,2)};
\endxy 
}

\newcommand{\condensate}{
	\smalllabels
	\xy
		(0,6)*{}="T";
		(0,-6)*{}="B";
		"B"*{};"T"**\dir{--}?(.6)*\dir{} + (2,-3)*{};
	\endxy
}

\newcommand{\longcondensate}{
	\smalllabels
	\xy
		(0,8)*{}="T";
		(0,-8)*{}="B"+(-2,3)*{};
		"B"*{};"T"**\dir{--}?(.6)*\dir{}; 
	\endxy
}

\newcommand{\condensatevertexket}{
	\smalllabels
	\xy
		(-4.24,4.24)*{}="a";
		(4.24,4.24)*{}="b";
		(0,-6)*{}="c";
		(0,0)*{}="mid" +(2,-1)*{};
		"mid"; "a" **\dir{--}?(.75)*\dir{} + (-2,0)*{};
		"mid"; "b" **\dir{--}?(.75)*\dir{} + (2,0)*{};
		"c"; "mid"**\dir{--}?(.7)*\dir{}+(-2,-3)*{};
	\endxy 
}

\newcommand{\condensatebraket}{
	\smalllabels
	\xy
		(-4.24,0)*{}="a";
		(4.24,0)*{}="b";
		(0,-10)*{}="c";
		(0,-4)*{}="mid" +(2,-1)*{};
		"mid"; "a" **\dir{--}?(.75)*\dir{} + (-2,0)*{};
		"mid"; "b" **\dir{--}?(.75)*\dir{} + (2,0)*{};
		"c"; "mid"**\dir{--}?(.7)*\dir{}+(-2,-3)*{};
		(-4.24,0)*{}="a'";
		(4.24,0)*{}="b'";
		(0,10)*{}="c'";
		(0,4)*{}="mid'" +(2,-1)*{};
		"mid'"; "a'" **\dir{--}?(.75)*\dir{} + (-2,0)*{};
		"mid'"; "b'" **\dir{--}?(.75)*\dir{} + (2,0)*{};
		"c'"; "mid'"**\dir{--}?(.7)*\dir{}+(-2,-3)*{};
	\endxy 
}

\newcommand{\condensatebraidket}{
	\smalllabels
	\xy
		(-4.24,4.24)*{}="b" +(-2,-1)*{};
		(4.24,4.24)*{}="a"+(2,-1)*{};
		(0,1)="cross";
		(0,-7)*{}="c";
		(0,-2.5)*{}="mid" +(2,-1)*{};
		"c"; "mid"**\dir{--}?(.7)*\dir{}+(-2,-3)*{};
		"mid";"cross"**\crv{"mid"+(-2,2)};
		"cross";"a"**\dir{--}?(.9)*\dir{};
		"mid";"cross"+(.7,-.7)**\crv{"mid"+(2,2)};
		"cross"+(-.7,.7);"b"**\dir{--}?(.9)*\dir{};
	\endxy 
}

\newcommand{\condensatedoublebraidket}{
	\smalllabels
	\xy
		(0,-12)="B";
		(0,-6)="M1";
		(-4,-3)="L1";
		(4,-3)="R1";
		(0,0)="M2";
		(-4,3)="L2";
		(4,3)="R2";
		(0,6)="M3";
		(-4,9)="L3";
		(4,9)="R3";
		"B";"M1"**\dir{--};
		"M1";"L1"**\dir{--};
		"M1";"R1"**\dir{--};
		"L1";"R2"**\dir{--}?(.5)*{\hole}="hole1";
		"R1";"hole1"**\dir{--};"L2"**\dir{--};
		"L2";"R3"**\dir{--}?(.5)*{\hole}="hole2";
		"R2";"hole2"**\dir{--};"L3"**\dir{--};
		
	\endxy 
}

\newcommand{\condensatetwototwoS}{
	\smalllabels
	\xy
		(-2,6)*{}="a" + (-2,-1)*{};
		(2,6)*{}="b"+ (2,-1)*{};
		(-2,-6)*{}="c"+ (-2,1)*{};
		(2,-6)*{}="d"+ (2,1)*{};
		(0,0)="e"+ (0,2)*{};
		(-2,2)*{}="e2"+(-1.5,-.5)*{};
		(2,-2)*{}="e1"+(1.5,.5)*{};
		"c";"a"**\dir{--};
		"c"+(0,3)*\dir{}+(0,8)*\dir{};
		"d";"b"**\dir{--};
		"d"+(0,3)*\dir{}+(0,8)*\dir{};
		"e1";"e2"**\dir{--}?(.7)*\dir{} ;		
	\endxy
}

\newcommand{\condensatetwototwoSanyonLvacR}[1]{
	\smalllabels
	\xy
		(-2,6)*{}="a" + (-2,-1)*{};
		(2,6)*{}="b"+ (2,-1)*{};
		(-2,-6)*{}="c"+ (-2,1)*{#1};
		(2,-6)*{}="d"+ (2,1)*{};
		(0,0)="e"+ (0,2)*{};
		(-2,2)*{}="e2"+(-1.5,-.5)*{};
		(2,-2)*{}="e1"+(1.5,.5)*{};
		"c";"a"**\dir{-};
		"c"+(0,3)*\dir{>}+(0,8)*\dir{>};
		"d";"e1"**\dir{};
		"e1";"b"**\dir{--};
		"d"+(0,3)*\dir{}+(0,8)*\dir{};
		"e1";"e2"**\dir{--}?(.7)*\dir{} ;		
	\endxy
}

\newcommand{\condensatetwototwoM}{
	\smalllabels
	\xy
		(-4,6)*{}="a"+(-1,-1)*{};
		(4,6)*{}="b"+(1,-1)*{};
		(-4,-6)*{}="c"+(-1,1)*{};
		(4,-6)*{}="d"+(1,1)*{};
		(0,0)*{}="e"+(-2,0)*{};
		(0,2)*{}="e2"+(2,0)*{};
		(0,-2)*{}="e1"+(2,0)*{};
		"e2";"a"**\dir{--}?(.7)*\dir{} ;
		"e2";"b"**\dir{--}?(.7)*\dir{} ;
		"c";"e1"**\dir{--}?(.7)*\dir{} ;
		"d";"e1"**\dir{--}?(.7)*\dir{} ;
		"e1";"e2"**\dir{--}?(.7)*\dir{};
	\endxy
}

\newcommand{\condensatetwototwoSanyonR}[2]{
	\smalllabels
	\xy
		(-2,6)*{}="a"; + (-2,-1)*{};
		(2,6)*{}="b"+ (2,-1)*{#1};
		(-2,-6)*{}="c"+ (-2,1)*{};
		(2,-6)*{}="d"+ (2,1)*{#2};
		(0,0)="e"+ (0,2)*{};
		(-2,2)*{}="e2"+(-1.5,-.5)*{};
		(2,-2)*{}="e1"+(1.5,.5)*{};
		"c";"a"**\dir{--};
		"c"+(0,3)*\dir{}+(0,8)*\dir{};
		"d";"b"**\dir{-};
		"d"+(0,3)*\dir{>}+(0,8)*\dir{>};
		"e1";"e2"**\dir{--}?(.7)*\dir{} ;		
	\endxy
}

\newcommand{\condensatetwototwoManyonR}[3]{
	\smalllabels
	\xy
		(-4,6)*{}="a"+(-1,-1)*{};
		(4,6)*{}="b"+(1,-1)*{#1};
		(-4,-6)*{}="c"+(-1,1)*{};
		(4,-6)*{}="d"+(1,1)*{#2};
		(0,0)*{}="e"+(2,0)*{#3};
		(0,2)*{}="e2"+(2,0)*{};
		(0,-2)*{}="e1"+(2,0)*{};
		"e2";"a"**\dir{--}?(.7)*\dir{} ;
		"e2";"b"**\dir{-}?(.7)*\dir{>} ;
		"c";"e1"**\dir{--}?(.7)*\dir{} ;
		"d";"e1"**\dir{-}?(.7)*\dir{>} ;
		"e1";"e2"**\dir{-}?(.7)*\dir{>};
	\endxy
}

\newcommand{\condensatetwototwoSanyonL}[1]{
	\smalllabels
	\xy
		(-2,6)*{}="a" + (-2,-1)*{#1};
		(2,6)*{}="b"+ (2,-1)*{};
		(-2,-6)*{}="c"+ (-2,1)*{#1};
		(2,-6)*{}="d"+ (2,1)*{};
		(0,0)="e"+ (0,2)*{};
		(-2,2)*{}="e2"+(-1.5,-.5)*{};
		(2,-2)*{}="e1"+(1.5,.5)*{};
		"c";"a"**\dir{-};
		"c"+(0,3)*\dir{>}+(0,8)*\dir{>};
		"d";"b"**\dir{--};
		"d"+(0,3)*\dir{}+(0,8)*\dir{};
		"e1";"e2"**\dir{--}?(.7)*\dir{} ;		
	\endxy
}

\newcommand{\condensateLonetotwoSanyonbirth}[1]{
	\smalllabels
	\xy
		(-2,6)*{}="a"; + (-2,-1)*{};
		(2,6)*{}="b"+ (2,-1)*{};
		(-2,-6)*{}="c"+ (-2,1)*{};
		(2,-6)*{}="d"+ (2,1)*{};
		(0,0)="e"+ (0,2)*{};
		(-2,2)*{}="e2"+(-1.5,-.5)*{#1};
		(2,-2)*{}="e1"+(1.5,.5)*{};
		"c";"e2"**\dir{--};
		"e2";"a"**\dir{-}+(0.1,-1.5)*\dir{>};
		"c"+(0,3)*\dir{}+(0,8)*\dir{};
		"e1";"b"**\dir{-}+(0.1,-3.5)*\dir{<};;
		"d"+(0,3)*\dir{}+(0,8)*\dir{};
		"e1";"e2"**\dir{-}?(.7)*\dir{>} ;		
	\endxy
}

\newcommand{\anyonLonetotwoS}[4]{
	\smalllabels
	\xy
		(-2,6)*{}="a"; + (-4,5)*{#2};
		(2,6)*{}="b"+ (2,-1)*{#3};
		(-2,-6)*{}="c"+ (-2,1)*{#1};
		(2,-6)*{}="d"+ (2,1)*{};
		(0,0)="e"+ (0,2)*{#4};
		(-2,2)*{}="e2"+(-1.5,-.5)*{};
		(2,-2)*{}="e1"+(1.5,.5)*{};
		"c";"e2"**\dir{-}?(.7)*\dir{>};
		"e2";"a"**\dir{-}+(0.1,-1.5)*\dir{>};
		"c"+(0,3)*\dir{}+(0,8)*\dir{};
		"e1";"b"**\dir{-}+(0.1,-3.5)*\dir{<};;
		"d"+(0,3)*\dir{}+(0,8)*\dir{};
		"e1";"e2"**\dir{-}?(.7)*\dir{>} ;		
	\endxy
}

\newcommand{\condensatetwototwoManyonL}[3]{
	\smalllabels
	\xy
		(-4,6)*{}="a"+(-1,-1)*{#1};
		(4,6)*{}="b"+(1,-1)*{};
		(-4,-6)*{}="c"+(-1,1)*{#2};
		(4,-6)*{}="d"+(1,1)*{};
		(0,0)*{}="e"+(2,0)*{#3};
		(0,2)*{}="e2"+(2,0)*{};
		(0,-2)*{}="e1"+(2,0)*{};
		"e2";"a"**\dir{-}?(.7)*\dir{>} ;
		"e2";"b"**\dir{--}?(.7)*\dir{} ;
		"c";"e1"**\dir{-}?(.7)*\dir{>} ;
		"d";"e1"**\dir{--}?(.7)*\dir{} ;
		"e1";"e2"**\dir{-}?(.7)*\dir{>};
	\endxy
}

\newcommand{\condensatetwotooneManyondeath}[1]{
	\smalllabels
	\xy
		(-4,6)*{}="a"+(-1,-1)*{};
		(0,6)*{}="b"+(1,-1)*{};
		(-4,-6)*{}="c"+(-1,1)*{#1};
		(4,-6)*{}="d"+(1,1)*{};
		(0,0)*{}="e"+(2,0)*{};
		(0,2)*{}="e2"+(2,0)*{};
		(0,-2)*{}="e1"+(2,0)*{};
		"e2";"a"**\dir{}?(.7)*\dir{} ;
		"e1";"b"**\dir{--}?(.7)*\dir{} ;
		"c";"e1"**\dir{-}?(.7)*\dir{>} ;
		"d";"e1"**\dir{-}?(.7)*\dir{>} ;
	\endxy
}

\newcommand{\condensatetwototwoSanyonvertexR}[3]{
	\smalllabels
	\xy
		(-2,6)*{}="a" + (-2,-1)*{#1};
		(2,6)*{}="b"+ (2,-1)*{#2};
		(-2,-6)*{}="c"+ (-2,1)*{};
		(2,-6)*{}="d"+ (2,1)*{#3};
		(0,0)="e"+ (0,2)*{};
		(-2,2)*{}="e2"+(-1.5,-.5)*{};
		(2,-2)*{}="e1"+(1.5,.5)*{};
		"c";"e2"**\dir{--};
		"e2";"a"**\dir{-}?(.7)*\dir{>};
		"c"+(0,3)*\dir{}+(0,8)*\dir{};
		"d";"b"**\dir{-};
		"d"+(0,3)*\dir{>}+(0,8)*\dir{>};
		"e1";"e2"**\dir{-}?(.7)*\dir{>} ;		
	\endxy
}

\newcommand{\condensatetwototwoManyonvertexR}[3]{
	\smalllabels
	\xy
		(-4,6)*{}="a"+(-1,-1)*{#1};
		(4,6)*{}="b"+(1,-1)*{#2};
		(-4,-6)*{}="c"+(-1,1)*{};
		(4,-6)*{}="d"+(1,1)*{#3};
		(0,0)*{}="e"+(2,0)*{};
		(0,2)*{}="e2"+(2,0)*{};
		(0,-2)*{}="e1"+(2,0)*{};
		"e2";"a"**\dir{-}?(.7)*\dir{>} ;
		"e2";"b"**\dir{-}?(.7)*\dir{>} ;
		"c";"e1"**\dir{--}?(.7)*\dir{} ;
		"d";"e1"**\dir{-}?(.7)*\dir{>} ;
		"e1";"e2"**\dir{-}?(.7)*\dir{>};
	\endxy
}

\newcommand{\condensatetwototwoSanyonbend}[1]{
	\smalllabels
	\xy
		(-2,6)*{}="a" + (-2,-1)*{};
		(2,6)*{}="b"+ (2,-1)*{};
		(-2,-6)*{}="c"+ (-2,1)*{#1};
		(2,-6)*{}="d"+ (2,1)*{};
		(0,0)="e"+ (0,2)*{};
		(-2,2)*{}="e2"+(-1.5,-.5)*{};
		(2,-2)*{}="e1"+(1.5,.5)*{};
		"c";"e2"**\dir{-};
		"e2";"a"**\dir{--};
		"c"+(0,3)*\dir{>}+(0,8)*\dir{};
		"d";"e1"**\dir{--};
		"e1";"b"**\dir{-};
		"d"+(0,3)*\dir{}+(0,8)*\dir{>};
		"e1";"e2"**\dir{-}?(.7)*\dir{>} ;		
	\endxy
}	

\newcommand{\condensatetwototwoManyonstraight}[1]{
	\smalllabels
	\xy
		(-4,6)*{}="a"+(-1,-1)*{};
		(4,6)*{}="b"+(1,-1)*{};
		(-4,-6)*{}="c"+(-1,1)*{};
		(4,-6)*{}="d"+(1,1)*{#1};
		(0,0)*{}="e"+(2,0)*{};
		(0,2)*{}="e2"+(2,0)*{};
		(0,-2)*{}="e1"+(2,0)*{};
		"e2";"a"**\dir{-}?(.7)*\dir{>} ;
		"e2";"b"**\dir{--}?(.7)*\dir{} ;
		"c";"e1"**\dir{--}?(.7)*\dir{} ;
		"d";"e1"**\dir{-}?(.7)*\dir{>} ;
		"e1";"e2"**\dir{-}?(.7)*\dir{>};
	\endxy
}

\newcommand{\condensatetwototwoManyonstraightMirror}[1]{
	\smalllabels
	\xy
		(-4,6)*{}="a"+(-1,-1)*{};
		(4,6)*{}="b"+(1,-1)*{};
		(-4,-6)*{}="c"+(-1,1)*{#1};
		(4,-6)*{}="d"+(1,1)*{};
		(0,0)*{}="e"+(2,0)*{};
		(0,2)*{}="e2"+(2,0)*{};
		(0,-2)*{}="e1"+(2,0)*{};
		"e2";"a"**\dir{--}?(.7)*\dir{} ;
		"e2";"b"**\dir{-}?(.7)*\dir{>} ;
		"c";"e1"**\dir{-}?(.7)*\dir{>} ;
		"d";"e1"**\dir{--}?(.7)*\dir{} ;
		"e1";"e2"**\dir{-}?(.7)*\dir{>};
	\endxy
}

\newcommand{\condensatetwototwoSanyonstraight}[1]{
	\smalllabels
	\xy
		(-2,6)*{}="a" + (-2,-1)*{};
		(2,6)*{}="b"+ (2,-1)*{};
		(-2,-6)*{}="c"+ (-2,1)*{};
		(2,-6)*{}="d"+ (2,1)*{#1};
		(0,0)="e"+ (0,2)*{};
		(-2,2)*{}="e2"+(-1.5,-.5)*{};
		(2,-2)*{}="e1"+(1.5,.5)*{};
		"c";"e2"**\dir{--};
		"e2";"a"**\dir{-};
		"c"+(0,3)*\dir{}+(0,8)*\dir{>};
		"d";"e1"**\dir{-};
		"e1";"b"**\dir{--};
		"d"+(0,3)*\dir{>}+(0,8)*\dir{};
		"e1";"e2"**\dir{-}?(.7)*\dir{>} ;		
	\endxy
}	

\newcommand{\condensatetwototwoManyonbirth}[1]{
	\smalllabels
	\xy
		(-4,6)*{}="a"+(-1,-1)*{#1};
		(4,6)*{}="b"+(1,-1)*{};
		(-4,-6)*{}="c"+(-1,1)*{};
		(4,-6)*{}="d"+(1,1)*{};
		(0,0)*{}="e"+(2,0)*{};
		(0,2)*{}="e2"+(2,0)*{};
		(0,-2)*{}="e1"+(2,0)*{};
		"e2";"a"**\dir{-}?(.7)*\dir{>} ;
		"e2";"b"**\dir{-}?(.7)*\dir{>} ;
		"c";"e1"**\dir{--}?(.7)*\dir{} ;
		"d";"e1"**\dir{--}?(.7)*\dir{} ;
		"e1";"e2"**\dir{--}?(.7)*\dir{};
	\endxy
}

\newcommand{\condensatetwototwoMcondensateLdown}[4]{
	\smalllabels
	\xy
		(-4,6)*{}="a"+(-1,-1)*{#1};
		(4,6)*{}="b"+(1,-1)*{#2};
		(-4,-6)*{}="c"+(-1,1)*{};
		(4,-6)*{}="d"+(1,1)*{#3};
		(0,0)*{}="e"+(2,0)*{#4};
		(0,2)*{}="e2"+(2,0)*{};
		(0,-2)*{}="e1"+(2,0)*{};
		"e2";"a"**\dir{-}?(.7)*\dir{>} ;
		"e2";"b"**\dir{-}?(.7)*\dir{>} ;
		"c";"e1"**\dir{--}?(.7)*\dir{} ;
		"d";"e1"**\dir{-}?(.7)*\dir{>} ;
		"e1";"e2"**\dir{-}?(.7)*\dir{>};
	\endxy
}

\newcommand{\condensatetwototwoMcondensateRup}[4]{
	\smalllabels
	\xy
		(-4,6)*{}="a"+(-1,-1)*{#1};
		(4,6)*{}="b"+(1,-1)*{};
		(-4,-6)*{}="c"+(-1,1)*{#2};
		(4,-6)*{}="d"+(1,1)*{#3};
		(0,0)*{}="e"+(2,0)*{#4};
		(0,2)*{}="e2"+(2,0)*{};
		(0,-2)*{}="e1"+(2,0)*{};
		"e2";"a"**\dir{-}?(.7)*\dir{>} ;
		"e2";"b"**\dir{--}?(.7)*\dir{} ;
		"c";"e1"**\dir{-}?(.7)*\dir{>} ;
		"d";"e1"**\dir{-}?(.7)*\dir{>} ;
		"e1";"e2"**\dir{-}?(.7)*\dir{>};
	\endxy
}

\newcommand{\condensatetwototwoSanyonbirth}[1]{
	\smalllabels
	\xy
		(-2,6)*{}="a" + (-2,-1)*{#1};
		(2,6)*{}="b"+ (2,-1)*{};
		(-2,-6)*{}="c"+ (-2,1)*{};
		(2,-6)*{}="d"+ (2,1)*{};
		(0,0)="e"+ (0,2)*{};
		(-2,2)*{}="e2"+(-1.5,-.5)*{};
		(2,-2)*{}="e1"+(1.5,.5)*{};
		"c";"e2"**\dir{--};
		"e2";"a"**\dir{-};
		"c"+(0,3)*\dir{}+(0,8)*\dir{>};
		"d";"e1"**\dir{--};
		"e1";"b"**\dir{-};
		"d"+(0,3)*\dir{}+(0,8)*\dir{>};
		"e1";"e2"**\dir{-}?(.7)*\dir{>} ;		
	\endxy
}	

\newcommand{\condensatetwototwoScondensateLdown}[4]{
	\smalllabels
	\xy
		(-2,6)*{}="a" + (-2,-1)*{#1};
		(2,6)*{}="b"+ (2,-1)*{#2};
		(-2,-6)*{}="c"+ (-2,1)*{};
		(2,-6)*{}="d"+ (2,1)*{#3};
		(0,0)="e"+ (0,2)*{#4};
		(-2,2)*{}="e2"+(-1.5,-.5)*{};
		(2,-2)*{}="e1"+(1.5,.5)*{};
		"c";"e2"**\dir{--};
		"e2";"a"**\dir{-};
		"c"+(0,3)*\dir{}+(0,8)*\dir{>};
		"d";"e1"**\dir{-}?(.7)*\dir{>};
		"e1";"b"**\dir{-};
		"d"+(0,3)*\dir{}+(0,8)*\dir{>};
		"e1";"e2"**\dir{-}?(.7)*\dir{>} ;		
	\endxy
}	

\newcommand{\condensatetwototwoScondensateRup}[4]{
	\smalllabels
	\xy
		(-2,6)*{}="a" + (-2,-1)*{#1};
		(2,6)*{}="b"+ (2,-1)*{};
		(-2,-6)*{}="c"+ (-2,1)*{#2};
		(2,-6)*{}="d"+ (2,1)*{#3};
		(0,0)="e"+ (0,2)*{#4};
		(-2,2)*{}="e2"+(-1.5,-.5)*{};
		(2,-2)*{}="e1"+(1.5,.5)*{};
		"c";"e2"**\dir{-}?(.7)*\dir{>};
		"e2";"a"**\dir{-};
		"c"+(0,3)*\dir{}+(0,8)*\dir{>};
		"d";"e1"**\dir{-}?(.7)*\dir{>};
		"e1";"b"**\dir{--};
		"d"+(0,3)*\dir{}+(0,8)*\dir{};
		"e1";"e2"**\dir{-}?(.7)*\dir{>} ;		
	\endxy
}

\newcommand{\condensatetwototwoScondensateMiddle}[4]{
	\smalllabels
	\xy
		(-2,6)*{}="a" + (-2,-1)*{#1};
		(2,6)*{}="b"+ (2,-1)*{#2};
		(-2,-6)*{}="c"+ (-2,1)*{#3};
		(2,-6)*{}="d"+ (2,1)*{#4};
		(0,0)="e"+ (0,2)*{};
		(-2,2)*{}="e2"+(-1.5,-.5)*{};
		(2,-2)*{}="e1"+(1.5,.5)*{};
		"c";"e2"**\dir{-}?(.7)*\dir{>};
		"e2";"a"**\dir{-}?(.7)*\dir{>};;
		"c"+(0,3)*\dir{}+(0,8)*\dir{};
		"d";"e1"**\dir{-}?(.7)*\dir{>};
		"e1";"b"**\dir{-}?(.7)*\dir{>};
		"d"+(0,3)*\dir{}+(0,8)*\dir{};
		"e1";"e2"**\dir{--}?(.7)*\dir{} ;		
	\endxy
}

\newcommand{\condensatetwototwoManyondeath}[1]{
	\smalllabels
	\xy
		(-4,6)*{}="a"+(-1,-1)*{};
		(4,6)*{}="b"+(1,-1)*{};
		(-4,-6)*{}="c"+(-1,1)*{#1};
		(4,-6)*{}="d"+(1,1)*{};
		(0,0)*{}="e"+(2,0)*{};
		(0,2)*{}="e2"+(2,0)*{};
		(0,-2)*{}="e1"+(2,0)*{};
		"e2";"a"**\dir{--}?(.7)*\dir{} ;
		"e2";"b"**\dir{--}?(.7)*\dir{} ;
		"c";"e1"**\dir{-}?(.7)*\dir{>} ;
		"d";"e1"**\dir{-}?(.7)*\dir{>} ;
		"e1";"e2"**\dir{--}?(.7)*\dir{};
	\endxy
}

\newcommand{\condensatetwototwoSanyondeath}[1]{
	\smalllabels
	\xy
		(-2,6)*{}="a" + (-2,-1)*{};
		(2,6)*{}="b"+ (2,-1)*{};
		(-2,-6)*{}="c"+ (-2,1)*{#1};
		(2,-6)*{}="d"+ (2,1)*{};
		(0,0)="e"+ (0,2)*{};
		(-2,2)*{}="e2"+(-1.5,-.5)*{};
		(2,-2)*{}="e1"+(1.5,.5)*{};
		"c";"e2"**\dir{-};
		"e2";"a"**\dir{--};
		"c"+(0,3)*\dir{>}+(0,8)*\dir{};
		"d";"e1"**\dir{-}?(.7)*\dir{>};
		"e1";"b"**\dir{--};
		"d"+(0,3)*\dir{}+(0,8)*\dir{};
		"e1";"e2"**\dir{-}?(.7)*\dir{>} ;		
	\endxy
}

\newcommand{\condensatetwototwoSanyonRight}[1]{
	\smalllabels
	\xy
		(-2,6)*{}="a" + (-2,-1)*{};
		(2,6)*{}="b"+ (2,-1)*{};
		(-2,-6)*{}="c"+ (-2,1)*{};
		(2,-6)*{}="d"+ (2,1)*{#1};
		(0,0)="e"+ (0,2)*{};
		(-2,2)*{}="e2"+(-1.5,-.5)*{};
		(2,-2)*{}="e1"+(1.5,.5)*{};
		"c";"e2"**\dir{--};
		"e2";"a"**\dir{};
		"c"+(0,3)*\dir{}+(0,8)*\dir{};
		"d";"e1"**\dir{-};
		"e1";"b"**\dir{-};
		"d"+(0,3)*\dir{>}+(0,8)*\dir{>};
		"e1";"e2"**\dir{--}?(.7)*\dir{} ;		
	\endxy
}

\newcommand{\condensatetwotooneRSanyondeath}[1]{
	\smalllabels
	\xy
		(-2,6)*{}="a" + (-2,-1)*{};
		(2,6)*{}="b"+ (2,-1)*{};
		(-2,-6)*{}="c"+ (-2,1)*{#1};
		(2,-6)*{}="d"+ (2,1)*{};
		(0,0)="e"+ (0,2)*{};
		(-2,2)*{}="e2"+(-1.5,-.5)*{};
		(2,-2)*{}="e1"+(1.5,.5)*{};
		"c";"e2"**\dir{-};
		"e2";"a"**\dir{};
		"c"+(0,3)*\dir{>}+(0,8)*\dir{};
		"d";"e1"**\dir{-};
		"e1";"b"**\dir{--};
		"d"+(0,3)*\dir{>}+(0,8)*\dir{};
		"e1";"e2"**\dir{-}?(.7)*\dir{>} ;		
	\endxy
}

\newcommand{\twotooneR}[4]{
	\smalllabels
	\xy
		(-2,6)*{}="a" + (-2,-1)*{};
		(2,6)*{}="b"+ (2,-1)*{#3};
		(-2,-6)*{}="c"+ (-2,1)*{#1};
		(2,-6)*{}="d"+ (2,1)*{#4};
		(0,0)="e"+ (0,2)*{#2};
		(-2,2)*{}="e2"+(-1.5,-.5)*{};
		(2,-2)*{}="e1"+(1.5,.5)*{};
		"c";"e2"**\dir{-};
		"e2";"a"**\dir{};
		"c"+(0,3)*\dir{>}+(0,8)*\dir{};
		"d";"e1"**\dir{-};
		"e1";"b"**\dir{-}?(.7)*\dir{>};
		"d"+(0,3)*\dir{>}+(0,8)*\dir{};
		"e1";"e2"**\dir{-}?(.7)*\dir{>} ;		
	\endxy
}

\newcommand{\condensatebubblemeetsanyon}[1]{
	\smalllabels
	\xy
		(-3,6)*{}="a"+(-1,-1)*{};
		(0.5,4)*{}="a1";
		(-3,3)*{}="a2";
		(4,6)*{}="b"+(1,-1)*{#1};
		(-3,-6)*{}="c"+(-1,1)*{};
		(0.5,-4)*{}="c2";
		(-3,-3)*{}="c1";
		(4,-6)*{}="d"+(1,1)*{};
		(0,0)*{}="e"+(2,0)*{};
		(4,2)*{}="e2"+(2,0)*{};
		(4,-2)*{}="e1"+(2,0)*{};
		"e2";"a1"**\dir{--}?(.7)*\dir{} ;
		"a2";"c1"**\dir{--}?(.7)*\dir{} ;
		"e2";"b"**\dir{-}?(.7)*\dir{} ;
		"c2";"e1"**\dir{--}?(.7)*\dir{} ;
		"d";"e1"**\dir{-}?(.7)*\dir{} ;
		"e1";"e2"**\dir{-}?(.7)*\dir{>};
		"a1"+(-1,1);"a2"+(0,1)**\crv{"a"};
		"c1"+(0,-1);"c2"+(-1,-1)**\crv{"c"};
	\endxy
}

\newcommand{\condensatebubblemeetsanyonT}[1]{
	\smalllabels
	\xy
		(-3,0)*{}="dx";	
		(0,.5)*{}="dy";
		(-1,3)*{}="a" + (-2,-1)*{};
		"a"+"dx"*{}="aa";
		(2,6)*{}="b"+ (2,-1)*{#1};
		(-1,-2)*{}="c"+ (-2,1)*{};
		"c"+"dx"*{}="cc";
		(2,-6)*{}="d"+ (2,1)*{};
		(0,0)="e"+ (0,2)*{};
		(-1,1)*{}="e2"+(-1.5,-.5)*{};
		(2,-2)*{}="e1"+(1.5,.5)*{};
		"c"+"dy";"a"-"dy"**\dir{--}?(.7)*\dir{};
		"cc"+"dy";"aa"-"dy"**\dir{--}?(.7)*\dir{};
		"a"+"dy";"aa"+"dy"**\crv{"a"+(0,3)&"aa"+(0,3)};
		"c"-"dy";"cc"-"dy"**\crv{"c"+(0,-3)&"cc"+(0,-3)};
		"c"+(0,3)*\dir{}+(0,8)*\dir{};		
		"d";"b"**\dir{-};
		"d"+(0,3)*\dir{>}+(0,8)*\dir{>};
		"e1";"e2"**\dir{--}?(.7)*\dir{} ;		
	\endxy
}

\newcommand{\condensatemeetsanyon}[1]{
	\smalllabels
	\xy%
		(-4.24,4.24)*{}="a";
		(4.24,4.24)*{}="b";
		(0,-6)*{}="c";
		(0,0)*{}="mid" +(2,-1)*{};
		"mid"; "a" **\dir{--}?(.75)*\dir{};
		"mid"; "b" **\dir{-}?(.75)*\dir{>} + (4,0)*{#1};
		"c"; "mid"**\dir{-}?(.7)*\dir{>}+(4,-3)*{#1};
	\endxy
}

\newcommand{\condensatemeetsanyonlong}[1]{
	\smalllabels
	\xy%
		(-6,6)*{}="a";
		(6,6)*{}="b";
		(0,-8)*{}="c";
		(0,0)*{}="mid" +(2,-1)*{};
		"mid"; "a" **\dir{--}?(.75)*\dir{};
		"mid"; "b" **\dir{-}?(.75)*\dir{>} + (2,0)*{#1};
		"c"; "mid"**\dir{-}?(.7)*\dir{>}+(2,-3)*{#1};
	\endxy
}

\newcommand{\condensatemeetsanyondoublebraidket}[1]{
	\smalllabels
	\xy
		(0,-12)="B"+(-2,1)*{#1};
		(0,-6)="M1"+(2,-1)*{};
		(-6,-3)="L1";
		(6,-3)="R1";
		(0,0)="M2";
		(-6,3)="L2";
		(6,3)="R2";
		(0,6)="M3";
		(-6,9)="L3"+(-2,-1)*{};
		(6,9)="R3"+(2,-1)*{#1};
		"B";"M1"**\dir{-}?(.7)*\dir{>};
		"M1";"L1"**\dir{--};
		"M1";"R1"**\dir{-};
		"L1";"R2"**\dir{--}?(.5)*{\hole}="hole1";
		"R1";"hole1"**\dir{-};"L2"**\dir{-};
		"L2";"R3"**\dir{-}?(.8)*\dir{>}?(.5)*{\hole}="hole2";
		"R2";"hole2"**\dir{--};"L3"**\dir{--}?(.8)*\dir{};
	\endxy 
}

\newcommand{\smatrix}[2]{
	\smalllabels
	\xy
		(-6,0)="L1";
		(-2,4)="T1";
		(2,0)="R1";
		(-2,-4)="B1"+(-3,-1)*{#1};
		(-2,0)="L2";
		(2,4)="T2";
		(6,0)="R2";
		(2,-4)="B2"+(3,-1)*{#2};
		"R1";"B1"**\crv{"R1"+(0,-2)&"B1"+(2,0)}?(.6)*{\hole}="lower";
		"B1";"L1"**\crv{"B1"+(-2,0)&"L1"+(0,-2)};
		"L1";"T1"**\crv{"L1"+(0,2)&"T1"+(-2,0)};
		"L2";"T2"**\crv{"L2"+(0,2)&"T2"+(-2,0)}?(.6)*{\hole}="higher";
		"T2";"R2"**\crv{"T2"+(2,0)&"R2"+(0,2)};
		"R2";"B2"**\crv{"R2"+(0,-2)&"B2"+(2,0)};
		"L2";"lower"**\crv{"L2"+(0,-2)"};
		"R1";"higher"**\crv{"R1"+(0,2)"};
		"B2";"lower"**\crv{"B2"+(-1,0)};
		"T1";"higher"**\crv{"T1"+(1,0)};
		"L2"+(0,-2);"L2"+(0,2)**\dir{}?(.7)*\dir{>};
		"R1"+(0,-2);"R1"+(0,2)**\dir{}?(.7)*\dir{>};
	\endxy
}

\newcommand{\sbar}[3]{{
	\smalllabels
	 \xy
		(-9,9)*{}="L1";
		(-3,9)*{}="L2";
		(-9,3)*{}="L3";
		(-3,3)*{}="L4";
		(-9,-3)*{}="L5";
		(-3,-3)*{}="L6";
		(-9,-9)*{}="L7";
		(-3,-9)*{}="L8";
		(9,9)*{}="R1";
		(3,9)*{}="R2";
		(9,3)*{}="R3";
		(3,3)*{}="R4";
		(9,-3)*{}="R5";
		(3,-3)*{}="R6";
		(9,-9)*{}="R7";
		(3,-9)*{}="R8";
		"L2"*{};"L1"**\crv{"L2"+(-1,5)&"L1"+(1,5)};
		"L7"**\dir{-};
		"L8"**\crv{"L7"+(1,-5)&"L8"+(-1,-5)};
		"R2"*{};"R1"**\crv{"R2"+(1,5)&"R1"+(-1,5)};
		"R7"**\dir{-};
		"R8"**\crv{"R7"+(-1,-5)&"R8"+(1,-5)};
		\vtwist~{"L4"}{"R4"}{"L6"}{"R6"};
		\vtwist~{"L6"}{"R6"}{"L8"}{"R8"};
		"L4"*{};"L2"**\dir{-}?(.5)*\dir{>} + (-3,0)*{#1};
		"R4"*{};"R2"**\dir{-}?(.5)*\dir{>} + (+3,0)*{#2};
		"R4"*{};"L2"**\dir{-}?(.7)*\dir{>} + (1.5,2)*{#3};
	\endxy}
}

\newcommand{\scondensate}[2]{
	\smalllabels
	 \xy
		(-9,9)*{}="L1";
		(-3,9)*{}="L2";
		(-9,3)*{}="L3";
		(-3,3)*{}="L4";
		(-9,-3)*{}="L5";
		(-3,-3)*{}="L6";
		(-9,-9)*{}="L7";
		(-3,-9)*{}="L8";
		(9,9)*{}="R1";
		(3,9)*{}="R2";
		(9,3)*{}="R3";
		(3,3)*{}="R4";
		(9,-3)*{}="R5";
		(3,-3)*{}="R6";
		(9,-9)*{}="R7";
		(3,-9)*{}="R8";
		"L2"*{};"L1"**\crv{"L2"+(-1,5)&"L1"+(1,5)};
		"L7"**\dir{-};
		"L8"**\crv{"L7"+(1,-5)&"L8"+(-1,-5)};
		"R2"*{};"R1"**\crv{"R2"+(1,5)&"R1"+(-1,5)};
		"R7"**\dir{-};
		"R8"**\crv{"R7"+(-1,-5)&"R8"+(1,-5)};
		\vtwist~{"L4"}{"R4"}{"L6"}{"R6"};
		\vtwist~{"L6"}{"R6"}{"L8"}{"R8"};
		"L4"*{};"L2"**\dir{-}?(.5)*\dir{>} + (-3,0)*{#1};
		"R4"*{};"R2"**\dir{-}?(.5)*\dir{>} + (+3,0)*{#2};
		"R4"*{};"L2"**\dir{--}?(.7)*\dir{} + (1.5,2)*{};
	\endxy
}

\newcommand{\trefoil}[1]{
	\smalllabels
	\xy
	(0,0)="M";
	(0,7)="T";
	(-8,-7)="L";
	(8,-7)="R";
	(-4,0)="l"+(-3,1)*{#1};
	(4,0)="r";
	(0,-7)="b";
	"T";"l"+(0,1)**\crv{(-4,7)};	
	"T";"r"**\crv{(4,7)};	
	"r";"b"+(1,1)**\crv{(4,-3)};	
	"L";"b"+(-1,-1)**\crv{(-6,-9)};	
	"L";"l"**\crv{(-10,-3)};
	"l";"r"+(-1,.5)**\crv{(0,2)};
	"R";"r"+(1,-.5)**\crv{(10,-3)};
	"R";"b"**\crv{(6,-9)};
	"b";"l"+(0,-1)**\crv{(-4,-3)};
	"L"+(0,4.5);"l"**\dir{}?(.6)*\dir{>}
\endxy
}

\newcommand{\qdimtwist}[2]{
	\smalllabels
	\xy
		(-4,0)="L"+(-2,0)*{#1};
		(4,2)="Rt";
		(4,-2)="Rb";
		(8,0)="R";
		(0,7)="T";
		(0,-7)="B";
		"L";"T"*\crv{"L"+(0,2)&"T"+(-2,0)};
		"T";"Rt"*\crv{(4,7)&(4,3)};
		"Rb";"B"*\crv{"Rb"+(0,-2)&"B"+(2,0)};
		"B";"L"*\crv{"B"+(-2,0)&"L"+(0,-2)};
		"L"+(.05,1)*\dir{#2};
		"Rb";"R"*\crv{"R"+(-2,4)};
		"R";"R"+(-2,-2)*\crv{"R"+(0,-2)};
	\endxy
}

\newcommand{\projector}[1]{
	\smalllabels
	\xy
	(0,-12)="B"+(2,1)*{#1};
	(0,-1.5)="Bh1";
	(0,.5)="Bh2";
	(0,10)="T"+(2,-1)*{};
	(0,-6)="B'";
	(-4,-2)="b";
	(4,1.8)="r";
	(1,3.8)="rh1";
	(-2,4.8)="rh2";
	(0,8)="l";
	(-4,6)="t";
	"B";"Bh1"**\dir{-}?(.3)*\dir{>};
	"Bh2";"T"**\dir{-}?(.7)*\dir{>};
	"B'";"b"**\dir{--};
	"b";"r"**\dir{--};
	"l";"t"**\dir{--};
	"r";"rh1"**\dir{--};
	"rh2";"t"**\dir{-};
	\endxy
}

\newcommand{\halfbraidcond}[1]{
	\smalllabels
	\xy
	(0,-12)="B"+(2,1)*{#1};
	(0,-1.5)="Bh1";
	(0,.5)="Bh2";
	(0,10)="T"+(2,-1)*{};
	(0,-6)="B'";
	(-4,-2)="b";
	(4,1.8)="r";
	(0,3.8)="rh1";
	(-2,4.8)="rh2";
	(0,8)="l";
	(-4,6)="t";
	"B";"Bh1"**\dir{-}?(.3)*\dir{>};
	"Bh2";"T"**\dir{-}?(.7)*\dir{>};
	"B'";"b"**\dir{--};
	"b";"r"**\dir{--};
	"r";"rh1"**\dir{--};
	\endxy
}

\newcommand{\vlcinvariancecondensateA}[3]{
	\smalllabels
	\xy
		(-6,6)*{}="a" + (-2,-1)*{#1};
		(2,6)*{}="b"+ (2,-1)*{#2};
		(-2,-10)*{}="c"+ (-2,1)*{#3};
		(-2,-7)*{}="cc";
		(2,-3)*{}="d"+ (2,1)*{};
		(0,0)="e"+ (0,2)*{};
		(-2,-2)*{}="e2";
		(-6,2)*{}="e3";
		(2,2)*{}="e1"+(1.5,.5)*{};
		"c";"e2"**\dir{-}?(.8)*\dir{>};
		"e3";"a"**\dir{-};
		"e2";"e3"**\dir{-}**\dir{-}?(.7)*\dir{>};
		"e1";"b"**\dir{-};
		"d";"e1"**\dir{--};
		"cc";"d"**\dir{--};
		"d"+(0,3)*\dir{}+(0,8)*\dir{};
		"e2";"e1"**\dir{-}?(.7)*\dir{>} ;		
	\endxy
}	

\newcommand{\vlcinvariancecondensateB}[3]{
	\smalllabels
		\xy
		(-6,6)*{}="a" + (-2,-1)*{#1};
		(2,6)*{}="b"+ (2,-1)*{#2};
		(-2,-10)*{}="c"+ (-2,1)*{#3};
		(-2,-7)*{}="cc";
		(-6,-3)*{}="d"+ (2,1)*{};
		(0,0)="e"+ (0,2)*{};
		(-2,-2)*{}="e2";
		(-6,2)*{}="e3";
		(2,2)*{}="e1"+(1.5,.5)*{};
		"c";"e2"**\dir{-}?(.8)*\dir{>};
		"e3";"a"**\dir{-};
		"e2";"e3"**\dir{-}**\dir{-}?(.7)*\dir{>};
		"e1";"b"**\dir{-};
		"d";"e3"**\dir{--};
		"cc";"d"**\dir{--};
		"d"+(0,3)*\dir{}+(0,8)*\dir{};
		"e2";"e1"**\dir{-}?(.7)*\dir{>} ;		
	\endxy
}

\newcommand{\vlcinvariancecondensateC}[3]{
	\smalllabels
	\xy
		(0,-4)*{}="a";
		(0,-8)*{}="b"+(2,0)*{#3};
		(-4,0)*{}="c"+(-2,1)*{#1};
		(4,0)*{}="d"+(2,1)*{#2};
		(-4,4)*{}="e";
		(-4,8)*{}="f";
		(4,8)*{}="g";
		"b";"a"**\dir{-}?(.7)*\dir{>};
		"a";"c"**\dir{-}?(.7)*\dir{>};
		"c";"e"**\dir{-};
		"e";"f"**\dir{-}?(.7)*\dir{>};
		"a";"d"**\dir{-}?(.7)*\dir{>};
		"d";"g"**\dir{-}?(.7)*\dir{>};
		"e";"d"**\dir{--};
	\endxy
}

\newcommand{\tadpole}[2]{
\smalllabels
	\xy
		(-4,0)="L"+(-2,0)*{#1};
		(4,0)="R";
		(0,4)="T"+(-1.5,2)*{#2};
		(0,-4)="B";
		"L";"T"*\crv{"L"+(0,2)&"T"+(-2,0)};
		"T";"R"*\crv{"T"+(2,0)&"R"+(0,2)};
		"R";"B"*\crv{"R"+(0,-2)&"B"+(2,0)};
		"B";"L"*\crv{"B"+(-2,0)&"L"+(0,-2)};
		"L"+(.05,1)*\dir{>};
		"T";"T"+(0,4)**\dir{-}?(.7)*\dir{>};
	\endxy
}

\newcommand{\qdimvar}[1]{
\smalllabels
	\xy
		(-4,0)="L"+(-.5,3)*{#1};
		(4,0)="R";
		(0,4)="T";
		(0,-4)="B";
		"L";"T"*\crv{"L"+(0,2)&"T"+(-2,0)};
		"T";"R"*\crv{"T"+(2,0)&"R"+(0,2)};
		"R";"B"*\crv{"R"+(0,-2)&"B"+(2,0)};
		"B";"L"*\crv{"B"+(-2,0)&"L"+(0,-2)};
		"L"+(.05,1)*\dir{>};
	\endxy
}

\newcommand{\XaboveYcond}{
	\smalllabels
	\xy
 (0,-2)="X";
  (0,2)="Y";
  (5,0)="l";
  (0,5)="h";
  (0,2.5)="d";
  "X";"X"-"l"**\dir{-};"X"-"l"-"h"**\dir{-};"X"+"l"-"h"**\dir{-};"X"+"l"**\dir{-};"X"**\dir{-};"X"-"d"*{X};
   "Y";"Y"-"l"**\dir{-};"Y"-"l"+"h"**\dir{-};"Y"+"l"+"h"**\dir{-};"Y"+"l"**\dir{-};"Y"**\dir{-};"Y"+"d"*{Y};
   "X";"Y"**\dir{--};
  \endxy
}

\newcommand{\XaboveY}{
	\smalllabels
	\xy
 (0,-2)="X";
  (0,2)="Y";
  (5,0)="l";
  (0,5)="h";
  (0,2.5)="d";
  "X";"X"-"l"**\dir{-};"X"-"l"-"h"**\dir{-};"X"+"l"-"h"**\dir{-};"X"+"l"**\dir{-};"X"**\dir{-};"X"-"d"*{X};
   "Y";"Y"-"l"**\dir{-};"Y"-"l"+"h"**\dir{-};"Y"+"l"+"h"**\dir{-};"Y"+"l"**\dir{-};"Y"**\dir{-};"Y"+"d"*{Y};
  \endxy
}

\newcommand{\XbesidesYcond}{
	\smalllabels
	\xy
 (-8,2.5)="X";
  (8,-2.5)="Y";
  (5,0)="l";
  (0,5)="h";
  (0,2.5)="d";
  "X";"X"-"l"**\dir{-};"X"-"l"-"h"**\dir{-};"X"+"l"-"h"**\dir{-};"X"+"l"**\dir{-};"X"**\dir{-};"X"-"d"*{X};
   "Y";"Y"-"l"**\dir{-};"Y"-"l"+"h"**\dir{-};"Y"+"l"+"h"**\dir{-};"Y"+"l"**\dir{-};"Y"**\dir{-};"Y"+"d"*{Y};
   "X"+"l" +(0,-2);"Y"-"l" +(0,2)**\dir{--};
  \endxy
}

\newcommand{\XbesidesY}{
	\smalllabels
	\xy
 (-8,2.5)="X";
  (8,-2.5)="Y";
  (5,0)="l";
  (0,5)="h";
  (0,2.5)="d";
  "X";"X"-"l"**\dir{-};"X"-"l"-"h"**\dir{-};"X"+"l"-"h"**\dir{-};"X"+"l"**\dir{-};"X"**\dir{-};"X"-"d"*{X};
   "Y";"Y"-"l"**\dir{-};"Y"-"l"+"h"**\dir{-};"Y"+"l"+"h"**\dir{-};"Y"+"l"**\dir{-};"Y"**\dir{-};"Y"+"d"*{Y};
  \endxy
}

\newcommand{\twototwoMcondMbubbleL}[2]{
	\smalllabels
	\xy
		(-4,8)*{}="a" + (-2,-1)*{#1};
		(4,8)*{}="b"+ (2,-1)*{#2};
		(-4,-8)*{}="c"+ (-2,1)*{};
		(4,-8)*{}="d"+ (2,1)*{};
		(0,0)="e"+ (0,2)*{};
		(-4,0)*{}="e2"+(-1.5,-.5)*{};
		(4,-4)*{}="e1"+(1.5,.5)*{};
		"c";"e2"**\dir{-};
		"e2";"a"**\dir{-};
		"c"+(0,3)*\dir{>}+(0,10)*\dir{>};
		"d";"e1"**\dir{-};
		"e1";"b"**\dir{-};
		"d"+(0,3)*\dir{>}+(0,10)*\dir{>};
		"e1";"e2"**\dir{--}?(.7)*\dir{} ;		
		"a"+(0,-2)="b1";
		"b1"+(5,-2)="b2";
		"a"+(0,-7)="b3";
		"b1";"b2"**\dir{--};
		"b3";"b2"**\dir{--};
	\endxy
}

\newcommand{\twototwoMcondMbubbleR}[2]{
	\smalllabels
	\xy
		(-4,8)*{}="a" + (-2,-1)*{#1};
		(4,8)*{}="b"+ (2,-1)*{#2};
		(-4,-8)*{}="c"+ (-2,1)*{};
		(4,-8)*{}="d"+ (2,1)*{};
		(0,0)="e"+ (0,2)*{};
		(-4,0)*{}="e2"+(-1.5,-.5)*{};
		(4,-4)*{}="e1"+(1.5,.5)*{};
		"c";"e2"**\dir{-};
		"e2";"a"**\dir{-};
		"c"+(0,3)*\dir{>}+(0,10)*\dir{>};
		"d";"e1"**\dir{-};
		"e1";"b"**\dir{-};
		"d"+(0,3)*\dir{>}+(0,10)*\dir{>};
		"e1";"e2"**\dir{--}?(.7)*\dir{} ;		
		"b"+(0,-2)="b1";
		"b1"+(-5,-2)="b2";
		"b"+(0,-7)="b3";
		"b1";"b2"**\dir{--};
		"b3";"b2"**\dir{--};
	\endxy
}

\newcommand{\twototwoMcondMdouble}[2]{
	\smalllabels
	\xy
		(-4,8)*{}="a" + (-2,-1)*{#1};
		(4,8)*{}="b"+ (2,-1)*{#2};
		(-4,-8)*{}="c"+ (-2,1)*{};
		(4,-8)*{}="d"+ (2,1)*{};
		(0,0)="e"+ (0,2)*{};
		(-4,0)*{}="e2"+(-1.5,-.5)*{};
		(4,-4)*{}="e1"+(1.5,.5)*{};
		(-4,2)*{}="ee2"+(-1.5,-.5)*{};
		(4,-2)*{}="ee1"+(1.5,.5)*{};
		"c";"e2"**\dir{-};
		"e2";"a"**\dir{-};
		"c"+(0,3)*\dir{>}+(0,10)*\dir{>};
		"d";"e1"**\dir{-};
		"e1";"b"**\dir{-};
		"d"+(0,3)*\dir{>}+(0,10)*\dir{>};
		"e1";"e2"**\dir{--}?(.7)*\dir{} ;		
		"ee1";"ee2"**\dir{--}?(.7)*\dir{} ;		
	\endxy
}

\newcommand{\conddeathL}[1]{
	\smalllabels
	\xy
		(-4,-4)*{}="A"+(-1.5,1.5)*{#1};
		(4,-4)*{}="B"+(1.5,1.5)*{};
		(0,4)*{}="t";
		"t" + (-6,+4)="C";
		"t";"A"**\crv{"t"+(-4,-.1),}?(.8)*\dir{<}?(.6);"C"**\dir{--};
		"t";"B"**\crv{"t"+(4,-.1),};?(.8)*\dir{};
	\endxy 
}

\newcommand{\conddeathR}[1]{
	\smalllabels
	\xy
		(-4,-4)*{}="A"+(-1.5,1.5)*{#1};
		(4,-4)*{}="B"+(1.5,1.5)*{};
		(0,4)*{}="t";
		"t" + (-6,+4)="C";
		"t"+(-3,-1.2);"A"**\crv{"t"+(-4,-2),}?(.8)*\dir{<};
		"t"+(-.5,0);"B"**\crv{"t"+(4,-.1),};?(.8)*\dir{}?(.9);"C"**\dir{--};
	\endxy 
}

\renewcommand{\[}{\begin{equation}}
\renewcommand{\]}{\end{equation}}

\newcommand{\braket}[2]{\left\langle #1\,\right|\left.#2 \right\rangle}

\newcommand{\x}{\times}
\renewcommand{\mod}[1]{{\ (\mathrm{mod}\ #1})}

\newcommand{\cat}[1]{\mathcal{#1}}
\newcommand{\sA}{\mathcal{A}}
\newcommand{\sT}{\mathcal{T}}
\newcommand{\sU}{\mathcal{U}}

\newcommand{\QD}{\mathcal{D}}

\newcommand{\F}[3]{[ F^{#1}_{#2}]_{#3}}
\newcommand{\Fvar}[4]{[( F^{#1}_{#2})^{#4}]_{#3}}

\newcommand{\phase}[1]{\varkappa_{#1}}

\newcommand{\q}[1]{[#1]_q}

\DeclareMathOperator{\qTr}{\widehat{Tr}}

\makeatletter
\renewcommand*\env@matrix[1][*\c@MaxMatrixCols c]{%
  \hskip -\arraycolsep
  \let\@ifnextchar\new@ifnextchar
  \array{#1}}
\makeatother

\newcommand{\qsixj}[6]{\left\{\begin{array}{c c | c} 	#1	&#2	&#5\\	#3&	#4	&#6\end{array} \right\}  }

\newcommand{\vlc}[6]{\begin{bmatrix}[cc|l] #1 & #2 & #3 \\ #4 & #5 & #6 \end{bmatrix}}

\newcommand{\nn}{\nonumber}

\newcommand{\bra}[1]{\langle #1|}
\newcommand{\ket}[1]{|#1\rangle}

\newcommand{\be}{\begin{equation}}
\newcommand{\ee}{\end{equation}}

\begin{document}


\title{Diagrammatics for Bose condensation in anyon theories}


\author{I.S. Eli\"{e}ns}
\email{i.s.eliens@uva.nl}
\affiliation{Institute for Theoretical Physics, University of Amsterdam, Science Park 904, P.O.Box 94485, 1090 GL Amsterdam, The Netherlands}
\author{J.C. Romers}
\affiliation{Rudolf Peierls Centre for Theoretical Physics, University of Oxford, 1 Keble Road, OX1 3NP, United Kingdom}
\author{F.A. Bais}
\affiliation{Institute for Theoretical Physics, University of Amsterdam, Science Park 904, P.O.Box 94485, 1090 GL Amsterdam, The Netherlands}
\affiliation{Santa Fe Institute, Santa Fe, NM 87501, USA}


\date{\today}

\begin{abstract}
We reformulate the topological symmetry breaking scheme for
 phase transitions in systems with anyons in a grahical manner.
 A  new set of quantities called vertex lifiting
coefficients  (VLCs) is introduced and used to specify the  the full
operator content of the broken phase.  First, it is shown how the assumption
that a set of charges behaves like the vacuum of a new theory
naturally leads to diagrammatic consistency conditions for a
condensate. This recovers the notion of a condensate used in earlier
aproaches and uncovers the connection to pure mathematics.  
The VLCs  are needed to solve the consistency conditions and  establish the mapping of the
fusion and splitting spaces of the broken theory into the parent phase.
This enables one to calculate the full set of topological data ($S$-,
$T$-, $R$- and $F$-matrices) for the condensed phase and closed form
expressions  in terms of the VLCs are
provided. We furthermore  furnish a cocrete recipe  to lift arbitrary
diagrams directly from the condensed phase to the original phase using
only a limited number of VLCs and we describe a method for the
explicit calculation of VLCs for a large class of bosonic condensates.
This allows for the explicit calculation of condensed-phase diagrams
in many physically relevant cases and representative examples are
worked out in detail.
\end{abstract}


\maketitle


\section{Introduction}
The classification of possible phases of matter is at the heart of condensed matter physics.
Conventionally this is linked to the notion of symmetry breaking characterized by a non-vanishing  vacuum expectation value of certain local order parameters. Examples include the superconducting gap function or the magnetization vector.

In the past decade it has become evident that this is not the whole story. Different integer quantum
Hall phases for example exhibit different macroscopic physics (in particular, their Hall conductivity)
while all of them appear like a structureless electron liquid on the microscopic level \cite{PhysRevLett.45.494}.
In a seminal paper \cite{Thouless:1982zz}, Thouless et al. argued that the distinction between these phases cannot be made with a 
local order parameter, but can be captured by a non-local quantity --- a topological invariant that is
obtained by integrating Berry flux over the Brillouin zone. Hence the names `topological orders' or 
`topological phases of matter' which are now commonly used to describe such phases that cannot be captured by
local order parameters.

A lot of progress has been made identifying such topological order parameters for different
systems. For free fermion systems this has led to an elegant `periodic table'-like classification \cite{Kitaev:2009mg,Ryu:2010zza},
containing phases of matter such as the quantum spin Hall state \cite{PhysRevLett.95.226801,PhysRevLett.96.106802},
the integer quantum Hall state and the $p+ip$ superconductor.
More generally the quest for observables that identify the type of topological order has spawned among others such quantities as topological entanglement entropies and spectra, whereas we have in previous work\cite{Bais:2011iu} proposed the topological $S$-matrix.

One particular class of interacting topological phases has received an
extraordinary amount of attention: those with anyonic excitations
in (2+1)-dimensions, not the least because they have been proposed as a way to realize fault-tolerant
quantum computing \cite{Kitaev:1997wr}. From a field theory point of
view two families of models have been identified:
Chern-Simons theories, which are closely related to the mathematics of knot theory and the Jones polynomial \cite{Witten:1988hf},
and discrete gauge theories \cite{Bais:1991pe}, of which Kitaev's toric code\cite{Kitaev:1997wr} is a close relative. Different lattice realizations 
of (non-chiral varieties of) these theories have been constructed, for example Levin-Wen models \cite{Levin:2004mi, Burnell20102550} 
or discrete lattice gauge theories \cite{Bais:2011iu}.

The study of the phase structure of an anyon model can be pursued along two complementary directions. First, one
could start from a lattice model and add perturbing terms to its Hamiltonian. These terms
can drive a phase transition in the system which can be studied using Monte Carlo methods \cite{Tupitsyn:2008ah,Bais:2011iu},
perturbative expansions \cite{PhysRevLett.110.147203} or mappings to exactly solvable models \cite{PhysRevB.84.125434}. 

Some  of these phase transitions can be attributed to the formation of a Bose condensate in the original theory.   Much of the physics of these transitions is independent of the 
underlying lattice realization and this opens up a second approach
taking the  knowledge of the topological quantum field theory (TQFT)
as a starting point.  It turns out that from that topological data one can indeed determine the low-energy effective TQFT in the presence of the bosonic condensate --- and this is the program we pursue further in this paper.

For discrete gauge theories (DGTs), where the particle sectors can be understood as irreducible representations (irreps) of an underlying
quantum group \cite{Bais:1991pe} (the quantum double $D(H)$ of a finite group $H$), it amounts to a Higgs-type effect and as such is really a case of symmetry breaking: the quantum group gets reduced and the low-energy TQFT is given by the irreps of this smaller quantum group \cite{Bais2003}.
For CS theories the situation is more complex:  these too can be understood in terms of quantum groups \cite{Slingerland2001229}
 --- quantum deformations of $SU(N)$ --- but identifying the TQFT after Bose condensation is more involved.
In Ref.~\onlinecite{Bais2009} this problem was tackled, not by looking
at the group explicitly, but rather by studying the breaking on the
level of  the fusion algebra of the excitations. It was shown that
by demanding that the Bose condensate acts as a vacuum sector and requiring commutativity and associativity
of the fusion algebra, an effective theory in the condensed phase can
often be identified.
However, what was found was a consistent fusion algebra  --- but not
the full set of all TQFT data for the condensed phase. In particular,
the $F$- matrices that implement an associator on the level of the
fusion spaces were not derived in this framework. Also were there some
cases where the identification of the broken phase remained ambiguous.

The present work aims to fill in that gap. We show how the fusion spaces in the condensed theory and its parent theory
are related. For this purpose, we introduce a new set of numbers called vertex lifting coefficients (VLCs) and
we show how the full set of topological data in the Bose condensed phase can be expressed in terms of the 
topological data of the original phase and these VLCs. We also provide a scheme to calculate these numbers
explicitly for a number of examples. As a byproduct all ambiguities
are in principle removed.

From a physics point of view, this work is relevant for a number of
reasons. First, it is theoretically
pleasing to extend the method of  
Ref.~\onlinecite{Bais2009} to include all topological data and allow
for the expression of all operators using only data of the unbroken
theory. This  provides a much deeper understanding of
Bose condensation in the context of topological phases.
It turns out that fixing  a more precise set of consistency conditions
has subtle  consequences. For example,  not all condensates
that one would naively consider  are actually allowed and we show in
an  example how this follows from an easy consistency check.
In the same time  many  links to other topics become clear, in pure
maths as discussed below, but more importantly also to other
important notions  in the study of topological phases,
such as  Levin-Wen models\cite{Levin:2004mi} and patterns of long-range entanglement\cite{2010_Chen_PRB_82},
where the same structures as here appear. 

A particular benefit of the current approach is that all operators
are lifted to the Hilbert space of the parent phase making. This makes the
connection to lattice studies much simpler. In earlier work \cite{Bais:2011iu}---a
numerical study of discrete gauge theories on a lattice---we found
that the correct identification of observables in the broken phase
needs coefficients we now identify as the VLCs. This was solved in an
\emph{ad hoc} manner at the time, but the VLCs clearly provide the
correct quantities for the general case and should show up in similar studies.

Determining the $F$- and $R$-matrices of
a theory from scratch, starting from the fusion algebra is a daunting task: it involves solving the pentagon and hexagon equations and has only been done for
theories with a handful of different particles (see
Ref.~\onlinecite{Bonderson20082709}).
At the same time  the $F$-symbols serve as the input for Levin-Wen
lattice models\cite{Levin:2004mi} which play an important
role in the study of topological phases.
  These models of effective
string-net degrees of freedom  provide fixed point
Hamiltonians and wavefunctions for a large class of topological
orders. The present work presents a route for obtaining a whole
family of consistent fusion data starting from a parent theory by
condensing the  bosons in the theory.  
This forms an alternative route to quantum group methods
\cite{Ardonne2010,1988_Kirillov_inProceedings} and can give access to
more exotic theories.

An important  open question is to what extend the universal properties
of the critical points separating topological phases are determined by
the topological order on either side. Microscopic studies indicate
interesting universality classes
\cite{BurnellSimonSlingerland,PhysRevB.84.125434,2014_Schulz_PRB_89,2013_Schulz_PRL_110},
but the inherent nonlocality of the orderparameters obstructs the
adaptation of conventional field theoretical methods to study
\emph{e.g} critical exponents. It remains to be seen if the algebraic
structure of  TQFT  can  be exploited to build a theory for the
critical behavior, but we envision the present work to play an
important role in future progress.  

In a similar category fall questions concerning  universal topological quantum
computation (UTQC) in the condensed phase. This and similar questions clearly needs knowledge of
the operators of the broken theory and not just the particle content.
More model independent considerations also fall in the general scope of the present work.

The formalism presented in this work is  easily adapted to other contexts. 
As an example we point  the extensive use of the notions we present from
a preprint of the present work in a recent formulation for a  unified framework for topological phases with
symmetry\cite{2014arXiv1402.3356G}.

A rigorous  mathematical analysis of some of what we present here, goes back to work of Kirillov and Ostrik \cite{KirillovOstrik2002,2013arXiv1307.8244K}. 
The consistency conditions on the condensate which we formulate turn
out to be  equivalent to the definition of a commutative seperable
Frobenius algebra in the context of unitary braided tensor
categories. In Ref.~\onlinecite{2013arXiv1307.8244K}  Kong identifies
this as the relevant mathematical structure for anyon condensation by performing a bootstrap analysis based on physically justifiable assumptions, and as such forms an interesting bridge between the physical and mathematical literature. 
In the mathematics or mathematical physics literature, the study of (commutative) Frobenius algebras in the context of tensor categories has received considerable attention, both abstractly~\cite{2006_Froehlich_AM_199,2003_Mueger_JPAA_180} and in relation to boundaries and defects in CFTs and TQFTs~\cite{2013_Fuchs_CMP_321,2011_Fuchs_NPB_843,2003_Fuchs_FP_51}.  In the work presented here we develop a diagrammatic formalism based on the clear physical picture of condensation, reducing the heavy language of the underlying mathematics to a minimum. This exceeds earlier approaches in being both simple and intuitive and in the same time completely general and computationally powerful. Indeed, this is much facilitated by the introduction of the VLCs, which to our knowledge have up to now not been exploited in the literature.

The remainder of this paper is organized as follows. In
Sec.~\ref{sec.cond} we give a more extensive introduction to anyons in
the presence of a Bose condensate. We outline the general properties
of the condensation transition and the topological symmetry breaking
scheme for condensation. We then discuss 
the graphical formalism for  anyon models and  the
relation of anyon condensation to other topics in physics and mathematics.
In Sec.~\ref{sec:diagrams} we go into the details of topological symmetry breaking. First we recap how the topological order of the broken phase can be obtained on the level of the particle sectors. Then we reformulate the scheme in a graphical language leading to the  full diagrammatics for Bose condensation in anyon theories.  It is shown that the fusion properties of the condensate play an important role and we formulate the precise conditions on the condensate in order to obtain a consistent theory. It is discussed how the characteristic features of topological symmetry breaking phase transitions---identification, splitting and confinement---appear in this language. Furthermore, we introduce the  vertex lifting coefficients (VLCs) as the crucial ingredient to lift the vertices of the theory. This is the data needed to completely  characterize the topological properties  of the phase transition including the mapping between the operator spaces of the theory. It is explained how they allow one to calculate the full set of topological
data of the broken phase, including $F$- and $R$-matrices.
Section~\ref{sec.diaglift} discusses precisely how to lift arbitrary diagrams from the broken phase to the original phase such that general observables from the broken phase can be computed. Important is the explicit occurrence of vacuum exchange lines (VELs) to incorporate the interaction with the condensate. The quantum dimension diagram and the topological $S$-matrix are discussed as simple applications of this protocol. 
The actual calculation of the VLCs is presented in Section~\ref{sec.calc_vlcs}. Two classes of condensates are worked out in full: one component 
condensates with and without a triple-vacuum vertex. An example of each case is worked out in detail. We also show from the consistency conditions that an expected condensate in five layers of Fibonacci anyons actually will not occur. After
that, we conclude and present some ideas for future research. In the Appendix a short but complete introduction into anyon models, as well as the topological data for
$SU(2)_k$ theories and  a full list of consistency conditions for the VLCs is presented.

%
%
%
%
%
%
%
%
%
%
%
%
%
%

\section{\label{sec.cond}Anyons and Bose condensation}
%

In this paper we study  Bose condensates in a system with anyonic particle-like excitations. In other words, by some mechanism that is beyond the scope of the present work one or more bosonic sectors in phase $\sA$ have gained a vacuum expectation value and we want to determine the effective physics in the presence of the condensate. 
This determination is done in two steps:
\begin{enumerate}
\item[I.]\label{item:1} We first achieve consistent fusion resulting
  in an algebra called $\cat{T}$.
\item[II.]\label{item:2} Then we project out sectors that are confined due to nontrivial braiding with the condensate resulting in a braided theory for the bulk in the broken phase called $\cat{U}$.
\end{enumerate}
 This scheme, known as topological symmetry breaking (TSB), can thus be abbreviated as
\begin{equation}
	\cat{A} \longrightarrow \cat{T} \longrightarrow \cat{U}.
\end{equation}
A detailed account can be found in Ref.~\onlinecite{Bais2009}, with
many worked out examples of the resulting mapping on the level of the
particle sectors of the theory.  An important question left unanswered
in this previous work is how the TSB scheme  extends to the
topological Hilbert space and operators of the theory. Our current aim
is to revisit the TSB scheme, explicitly including the topological Hilbert space and operators  in the discussion. 

The above scenario has a certain common ground with spontaneous symmetry breaking in gauge theories and the  Landau paradigm of second order phase transitions. 
In stark contrast, however, there is no local field that features as an order parameter for the broken phase. 
This should  not be surprising as it is precisely the  absence of local order parameters and the defiance of the Landau paradigm that have spurred the interest in topological phases in the first place.  

Due to the inherent non-locality of topological phases, conventional field theoretical methods fail to study topological phase transitions but the algebraic structures present in TQFT allow one to make progress. 

Quantum groups play an important role this respect. They can account
for the exotic fusion and braiding properties of anyons through their
representation theory and it is tempting view the the anyons as
representation modules of a quantum group, even if there are no
apparent internal degrees of freedom.  This suggests the mechanism of
quantum group symmetry breaking for  phase transitions. Indeed, this
is precisely the idea  underlying the formalism  in
Ref.~\onlinecite{Bais2009} and the present work. It turns out,
however, that in practice one can forget about the underlying symmetry
algebra and work directly on the level of the
excitations. Formally, this means that will work with braided
tensor categories and modular tensor categories (MTCs) in particular,
which encode the algebraic structure present in TQFTs and rational CFTs\cite{1989_Moore_CMP_123}. 

The heavy mathematical language customarily used in the study of MTCs
is  not very appealing for applications in physics. In the spirit of
Kitaev's paper \citep{Kitaev2006} we will therefore get rid of many of
the formalities but use a graphical formalism containing all the
important structure (also called \emph{anyon models} \cite{Bonderson:2007zz}).

\subsection{Graphical formalism for anyon theories}

Let us outline the main aspects of the graphical formalism at this point. A concise but complete 
introduction can be found in Appendix~\ref{sec.rules}. For more details we refer the reader to Refs.~\onlinecite{Kitaev2006,Bonderson:2007zz,Bonderson20082709,Preskill2004}. 

The  idea is to use diagrams to represent the operators that implement  the fusion and braiding of anyons, together with rules for  evaluation. Each anyonic charge $a\in \mathcal{A}$ (we generally use indices $a,b,c,\ldots$ for anyon types of $\mathcal{A}$) can label a directed line segment which can be thought of as the world line of an anyon with charge $a$. Three lines can meet in trivalent vertices representing the fusion or splitting of the anyons. Furthermore, lines can be crossing over or underneath each other representing the braiding of anyons. The elementary building blocks of any diagram are therefore
\begin{equation}
	\anyon{a},\quad \vertexket{a}{b}{c}{}{},\quad \vertexbra{a}{b}{c}{}{},\quad \braid{a}{b},\quad \antibraid{a}{b}
\end{equation}
Vertices are only nonzero when the combination $(a,b,c)$ is  allowed by the fusion rules
\begin{equation}
	a\times b = \sum_{c} N_{ab}^c c
\end{equation}
(\emph{i.e.} $N_{ab}^c \neq 0$).
The fusion coefficients $N_{ab}^c$ are nonnegative integers in general, but we consider only $N_{ab}^c = 0,1$ (otherwise an extra label on the vertex should distinguish the different fusion states). 

The topological Hilbert space spanned by a collection of anyons has a
dimension equal to the number of inequivalent ways one can fuse all
the anyons back to the vacuum (assuming  overall  charge neutrality).
For the case of a large number of anyons of type $a$, the dimension per anyon asymptotically approaches  the largest eigenvalue $d_a$ of the fusion matrix $N_a$ (with coefficients $(N_{a})_{bc} = N_{ab}^c$), and $d_a$ is called the quantum dimension of $a$. 
 It also has the diagrammatic expression
\begin{equation}\label{eq:qdimdiagram}
	d_a\quad =\quad \qdim{ {a }}{>}
\end{equation}
Every anyonic charge in the theory has a unique conjugate charge with which it may annihilate $a\times \bar{a} = 0 +\ldots$ where $0$ is the unique vacuum charge.

Vacuum lines in diagrams are generally left invisible, or when made explicit we draw them dotted. The reason is of course that vacuum lines can be added and removed at will since the vacuum has trivial fusion and braiding. The precise graphical conditions defining the triviality of the vacuum will play an important role in this work.

Since we want to study condensates of bosons within anyon theories, we will be  particularly interested in anyons that have trivial spin. The topological spin factor $\theta_a = \exp (i 2\pi h_a)$  can be used in diagrams to remove twists corresponding to a $2\pi$ rotation of the anyon (in that sense the world lines should be thought of as ribbons rather than lines). 

The original TSB scheme~\cite{Bais2009} 
was developed such that only knowledge of the fusion coefficients $N_{ab}^c$, quantum dimensions $d_a$ and spin $h_a$ is required. Accordingly, one is only able to reconstruct the analogous data for the broken phase.

The full topological data of an anyon model is captured by the so called $F$-symbols and $R$-symbols, encoding the fusion braid properties respectively. There are two equivalent ways of defining $F$-symbols  diagrammatically
\begin{equation}
\label{eq:Fmove}
\begin{split}
   \onetothreeL{a}{b}{c}{d}{e}{}{}  &= 
        \sum_{f} 
        \F{abc}{d}{ef} \onetothreeR{a}{b}{c}{d}{f}{}{}\;\;,\\
        \twototwoS{a}{b}{c}{d}{e}{}{} &= \sum_{f} \F{ab}{cd}{ef}\twototwoM{a}{b}{c}{d}{f}{}{}.
\end{split}
\end{equation}
which relate to one another via
\begin{equation}
	  \F{ab}{cd}{ef} =\sqrt{\frac{d_e d_f}{d_a d_d}} \F{ceb}{f}{ad}^*\;\; .
\end{equation}
An asterisk generally denotes complex conjugation (theories are assumed to be  unitary 
\footnote{Unitarity means that one can take the adjoint of a diagram by mirroring it in the horizontal plane, reversing orientation of the arrows and taking the complex conjugate of all coefficients. See Appendix~\ref{sec.rules}.}
throughout this paper).

The $R$-symbols implement the following diagrammatic equality
\begin{equation}\label{eq:Rmove}
  \braidket{a}{b}{c}{} = R^{ab}_{c} \vertexket{b}{a}{c}{}\;,
\end{equation}
All other topological quantities can be expressed in terms of the $F$- and $R$-symbols, for example
\begin{align}
	  d_a &= |\F{a\bar{a}a}{a}{00}|^{-1},\nonumber\\
	  N_{ab}^c &=  \sqrt{d_a d_b\over d_c}\F{ab}{ab}{0c},\\
	  \theta_a &= \F{0a}{a0}{\bar{a}a}{R^{\bar{a}a}_0}^*.\nonumber
\end{align}
Anyon models can to a great extend be understood as conventional quantum mechanics on the topological Hilbert space. The states of a collection of anyons with labels  $a_1,\dots, a_N$ can be represented graphically as (for $N=4$)
\begin{equation}
	\ket{\psi}=\sum_{c } {\psi^{a_1a_2a_3a_4}_{c} \over (d_{a_1} d_{a_2} d_{a_3} d_{a_4} d_{c})^{1/4}}. \fouranyonstate{a_1}{a_2}{a_3}{a_4}{c}{}{},
\end{equation}
 We denote the vector space (topological Hilbert space) spanned by these states as $V_0^{a_1\ldots a_N}$. Note that the $F$-symbols implement a change of basis in this  space. More generally, the spaces of operators of the theory can be denoted $V_{a_1\ldots a_N}^{b_1 \ldots b_M}$ and are spanned by the diagrams with $N$ charge lines labeled
$(a_1,\ldots,a_N)$ sticking out on the bottom and $M$ charge lines with labels $(b_1,\ldots,b_M)$ sticking out on top. Diagrams denote the same operator when they differ by a sequence of $F$-moves \eqref{eq:Fmove}, $R$-moves \eqref{eq:Rmove} and removing bubbles \eqref{eq:qdimdiagram}. The elementary fusion and splitting spaces $V_{ab}^c$ and $V^{ab}_c$ are special cases 
with dimension $N_{ab}^c$.  When $N_{ab}^c=0$ one  has $V_{ab}^c=V^{ab}_c=0$ so any diagram with the corresponding vertex is necessarily zero.

Let us conclude this subsection with a remark on the Frobenius-Schur indicator of a charge $a$. This quantity, denoted $\phase{a}$,   satisfies the diagrammatic relation
\begin{equation}
	\twototwoSvacLvacR{a}{\bar{a}} = \phase{a}\quad\anyon{a}\;\;.
\end{equation}
and so can be defined as $\phase{a} = \F{0a}{a0}{\bar{a}a}$. It is a phase which can generally be put to 1 by picking a ``gauge'' (a phase freedom in the definition of the vertices), but for self-dual charges it is a ``gauge invariant'' sign $\phase{a} = \pm 1$. We will assume conventions where the $\phase{a}=1$ for non self-dual $a$.  

 The Frobenius-Schur indicator has received much less attention in the
 literature than, say, the topological spin.  Remarkably, we find in
 the present context that it has to be trivial for bosons in the
 condensate in order to fulfill  our fusion consistency conditions and
 seems as (or even more) important than the topological spin for condensation.

\subsection{TSB beyond Bose condensation}

The physics of Bose condensation in anyon theories has many ties to topics in
mathematics and physics of current interest. We give a brief overview
of these connections here.

In Ref.~\onlinecite{Bais2009}  a detailed discussion can be found
relating topological (quantum group) symmetry breaking to
constructions in conformal field theory such as conformal embeddings,
the coset construction and orbifolds. It is argued that breaking down
the quantum group by condensation of a bosonic sector is dual to enlarging the (local)
chiral algebra. As  many aspects of the aforementioned CFT constructions can be
understood in terms of chiral algebra extensions, this provides a
physical reinterpretation of these constructions in terms of  Bose
condensation.

The  mathematics underlying this  paper has  been introduced in
Ref.~\onlinecite{KirillovOstrik2002} where a connection was made
 to the McKay correspondence\cite{1979_McKay_PureMath_37}. This is important to the
 classification problem for
 CFTs\cite{1987_Pasquier_NPB_285,1987_Capelli_CMP_113}
 , and possibly even
   mostreous moonshine\cite{1997_Conway_BLM_11} or generalizations thereof\cite{2011_Eguchi_EM_20,2012_Cheng_arXiv_1204.2779}, but we will not dwell on such abstract
topics here.
The same  structures have also been used in the
description of boundary conditions and topological defects in CFT \cite{2010maph.conf..608F,2013_Fuchs_CMP_321,
2011_Fuchs_NPB_843,2003_Fuchs_FP_51}. On the other hand there have
been studies on the effects of nontrivial boundaries in lattice realizations of
topological order based on Kitaev's quantum double model
\cite{2008_Bombin_PRB_78,2011_Beigi_CMP_306} and Levin-Wen models
\cite{2012CMaPh.313..351K}. The latter works consistently find that
nontrivial boundary conditions  can be a source of
condensation with consequences very similar to  the TSB scheme. That the two topics are intimately related also shows
from the connection between condensation and the protection of gapless
edge modes\cite{PhysRevX.3.021009,2014arXiv1407.5790B} which linked to
the absence of a Lagrangian  subgroup in the fusion algebra, the
definition of which  clearly is a special case of our
definition for a condensate  (see also~\onlinecite{2013arXiv1307.8244K}). 

It is
an intriguing fact that lattice formulation of topological order like  Levin-Wen  models
 always realize non-chiral
theories. In the simplest case where one starts with the $F$-symbols
of a fully braided MTC/anyon model $\mathcal{A}$ as input for the model, the topological
order is described by the double $\overline{\mathcal{A}}\times
\mathcal{A}$. If one however starts with a theory that admits
consistent fusion but no consistent braiding such as $\mathcal{T}$,
one ends up with the so called quantum double or Drinfeld center
$\mathcal{Z}(\mathcal{T})$. Apparently the model knows if a certain
set of $F$-symbols  \emph{admits} a consistent set of $R$-symbols or
not. The properties of the quasi-particles and how the resulting
topological order emerges are less obvious in this case. This question
has received additional attention recently \cite{2014_Wen_PRB_89}. In
this context explicit examples of $F$-symbols that do not allow a
compatible braiding structure are quite interesting, for example for
computational studies. The  TSB scheme produces naturally a nonbraided
theory, namely $\mathcal{T}$, and the present work provides the
necessary tools to calculate these $F$-symbols explicitly. 

There is a list of topics in the context of topological phases in condensed matter for which the precise connection with TSB is worth exploring further. For example the interplay with conventional symmetries and symmetry breaking, in particular for symmetry protected 
and symmetry enriched topological phases\cite{2011_Xie_PRB_83,2013_Essin_PRB_87,2013_Hung_PRB_87,2009_Zheng_PRB_80,2012_Levin_PRB_86,2014_Wen_PRB_889, 2012_PRB_Pollman_85,2013_Chen_PRB_87,2013arXiv1308.4673H,2014arXiv1402.3356G}.  

Recently, the formalism of condensate induced transitions has been
used in a completely different context, namely to find solvable
one-dimensional spin Hamiltonians of which the CFTs for the critical
behaviour follow the topological symmetry breaking scheme
\cite{2014_Maanson_PRB_88,2014_Lahtinen_PRB_89}.

 \section{\label{sec:diagrams} Diagrammatics for Bose condensates}

In this section we construct the diagrammatic theory for Bose condensates in a system of anyons, but first we review the original TSB scheme.

\subsection{Topological symmetry breaking revisited}\label{subsec:TSBold}

The aim of the TSB protocol is to find the effective theory $\mathcal{T}$  with particle labels $t,r,s,\ldots$ together with the branching or restriction coefficients $n^t_a$ under the assumption that one or more bosonic sectors $\gamma$ of the unbroken phase $\mathcal{A}$ form a condensate. The branching coefficients  implement a map characterizing the phase transition which we call restriction:
\begin{equation}
	a\to \sum_t n^t_a t \qquad \text{(restriction)}
\end{equation} 
The name restriction originates from thinking about $a$ as a representation of some algebra $A$ that branches into smaller representations if the algebra that acts on the representation is restricted to some subalgebra $T \subset A$. The same coefficients provide the adjoint of the restriction map, which we call the lift
\begin{equation}
	t \to \sum_a n^t_a a \qquad\text{(lift)}
\end{equation}
The question of which particles can actually condense is highly nontrivial  in the context of notrivial braid statistics. Even if a particle has trivial spin $\theta_\gamma =1$, a macroscopic state may not be invariant under interchange of the particles as the spin statistics theorem does not hold in this context.
 In Ref.~\onlinecite{Bais2009} the following conditions were proposed
\begin{enumerate}
\item[(i)]\label{oldTSB1} a boson $\gamma \in \mathcal{A}$ has trivial spin $\theta_\gamma =1$
\item[(ii)]\label{oldTSB2} $\gamma$ has partial or completely trivial self monodromy
\end{enumerate} 
The latter condition is understood as the existence of a fusion channel $f\in \gamma\times\gamma$ which has trivial spin itself, $\theta_f = 1$. These conditions are meant to ensure that for every number $N$ of the condensed boson  $\gamma$ there should be a state in the topological Hilbert space $V^{\gamma^{\times N}}$ that is invariant under monodromy of the particles. In the present work we refine these requirements.

The boson that condenses should  have  $\gamma \to \varphi + \ldots$ under the restriction map, where $\varphi$ is the vacuum of the new theory. It is postulated that fusion commutes with the restriction map
\begin{equation}
	a\times b = \sum_c N_{ab}^c c \Rightarrow \left( \sum_r n^r_a r\right)\times \left( \sum_s n^s_b s\right) =\sum_{c,t} N_{ab}^c n^t_c t 
\end{equation}
From this simple assumption and properties such as the uniqueness of the vacuum one can deduce that
\begin{align}
	0&\to \varphi\nonumber\\
	\bar{a}&\to \sum_t n^t_a \bar{t}\\
	d_a &= \sum_t n^t_a d_t \nonumber
\end{align}
The  game of the TSB protocol is now figuring out the particle content $\{t,r,s,\ldots\}$ and fusion rules of $\mathcal{T}$ and the restriction coefficients $n^t_a$ by making smart use of these properties. 

At this stage we denote the theory with $\mathcal{T}$ since it may still include sectors that braid nontrivially with the condensate.
These will actually not appear in the bulk as free particles as they will pull strings (domain walls) in the condensate. They can still appear on the boundary as massive excitations \cite{2014arXiv1407.5790B}. We will call these charges confined.

The next step  to project confined sectors out of the theory resulting in the theory $\mathcal{U}$ which provides the actual description of the bulk phase and has well defined braiding. A simple criterion to see which particles are confined and which ones are not is given by the lift: If all the sectors $a$ in the lift $t$ have the same spin $\theta_a$ the sector $t$ will be unconfined and survives in $\mathcal{U}$
\begin{eqnarray}
	&\text{$u\in \mathcal{U}\subset\mathcal{T}$ (unconfined)}&\nonumber\\
	& \Leftrightarrow&\\
	& \text{$\theta_a= \theta_{a'} \equiv \theta_u$ for all $a,a'$ with $n^u_a \neq 0\neq n_{a'}^u$}&\nonumber
\end{eqnarray}
 (we use $u,v,w,\ldots$ for the particle labels if we want to
 emphasize that the sectors belong to $\mathcal{U}$).

 The setup outlined above can identify the fusion coefficients,
 quantum dimensions and topological spins of the $\mathcal{U}$ theory
 together with the resctriction coefficients
 $n_a^t$  in many cases. However it is hard to make further progress. In order to put
 the TSB scheme on firmer ground and discuss the faith of the
 operators of the theory, we have to take a step back and reformulate
 the whole scheme diagrammatically.  

\subsection{Diagrammatics: conditions on the condensate}

In reformulating the TSB  scheme diagrammatically our strategy will be to work exclusively in the $\mathcal{A}$ theory for which we assume we know the  $F$- and $R$-symbols and hence how to compute diagrams. It is very convenient to identify $t$ with its lift
\begin{equation}
	t = \sum_a n^t_a a
\end{equation}
as a superposition of charges in $\mathcal{A}$. 

Assume that a number of bosons $\gamma_1,\ldots,\gamma_n$ condenses. In particular, the superposition of all the bosons $\gamma_j$  and the original vacuum $0$ is nothing other than the lift of the new vacuum $\varphi$. We will also call $\varphi$ the condensate. 
\begin{equation}
	\varphi= 0 + \gamma_1 + \ldots +\gamma_n \qquad \text{(condensate)}
\end{equation}
(we use  $\gamma_0 = 0$ in some formulas).

The whole construction described in this paper follows from the requirement that the condensate satisfies all the properties of the vacuum within a certain subclass of diagrams in $\mathcal{A}$. This provides the lift of the full $\mathcal{T}$ theory---including diagrams---into the $\mathcal{A}$ theory, which is the essential extension w.r.t. to the TSB protocol from Ref.~\onlinecite{Bais2009}.

We associate a dashed charge line to the condensate,
\[
	\condensate\quad=\qquad
	\anyon{\varphi}\quad= \quad\vac{0} \;+\;
	\sum_{i=1}^n \quad\anyon{\gamma_i},
\]
Since the vacuum is self-conjugate we require $\bar{\varphi} = \varphi$ \emph{i.e.} $\gamma\in\varphi \leftrightarrow \bar{\gamma}\in\varphi$, so there is no need to give the condensate line a direction.
 We note here that the quantum dimension of the condensate $q = d_\varphi=\sum_j d_{\gamma_j}$, also called \emph{quantum embedding index}, will turn out to be an important number associated to the condensation. In particular it often shows up in the proper normalization. For example removing a disconnected condensate bubble  from a diagram should lead to a factor of $1$ while we find a factor of $q$. One may think that we should normalize the condensate charge line by a factor of $1/q$. In stead, we will put the correct normalization in place in the condensate vertex
 \begin{equation}\label{eq:condvertex}
	\condensatevertexket = {1\over \sqrt{q}}\sum_{ijk} \phi_{k}^{ij} \vertexket{\gamma_i}{\gamma_j}{\gamma_k}{}{}
\end{equation}
and interpret a condensate bubble as a condensate ``splitting'' and a condensate ``fusion'' vertex  stacked on top of eachother and top and bottom line projected to the original vacuum 0. We will come back to normalization issues in Sec.~\ref{sec.diaglift}.

Note the appearance of the coefficients $\phi_{k}^{ij} $ in Eq.~\eqref{eq:condvertex}. This is the  first appearance of vertex lifting coefficients (VLCs), which play an essential role in our construction and will be discussed extensively  when we properly define vertices in the condensed phase.  

Now which conditions should the condensate satisfy? A straightforward requirement would be that $\theta_{\gamma_i} =1$ for all the $\gamma_i$, in other words that the condensate consistes of bosons. However, as discussed before,  the braiding properties of charges with trivial spin may still be very nontrivial especially for non-abelian anyons. The following diagram equality holds for  vacuum vertices
\[\label{eq:braidconditioncondensate}
	\condensatebraidket\quad = \quad \condensatevertexket\;\;.
\]
This is  the braiding condition on the condensate that we will require. Although deceptively simple looking, one should keep in mind that both left and right hand side expand as a possibly large superposition which should agree term by term.  That leads to the condition
 \[\label{eq:braidconditionRsymb}
 	R_{\gamma_k}^{\gamma_i \gamma_j} \phi^{ij}_k = \phi^{ji}_k\;\;.
 \]
 By choosing $k=0$ we find  $R^{\gamma\gamma}_0 = \phase{\gamma} \theta_{\gamma}^* =1$. Here we did assume that $\phi^{ij}_0 = \delta_{\gamma_j\bar{\gamma}_i}$ which will be justified later. This immediately shows that indeed the condensed sectors $\gamma\in \varphi$ have to be bosonic $\theta_\gamma =1$ at least when they are not self-dual. At this point, it seems that the consistency condition allows for fermionic self-dual particles in the condensate. However, we did not yet discuss the fusion properties of the condensate.

To reproduce all properties of a vacuum,  we should be able to reconnect condensate lines freely,
\[\label{eq.condensatecondition}
	\condensatetwototwoS\qquad =\qquad \condensatetwototwoM.
\]
This fusion condition  is equivalent to the definition of a Frobenius algebra in the context of tensor categories. Relation \eqref{eq:braidconditioncondensate} is known as the commutativity of the algebra. 

The fusion condition \eqref{eq.condensatecondition} is in some respects more fundamental than the braid condition \eqref{eq:braidconditioncondensate}---one may define a Frobenius algebra that is not commutative.   A ``condensate'' that does not satisfy \eqref{eq:braidconditioncondensate} still makes sense mathematically and is relevant for one-dimensional topological phases\cite{2012CMaPh.313..351K,2013arXiv1307.8244K}. We refer the reader who is interested in the mathematics of algebras in tensor categories to Ref.~\onlinecite{2013arXiv1307.8244K} and references therein for further details.

A  consequence of Eq.~\eqref{eq.condensatecondition} is that condensed bosons have a trivial Frobenius-Schur indicator. Taking the upper left and lower right charge line  to be labeled by the original vacuum, we find
\begin{equation}
\twototwoSvacLvacRvar{\gamma}  = \quad\anyon{\gamma}
\end{equation}
for all $\gamma\in\varphi$, so $\phase{\gamma}=1$. Hence we see that actually the braid condition  and the fusion conditions together only allow for a condensate with $\theta_\gamma =1,\phase{\gamma}=1$ also for self-dual charges so we find that  indeed all condensed charges are bosonic.

The braid condition \eqref{eq:braidconditioncondensate} is the strictest condition we can put on the braiding properties on the condensate vertex. Condition~{(ii)} in Subsection~\ref{subsec:TSBold} only requires a form of trivial monodromy, as opposed to a single braid exchange. 
One can also propose the following condition
\[\label{eq:monodromyconditioncondensate}
	\condensatedoublebraidket = \condensatevertexket\;\;.
\]
This equation is a consequence of Eq.~\eqref{eq:braidconditioncondensate}.
On its own it yields a condition on the spin factors of the condensed charges
\[
	\frac{\theta_{\gamma_k}}{\theta_{\gamma_i}\theta_{\gamma_j}} = 1.
\]
Again, by choosing $k=0$ we find $\theta_{\gamma_i}\theta_{\bar{\gamma_i}} = \theta_{\gamma_i}^2 =1 \Rightarrow \theta_{\gamma_i} = \pm 1 $.  
 This actually seems to allow for fermionic condensates with $\theta_{\gamma}=-1$. We will however stick to condition \eqref{eq:braidconditioncondensate} in the present paper.

It is interesting to see what happens if we stack several condensate vertices together. In particular, the diagram
\begin{equation}\label{cond-state}
	\fouranyonstatecond
\end{equation}
denotes a superposition of states with mixed particle-number in the
topological Hilbert space of the theory which will not notice any
``stirring''.  It is tempting to view \eqref{cond-state} as something
like the groundstate of the condensed phase.  It would be interesting
to see if the coefficients $\phi_{k}^{ij}$ have a role to play in
microscopic realizations of TSB phase transitions and ground state wavefunctions.

\subsection{Particle spectrum in condensed phase}
%

In the presence of a condensate, the charges $a$ in $\sA$ can no longer be viewed as the elementary excitations in the system, as they may fuse freely with the new vacuum $\varphi$ and this fusion product in general contains multiple charges.  This observation hints on how to proceed with the next step of the TSB protocol, namely determining the particle content of the unbraided theory $\mathcal{T}$.

From the fusion properties of the condensate we find
\begin{equation}
	\varphi \times a = \sum_b  N_{ba}^c n^\varphi_b c = \sum_{t;c} n^t_a n^t_c c = \sum_t n^t_a t
\end{equation}
so we see that taking the fusion product with the lift of the new vacuum generates a superposition precisely containing all the lifts of the restriction of $a$. 
This, in fact,  constitutes a major simplification in finding the branching coefficients $n^t_a$ as compared to the strategy in Ref.~\onlinecite{Bais2009}. The problem of finding the branching coefficients reduces to  writing out the table $\varphi \times a$ for the conjectured $\varphi$ and group the right hand side in ``irreducible blocks''.  This turns out to be a very simple exercise in many concrete examples.

 When charges of  $\cat{A}$ can no longer be distinguished in the condensed phase we say that they get \emph{identified}. This now corresponds to the following condition for identification
\begin{equation}
	a \sim b \Leftrightarrow a,b\in t \qquad \text{(identification)}
\end{equation} 
It will not always  be the case that $t$ is simply equal to a fusion product of $a$ with the condensate. It may happen that $ \varphi\times a = t_1 + \dots + t_n$. In that case we say that a splits 
\begin{equation}
	\varphi\times a = t_1+t_2 \qquad \text{(splitting)}
\end{equation}
We will see that in this case the $t_i$ are precisely all the $\cat{T}$ charges in which $a$ appears.The full structure of splittings and identifications, \emph{i.e.} the branching coefficients $n^t_a$, can in the end of the day thus being read of from the table
 \begin{align}
 	\varphi \times a & = \sum_t n_a^t t
 \end{align}
In Section~\ref{sec.calc_vlcs} we provide two examples of Bose condensation ($SU(2)_4 \, \to \, SU(3)_1$ and $SU(2)_{10}\,\to\,SO(5)_1$) where we use the above procedure to determine the spectrum in the Bose condensed phase.

\subsection{Fusion spaces from $\mathcal{T}$ to $\mathcal{A}$: vertex lifting coefficients (VLCs)}
 Let us now study the effect of the Bose condensate on the level of the fusion spaces of the theory. 

Our main result  is that lifting the vertices in $\sT$ to a superposition of vertices in   $\sA$  
\[\label{eq.vlcs}
	\vertexket{r}{s}{t}{} = \sum_{a,b,c} \vlc{r}{s}{t}{a}{b}{c} \vertexket{a}{b}{c}{},
\]
gives a complete embedding of all diagrams of   $\mathcal{T}$ into the theory $\mathcal{A}$. As such it completes the lift of the particle sectors which is discussed in Ref.~\onlinecite{Bais2009} to include the full content of the theory. Here the $\vlc{r}{s}{t}{a}{b}{c}$ are a set of complex coefficients that we call vertex lifting coefficients (VLCs). Together with the branching coefficients $n^t_a$ they completely specify the topological content of the transition $\sA \rightarrow \mathcal{T} \rightarrow\mathcal{U}$. In Section~\ref{sec.calc_vlcs} we go into the details of how to actually calculate these numbers, which is a non-trivial task that amounts to solving and checking a large number of consistency conditions.

Note that once these numbers are known, it is a straightforward matter to express the topological data of the broken phase in terms of the topological data of the unbroken phase. At the end of this section we will give  explicit formulae for the $F$- and $R$-matrices (see Eq.~\eqref{eq.Fsymb} and \eqref{eq.Rsymb}). However, first we will study two important special cases of VLCs, that involve the condensate as three respectively one of the $\sT$-fields.

\subsubsection*{Condensate vertex}
For a given condensate $\varphi = 0+\gamma_1 + \dots + \gamma_n$ the problem starts with checking if the consistency conditions Eq.~\eqref{eq:braidconditioncondensate} and \eqref{eq.condensatecondition} can be satisfied. This amounts to finding VLCs for the condensate,
\begin{equation}\label{eq:phivlcs}
	\phi^{ij}_k \equiv \sqrt{q} \vlc{\varphi}{\varphi}{\varphi}{\gamma_i}{\gamma_j}{\gamma_k}.
\end{equation}
From the fundamental fusion condition,
\[\label{eq.condensateconditionagain}
	\condensatetwototwoS\qquad =\qquad \condensatetwototwoM,
\]
by transforming the left hand side using the $F$-symbols, one obtains the equality
\begin{equation}
	\sum_{m} {\phi^{km}_{i}}^* \phi^{mj}_{l} \F{\gamma_i\gamma_j}{\gamma_k\gamma_l}{\gamma_m \gamma_{m'}} = \phi^{ij}_{m'} {\phi^{kl}_{m'}}^*.
\end{equation}
We can now infer properties of the VLCs for the condensate by specializing the above equation to certain choices of $i,j,k,l$ and ${m'}$ (written this way the $m$ has to be summed over, although only $m$ that are allowed by fusion contribute). The $\phi^{00}_{0},~\phi^{i0}_{i},~\phi^{0i}_{i},~\phi^{i\bar{\imath}}_{0}$  (with the notation $\gamma_{\bar{\imath}} = \bar{\gamma_i}$) are not constrained by this equation. We are free to choose the obvious normalization
\[
	\condensatebraket\quad =\quad \longcondensate\quad,
\]
which imposes  $|\phi^{00}_{0}| = |\phi^{i\bar{\imath}}_{0}| = 1$ (note the normalization factor $\sqrt{q}$ in Eq.~\eqref{eq:phivlcs} for the $\phi^{ij}_k$ as compared to the bare VLCs). One can in fact freely put
$\phi^{00}_{0},~\phi^{i0}_{i},~\phi^{0i}_{i},~\phi^{i\bar{\imath}}_{0} = 1$ by the gauge freedom of the VLCs discussed later (see Eq.~\eqref{eq:vlcgauge}).

\subsubsection*{$\sT$ particle propagators}
The charge lines involving $\mathcal{T}$ anyons straightforwardly lift to the original phase, but we should always think of the interaction with the condensate.
 We can compute the associated VLCs using the  relations
\[\label{eq.particlecondition-full}
\condensatetwototwoSanyonR{t}{t}\quad=\quad \condensatetwototwoManyonR{t}{t}{t},\qquad \condensatetwototwoSanyonL{t}\quad=\quad \condensatetwototwoManyonL{t}{t}{t}.
\]
We will refer to these diagrams as propagators as they can be read as the  propagation of a $t$ particle in the condensed phase.
 
Using relations \eqref{eq.particlecondition-full}  we obtain
\[
	t^{\gamma a}_b \equiv \sqrt{q} \vlc{\varphi}{t}{t}{\gamma}{a}{b},\quad 	t^{a\gamma }_b \equiv \sqrt{q} \vlc{t}{\varphi}{t}{a}{\gamma}{b}
\]
(Again we pull out the normalization factor $\sqrt{q}$; with this definition we also have $t^{0a}_{b} = \delta_{ab}$ when $a,b\in t$ and $0$ otherwise.) In fact, the equation on the right in \eqref{eq.particlecondition-full} is trivially satisfied if we define
\begin{equation}\label{eq:tRchoice}
	t^{a\gamma}_{b} \equiv R^{\gamma a}_{b} t^{\gamma a}_{b}
\end{equation}
or diagrammatically
\begin{equation}
	\newvacRket{}{}{t} \equiv \condensateRbraidket{}{}{t}{}.
\end{equation}
Of course this constitutes a choice: we could as well have taken the inverse braid relation. 

Let us also define 
\[
	t^{ a\bar{b}}_{\gamma} \equiv \sqrt{q} \vlc{t}{\bar{t}}{\varphi }{a}{\bar{b}}{\gamma}.
\]
There are numerous  relations between  VLCs. Using
\[\label{eq.tbarcondition}
		\condensatetwototwoSanyondeath{t} =\condensatetwototwoManyondeath{t},\quad 		\condensatetwototwoSanyonbirth{t} =\condensatetwototwoManyonbirth{t},
\] 
 and unitarity we find
 \begin{align}\label{eq:tabrels}
 t^{a\bar{b}}_\gamma &= \F{a\bar{b}}{\gamma 0}{b\gamma} {t^{\gamma b}_{a} }^*\;\;,\\
 t^{a\gamma}_{b} &= \F{0\gamma}{\bar{a}b}{a\gamma}^* \F{\bar{a}b}{\gamma 0}{\bar{b}\gamma}^* \bar{t}^{\gamma\bar{b}}_{\bar{a}}\;\;.
 \end{align}
After determination of the complete set of $t^{\gamma a}_{b}$ in fact all the $t^{a\gamma}_b$ and $t^{a\bar{b}}_{\gamma}$ follow immediately. In Appendix \ref{ch.vlcprops}, we present a full list of diagrammatic relations and their symbolic equivalents, like \eqref{eq.tbarcondition} and special cases thereof. 

A particularly useful relation is the orthogonality condition
\[\label{eq.orthogonality}
	\sum_{a,b}\left( {  d_a d_b d_t \over  d_r d_s d_c } \right)^{1 \over 2} \vlc{r}{s}{t}{a}{b}{c} \vlc{r}{s}{t'}{a}{b}{c}^* = \delta_{tt'} \quad \forall \quad c .
\]
This is derived by working out the diagrammatic inner product in terms of VLCs:
\begin{align}
	\vertexbraket{t'}{r}{s}{t}{}{} &= \sum_{a,b,c}\vlc{r}{s}{t}{a}{b}{c}
	\vlc{r}{s}{t'}{a}{b}{c}^*	\vertexbraket{c}{a}{b}{c}{}{} \nonumber \\
&= \sum_{a,b,c} \left( d_a d_b \over d_c\right)^{1\over 2} \vlc{r}{s}{t}{a}{b}{c} \vlc{r}{s}{t'}{a}{b}{c}^* \;\; \anyon{c}\\
\intertext{ but also}
\vertexbraket{t'}{r}{s}{t}{}{} &=\delta_{tt'} \left( { d_r d_s \over d_t} \right)^{1 \over 2} \;\; \anyon{t} = \delta_{tt'} \left( { d_r d_s \over d_t} \right)^{1 \over 2} \sum_{c \in t} \;\; \anyon{c}.
\end{align}
Comparing the coefficients in front of the $c$-charge lines we obtain \eqref{eq.orthogonality}.

Writing down a diagrammatic equality in $\cat{T}$ and checking its content in $\cat{A}$ is a powerful means to reveal information. By inspecting
\begin{equation}
	\condensatetwototwoSanyonbend{t} = \phase{t} \condensatetwototwoManyonstraightMirror{t}
\end{equation}
for example, we find that $\phase{t} = \phase{a}$ for all $a\in t$, so the restriction  respects the Frobenius-Schur indicators.

Finally, let us note that the condensate slides freely past vertices, expressed \emph{e.g.} by
\[\label{eq.vertexvac}
	\condensatetwototwoSanyonvertexR{r}{s}{t}\qquad = \qquad \condensatetwototwoManyonvertexR{r}{s}{t},
\]
such that it indeed behaves as a proper vacuum in the new theory.

\subsubsection*{Topological data of the condensed phase}
Using the VLCs it is straightforward to obtain the full topological data of the $\mathcal{T}$ and $\sU$ phases.
We start form the defining relations of the $F$-symbols, namely
\begin{align}
	\onetothreeL{r}{s}{t}{u}{v}{}{}	 &= 
	\sum_{w} 
	\F{rst}{u}{vw} \onetothreeR{r}{s}{t}{u}{w}{}{}.
\end{align}
In principle these $F$-symbols in turn determine the the $F$-symbols with two legs up and two legs down, but we can also directly obtain those starting from
\begin{align}
		\twototwoS{r}{s}{t}{u}{v}{}{}	 &= 
	\sum_{w} 
	\F{rs}{tu}{vw} \twototwoM{r}{s}{t}{u}{w}{}{}.
\end{align}
Expanding both sides using VLCs  we obtain a relation featuring both the $F$-symbols of the $\cat{A}$ and of the $\cat{T}$ theory. Finally we  use the orthogonality relation \eqref{eq.orthogonality} to derive the closed expressions
\begin{widetext}
\begin{align}\label{eq.Fsymb}
	\F{rst}{u}{vw} &= \sum_{abcef} \vlc{r}{s}{v}{a}{b}{e}\vlc{v}{t}{u}{e}{c}{d}\vlc{s}{t}{w}{b}{c}{f}^*\vlc{r}{w}{u}{a}{f}{d}^* \sqrt{\frac{d_a d_b d_c d_u}{d_r d_s d_t d_d}}\F{abc}{d}{ef},\\
	\F{rs}{tu}{vw} &= \sum_{abcef} \vlc{v}{s}{u}{e}{b}{d}\vlc{t}{u}{w}{c}{d}{f}\vlc{r}{s}{w}{a}{b}{f}^*\vlc{t}{v}{r}{c}{e}{a}^* \sqrt{\frac{d_a d_b d_c d_d}{d_r d_s d_t d_u}}\left(\frac{d_w}{d_f}\right)\F{ab}{cd}{ef}.
\end{align}
\end{widetext}
These expressions are  valid for arbitrary $\mathcal{T}$-charges and provide consistent fusion data. The fusion coefficients are straightforwardly obtained as $N_{rs}^t = \sqrt{ \frac{d_r d_s}{d_t}} \F{rs}{rs}{\varphi t}$.

\subsection{Braiding and confinement in the condensed phase}
In this section we discuss braiding and confinement in the broken phase, which amounts to the difference between the algebras $\sT$ and $\sU$.  
%

Diagrammatically, we take the  condition for unconfined sectors to be
\[\label{eq.confinement}
	\condensatemeetsanyondoublebraidket{u} \quad=\quad \condensatemeetsanyonlong{u}\qquad \text{for $u$ in $\mathcal{U}$}.
\]
From the usual  monodromy equation (Eq.~\eqref{eq:monodromy} in the Appendix) we see that this is equivalent to $\theta_a = \theta_b$ for  $a,b\in u$ and yields no condition on the VLCs. The charges for which (\ref{eq.confinement}) holds form a subset of the charges of $\cat{T}$ that are closed under fusion, and the labels $u,v,w,\dots$ usually refer to this subalgebra $\cat{U}$. There is no difference in this confinement condition and that of Ref.~\onlinecite{Bais2009}.

The $\cat{U}$ theory has consistent braiding with $R$-symbols 
\begin{equation}\label{eq.Rsymb}
	R^{uv}_w =\left( \vlc{u}{v}{w}{a}{b}{c} / \vlc{v}{u}{w}{b}{a}{c}\right) R^{ab}_{c}.
\end{equation}
It is now clear how the present work provides an alternative route to obtaining the topological data for theories that are produced by TSB, at least when the $F$-and $R$-symbols of the original theory are known, such as is the case for $SU(2)_k$.

For diagrams that involve nontrivial braiding, one would expect that it is not allowed to do the evaluation on confined charges. Nicely enough, these are automatically projected to zero. The following projection property holds \cite{KirillovOstrik2002, 2013arXiv1307.8244K}:
\begin{equation}\label{eq.projector}
	\projector{t}= \delta_{t\in\cat{U}}\quad \anyon{t},
\end{equation}
(where $\delta_{t\in\cat{U}}=1 $ when $t\in \cat{U}$ and $0$ otherwise). For unconfined charges, this is a trivial consequence of the defining relation~\eqref{eq.confinement}. Expression \eqref{eq.projector} is equivalent to the equation
\begin{equation}
	\sum_a \sum_{i=0}^n |t^{\gamma_i a}_b|^2 \theta_a \sqrt{{d_a d_{\gamma_i}\over d_b}} = \theta_b \delta_{t\in\cat{U}}.
\end{equation}
 The left hand side turns out to be $b$-independent, which is only consistent with it not being zero when the $\theta_b$ are identical for all $b\in t$. This is precisely the confinement condition.

As we can always attach a condensate bubble at a point in the diagram, in diagrams with braiding we can usually slide the condensate around until a configuration like \eqref{eq.projector} is reached locally. This means \emph{e.g} that
\begin{equation}
	\qdimtwist{t}{>}\quad=\quad\trefoil{t} =0 \quad \text{for $t$ confined}.
\end{equation}
The  diagram on the left leads to the rather nice equation
\begin{equation}
	\sum_{a\in t} d_a \theta_a = 
	\begin{cases} 
	q d_t \theta_t \qquad &\text{ for $t$ unconfined}\\
	0 \qquad &\text{ for $t$ confined}\end{cases}\;\;.
\end{equation}

\section{\label{sec.diaglift}Evaluation of diagrams in the condensed phase}
Equations \eqref{eq.Fsymb} and \eqref{eq.Rsymb} give access to the full topological data of the condensed phase (bulk and boundary) through the VLCs, and using these, arbitrary diagrams can in principle be evaluated. However, in practice it is often much more natural to lift the diagrams directly and evaluate them using the data of the $\mathcal{A}$-theory. This has the advantage that generally only a small subset of the VLCs has to be known and, moreover, it emphasizes the physical picture and reveals interesting  relations that remain hidden in the more indirect route. 

In order to lift diagrams that contain more than one component  one has to draw vacuum exchange lines (VELs), that describe the interaction of the particles with the condensate.
The full topological symmetry breaking scheme may now be understood in terms of the following commutative diagram:
\[
\label{vlcdiagram}
\begin{CD}
\text{Spectrum of }\sA			@>\text{calculate}>F, \, R \text{ symbols}>		\text{Operators in }\sA \\
@V\text{TSB}VV																	@V\text{VEL}V\text{VLC}V\\
\text{Spectrum of }\sU			@>\text{calculate}>F, \, R \text{ symbols}>		\text{Operators in }\sU
\end{CD}
\]
We will now fill in the final gaps in the description of the vertical arrow on the right , providing the details of how general operators from $\cat{U}$ (and $\cat{T}$) are expressed in $\cat{A}$.

\subsection{A recipe for lifiting diagrams}
We first give the general recipe to lift diagrams and then comment on each of its steps. Thereafter we discuss the ingredient that had been missing up to now: when lifting diagrams the appearance of the condensate has to be made explicit.

\begin{enumerate}
\item Draw a diagram in the $\sU$ theory.
\item Normalize the full diagram by a factor $1/q$.
\item Dressing the diagram with vacuum exchange lines (VELs).
\item Lift diagram: replace charge lines labeled by $u,v,\dots$ by the superposition  charges $\sum_{a\in u} a$ in $\sA$, and
put in the VLCs as weights for all the vertices in the expansion.
\item Evaluate using the data of $\sA$.
\end{enumerate}

\subsubsection{Draw a diagram in $\mathcal{U}$} 
This step is self-explanatory. We will assume in this section that the diagram that needs evaluation is labeled by $\mathcal{U}$ charges as we focus on the bulk properties of the anyon model. But, as the theory projects diagrams that are not allowed for confined charges automatically to zero, general $\mathcal{T}$ charges are in fact allowed.   Diagrams for physical observables are usually given by a link, corresponding to a vacuum-to-vacuum expectation value. These diagrams can be read as the creation of a (set of) anyons, followed by some braiding pattern and the subsequent annihilation of the anyons.

\subsubsection{Normalize}
The proper normalization turns out to be an overall factor of $1/q$ voor vacuum-to-vacuum diagrams, where $q = d_\varphi$ is the quantum embedding index. An easy way to establish this is by considering the diagram for the quantum dimensions of the condensate. Clearly, this should evaluate to 1 while a naive evaluation leads to $q$. One can also say that there should always be an additional condensate line connected to the top and bottom since this is a vacuum-to-vacuum diagram in $\mathcal{U}$. Projecting these to the original vacuum $0$ also gives the correct factor $1/q$.

\subsubsection{Dressing of diagram with VELs}
In this step we make sure that the diagram is connected by attaching VELs (lines labeled by the condensate) between the otherwise disconnected components.  

\paragraph*{Necessity of VELs}
To see why this is necessary, remember that the conditions on the condensate all basically boiled down to one thing: the new vacuum line should act as a proper vacuum sector, {\em i.e.}  one should be able to freely attach vacuum lines to any diagram representing an operator expectation value. Condensate lines starting on and ending on the same single connected component of the diagram can always be removed by applying the vacuum consistency conditions for the broken phase, for example one has the sequence: \begin{equation}\label{eq.bubble}
	\condensatebubblemeetsanyon{u} \; =\; \condensatebubblemeetsanyonT{u} \;= \;\; \anyon{u}.
\end{equation}
For vacuum lines connecting two different components the situation is slightly more subtle. If there are two condensate lines connecting the components, then the endpoints of one of the condensate lines can only be shifted from one component to the other using the second vacuum line:
\begin{equation}
\twototwoMcondMbubbleL{u}{v} \quad=\quad  \twototwoMcondMdouble{u}{v} \quad =\quad \twototwoMcondMbubbleR{u}{v}.
\end{equation}
implying that it suffices to only draw single vacuum lines until all components are  directly or indirectly connected.

Consequently the necessity of the connecting condensate lines can be viewed in two ways. One way is that, once the connecting VELs are in place, we can freely connect other condensate lines as is appropriate for the vacuum. They can be annihilated by sliding them around until the endpoints meet and we remove the bubble using \eqref{eq.bubble}. The evaluation of the diagram remains the same. Here we assume that the lines in the diagram are labeled by unconfined charges so that all braidings with the condensate can be undone. Recall that the theory takes care of confined charges in a natural way by projecting diagrams that are not allowed to zero, so there is no loss of generality here. The other way to think about the necessity of the VELs is imagining that we must somehow take into account all imaginable vacuum lines in the diagram. Then, by the same logic, we can always effectively remove them as long the resulting diagram stays connected. The physical reason for the appearance of the VELs is clearly the fact that in the condensed phase, the excitations  will unavoidably interact with the condensate.

We call a diagram where all components are connected by condensate lines a \textit{dressed} diagram. In previous work\cite{Bais:2011iu}, we had already identified such nontrivial contributions to the $S$-matrix in a lattice model and dubbed them {\em vacuum exchange diagrams}.

\paragraph*{VELs and linking}
Charge conservation implies that when a diagram that can be separated in two  unlinked components, the connection by a VEL between the parts is in fact spurious, if we agree to normalize the components by $1/q$ separately. To see this,
first imagine two processes  $X$ and $Y$ separated in time, each  consisting of a set of  creation, braiding and annihilation processes. These are represented by two separate link diagrams drawn one above the other. The vertical VEL
that should be drawn according to the recipe can be removed without changing the evaluation of the diagram
\begin{equation}
	\XaboveYcond\quad =\quad \frac{1}{q}\left(\  \XaboveY\ \right).
\end{equation}
This fact is derived from the simple statement that tadpole diagrams evaluate to zero for nontrivial charges,
\begin{equation}
	\tadpole{a}{\gamma}  \quad =\   \delta_{\gamma, 0} \qdimvar{a}\; \;.
\end{equation}
The same reasoning holds for diagrams separated in space,
\begin{equation}
\XbesidesYcond\quad= \ \frac{1}{q}\left(\ \XbesidesY\  \right ).
\end{equation}
This type of diagram generally portrays operations on spatially separated groups of anyons that do not lead to entanglement between the groups. Hence, we see that the presence of the condensate does not induce entanglement by itself.

For confined charges, the only non-zero diagrams are planar and so here we do not need the VELs if we just put in a normalization $1/q$ for all separate components.

\subsubsection{Lift diagram}
This is the step where the present work provides the essential ingredients, such that diagrams can actually be computed. The diagram in $\cat{U}$ is expanded in the  $\cat{A}$ theory by putting in the corresponding superpositions for the labels $u,v,w,\dots$  at the charge lines
\begin{equation}
	\anyon{u} \quad \to \quad \sum_{a \in u}\;\; \anyon{a}, 
\end{equation}
and,  importantly, by putting in the VLCs as coefficients at all the appearing vertices
\begin{equation}
	\vertexket{u}{v}{w}{} \to \sum_{a,b,c} \vlc{u}{v}{w}{a}{b}{c} \vertexket{a}{b}{c}{}.
\end{equation}
 The vertices can either appear because the original $\cat{U}$ diagram contained vertices or because disconnected components are now connected by VELs and thus have a vertex with the condensate.

\subsubsection{Evaluate using $\mathcal{A}$ data}

Finally, we can evaluate the terms in the superposition by making use of the $F$-symbols and $R$-symbols of $\cat{A}$. This way, the protocol expresses any vacuum-to-vacuum diagram of the $\cat{U}$-theory in terms of VLCs and the $F$- and $R$-symbols of the $\cat{A}$ theory. Generally, operators going from a set of $\cat{U}$-anyons $u_1,\dots,u_{n}$ to  anyons $u'_1,\dots, u'_m$, result in a complicated superposition of operators of all the mixed combinations of the $a_j \in u_j$ to combinations of the $a'_i \in u'_i$.


\subsection{General observables and splitting}
Interesting observables are generally represented by diagrams of
knotted and linked loops, representing creation of anyon particle
anti-particle pairs, some braiding pattern, and annihilations of all
the anyons. One example of such a diagram is the topological
$S$-matrix. Charge conservations implies that the  VELs only give
non-zero vertices  with a boson $\gamma$ and only one charge
$a$. The non-trivial contributions thus involve the $t^{\gamma_i
  a}_{a}$ and $t^{a \gamma_i}_{a}$ which are only non-zero for $a$
that split when $\gamma\neq 0$. Hence only charges that split provide nontrivial VEL contributions.
The splitting of charges in the condensation is thus tightly linked
to interaction with the condensate and the correct outcome for observables in the condensed phase. 


\subsubsection{Two basic examples}

\paragraph*{Condensed phase quantum dimension}
As the quantum dimensions of the  theory are fundamental quantities, these are the first objects that should reproduced by our scheme. They correspond to the evaluation of a loop. To lift such a diagram to $\mathcal{A}$, all we need to do is take the superposition as no vertices appear and a loop has only one component. This way we obtain
\[
	d_t = \qdim{t}{>} = \frac{1}{q} \sum_{a\in t}\qdim{a}{>} = \frac{1}{q} \sum_{a\in t } d_a.
\]
This can be shown independently \cite{Bais2010}. From $\varphi\times a = t_1 + \dots + t_n$ is is then also evident that $d_a = \sum_i d_{t_i}$ when $a\to t_1 +\dots +t_n$ in the condensation. In other words, the quantum dimension is conserved.

\paragraph*{Condensed phase topological $S$-matrix}

It is convenient to forget about the normalization by the total quantum dimension of the topological $S$-matrix for the moment, and simply take
\[
	S_{ab} =  \smatrix{a}{b}\;\;.
\]
Following the general recipe, the $S$-matrix of the $\cat{T}$-theory is given by the diagram
\[
	S_ {st} =  \frac{1}{q}\ \scondensate{s}{t} \;\;\;.
\]
For convenience, we will specialize to the case $\varphi = 0+\gamma$. 
Identifying the  contribution from the non-trivial vacuum exchange lines yields the diagram
\[
	\tilde{S}_{ab}   \equiv \sbar{a}{b}{\gamma} = \sum_c \F{ab}{ab}{\gamma c} \frac{\theta_c}{\theta_a \theta_b} \sqrt{d_a d_b d_c}\;\;.
\]
This means that we may write 
\[
\label{eq.smatU}
	S_{st} =\frac{1}{ q^{2}}\sum_{a\in s}\sum_{b\in t} (S_{ab} + {s_a^{a\gamma}}^* t_b^{\gamma b}\, \tilde{S}_{ab}).
\]

\section{\label{sec.calc_vlcs}Calculation of VLCs}
In the previous Sections we defined the VLCs and gave some relations between them. Here we will go into the details of actually computing them by making use of diagrammatic equalities listed in  Appendix \ref{ch.vlcprops}. 

We focus on the study of one-component condensates of the form $\varphi = 0 + \gamma$, that either have or have not got a triple-vacuum vertex.
For these condensates the problem is completely solved and we derive the general formulas. These cases already contain all of the interesting physics (splitting, identification, confinement)  and show how intricately the VLCs are linked to the these properties.
We work out a representative example of both cases: the breaking of $SU(2)_4$ to $SU(3)_1$ and the breaking of $SU(2)_{10}$ to $SO(5)_1$. We also show that an expected condensate in a system with five layers of Fibonacci anyons does in fact not satisfy the consistency condition. 

\subsection{General scheme}
%
For simplicity we will assume a condensate of the form $\varphi = 0+\gamma$. The VLCs can be calculated in that case according to the following scheme. We believe that the  steps can be extended to  general condensates, but the technical details will become  more demanding.

First, one needs to calculate the $\phi^{ij}_k$ symbols that appear in the lift of the triple-vacuum vertex. To do this, one needs to find solutions to the diagrammatic equation~\eqref{eq.condensatecondition}, repeated here for clarity:
\[
	\condensatetwototwoS\qquad =\qquad \condensatetwototwoM.
\]
Then, one has to solve the $\sT$-propagators. It basically involves finding solutions to equations of the form
\[\label{eq:propagatorequations}
\condensatetwototwoSanyonR{t}{t}\quad=\quad \condensatetwototwoManyonR{t}{t}{t},\qquad \condensatetwototwoSanyonL{t}\quad=\quad \condensatetwototwoManyonL{t}{t}{t}, \dots ,
\]
where two of the four legs are the vacuum.  A full list of these consistency conditions is presented in Appendix \ref{ch.vlcprops}. For VLCs involving the condensate on one of the legs, relations \eqref{eq:tabrels} shows that we only need to solve for the $\vlc{\varphi}{s}{t}{\gamma}{a}{b} \equiv t^{\gamma a}_b/\sqrt{q}$. We show how to do this for the single component condensates in what follows.

When  the VLCs for the condensate and for the condensate hitting another charge are calculated the others can be obtained. We exploit the following equations to calculate the others:
\[\label{eq:vlcEVdiag1}
\vlcinvariancecondensateA{r}{s}{t} \;\; =  \quad \vertexket{r}{s}{t}{},
\]
\[\label{eq:vlcEVdiag2}
\vlcinvariancecondensateB{r}{s}{t} \;\; =  \quad \vertexket{r}{s}{t}{},
\]
\[\label{eq:vlcEVdiag3}
\vlcinvariancecondensateC{r}{s}{t} \;\; =  \quad \vertexket{r}{s}{t}{}.
\]
Expanding both sides of these equation using the VLCs yields the eigenvalue equations
\[\label{eq:vlcEVeq1}
\sum_{b',c'} \tensor*{A}{*^{b'}_b^{c'}_c} \; \vlc{r}{s}{t}{a}{b'}{c'} = \; \vlc{r}{s}{t}{a}{b}{c},
\]
\[\label{eq:vlcEVeq2}
\sum_{a',c'} \tensor*{B}{*^{a'}_a^{c'}_c} \; \vlc{r}{s}{t}{a'}{b}{c'} = \; \vlc{r}{s}{t}{a}{b}{c},
\]
\[\label{eq:vlcEVeq3}
\sum_{a',b'} \tensor*{C}{*^{a'}_a^{b'}_b} \; \vlc{r}{s}{t}{a'}{b'}{c} = \; \vlc{r}{s}{t}{a}{b}{c},
\]
with
\begin{align}
\tensor*{A}{*^{b'}_b^{c'}_c} &= \frac{1}{q}\sum_{\gamma \in \varphi} {s^{b'\gamma}_{b}}^* t^{c'\gamma}_{c} \F{c' \gamma}{a b}{b'c}^*
\sqrt{d_{c'} d_{\gamma} \over d_c},\\
\tensor*{B}{*^{a'}_a^{c'}_c} &= \frac{1}{q}\sum_{\gamma \in \varphi} {r^{ \gamma a'}_{a}}^* t^{\gamma c'}_{c} \F{a b}{\gamma c'}{a' c}
\sqrt{d_{c'} d_{\gamma} \over d_c},\\
\tensor*{C}{*^{a'}_a^{b'}_b} &= \frac{1}{q}\sum_{\gamma \in \varphi} {r^{a'\gamma}_{a}}^* s^{\gamma b}_{b'} \F{a b}{a' b'}{\gamma c}
\sqrt{d_{a'} d_{b'} \over d_c}.
\end{align}
The eigenvectors of these matrices with eigenvalue $1$ are therefore proportional to the VLCs. To properly normalize them, we use the three  orthogonality relation \eqref{eq:ortho1}, \eqref{eq:ortho2} and \eqref{eq:ortho3} in  Appendix \ref{ch.vlcprops}. 
Note that one needs to simultaneously solve \eqref{eq:vlcEVeq1}, \eqref{eq:vlcEVeq2}
and \eqref{eq:vlcEVeq3} to obtain a complete solution including relative phases, which is not an easy task. 

In fact, the VLCs are not completely fixed but retain a \emph{residual gauge freedom}. This is not to be confused with the gauge freedom present in the $F$-symbols related to a unitary transformation in the elementary splitting spaces of the theory: the $F$-symbols of $\sA$ are gauge fixed in all of our considerations. But still, we may redefine the VLCs according to 
\[\label{eq:vlcgauge}
	\vlc{r}{s}{t}{a}{b}{c} \to {\zeta^{r}_{a}\zeta^{s}_{b}\over \zeta^{t}_{c}} \vlc{r}{s}{t}{a}{b}{c}\;\;,
\]
where $\zeta^{r}_{a}$ are arbitrary phases but with the condition  $\zeta^{\bar{r}}_{\bar{a}}= {\zeta^{r}_{a}}^*$. In practice, we use this freedom to gauge fix the $t^{\gamma a}_b$. 

A final overall phase freedom for the VLCs with $r,s,t\neq\varphi$  remains, corresponding to the  freedom in the $\cat{T}$ theory to redefine elementary vertices by a phase.

\subsection{Single boson without triple-vacuum vertex}
%
The simplest class of condensates is clearly
\begin{equation}
\begin{split}
\varphi &= 0 + \gamma \\
\gamma \times \gamma &= 0\label{eq:simcurrcondfus}
\end{split}
\end{equation}
in other words, a simple current of order two. The absence of a triple-vacuum vertex means that the condensate vertex expands as
\[
\condensatevertexket =	\vacvertexket{}{}{} + \vacLket{}{\gamma}{\gamma} +
										 \vacRket{\gamma}{}{\gamma} +	\vacBket{\gamma}{\gamma}{}.
\]
 By viewing (\ref{eq.condensatecondition}) in the $V_{\gamma\gamma}^{\gamma\gamma}$ subspace, we find
\[
	\twoanyons{\gamma}{\gamma} = \twototwoMvacM{\gamma}{\gamma}{\gamma}{\gamma}.
\]
This implies that  $\gamma\times\gamma = 0$,  $d_\gamma = 1$  so that the absence of a triple-vacuum vertex allows no other fusion rules than \eqref{eq:simcurrcondfus} for $\gamma$. 

\subsubsection*{$\sT$-propagators}
Our first objective is to obtain the $t^{\gamma a}_b$. The trick is to start with the
\[ 
	 t^{\gamma a}_a =\sqrt{q}\vlc{\varphi}{t}{t}{\gamma}{a}{a}.
\]
 Note that these are gauge invariant quantities, only non-zero when $a$ splits.
The diagrammatic equations  \eqref{eq:propagatorequations} lead \emph{e.g.} to 
\begin{equation}\label{eq:tabsimple}
	t^{\gamma a}_{b} \F{0a}{\gamma b}{\gamma a} = {t^{\bar{\gamma} b}_{a}}^*,\quad |t^{\gamma a}_{b}|^2 = \F{\gamma a}{\gamma a}{0b} \;\;.
\end{equation}
 For $\gamma=\bar{\gamma}$ this leads to two solutions for $t^{\gamma a}_{a}$, 
 \begin{align}
	t_{\pm a}^{\gamma a} = &\pm \F{0a}{\gamma a}{\gamma a}^{*\frac{1}{2}}\;\;,
\end{align}
and since there is no gauge freedom left, these correspond to the different particles in the restriction of $a$. We will write these as $t_+,t_-$, so $a\to t_+ + t_-$ or $\varphi \times a = t_+ + t_-$. Note that splittings into more than two sectors do not occur for a single boson condensate when we assume there are no  fusion multiplicities $N_{ab}^c> 1 $. In general the maximum number of charges $a$ restricts to is limited to the number of bosons in the condensate times the maximum fusion coefficient. 
 
The modulus of the $t^{\gamma a}_b$ follows straight from the right hand side of  \eqref{eq:tabsimple}, while the left hand side relates $t^{\gamma a}_b$ with $t^{\gamma b}_a$ as $\gamma$ is self-conjugate. Often, we can now use the VLC gauge freedom to fix the phase of $t^{\gamma a}_b$ to unity and the phase of $t^{\gamma b}_{a}$ to $\F{0a}{\gamma b}{\gamma a}$. A complication arises when $t$ and $a,b$ are self-dual, because the gauge freedom \eqref{eq:vlcgauge} requires $\zeta^{\bar{t}}_{\bar{a}} = {\zeta^t_a}^*$. For this case, the general equation 
\begin{equation}\label{eq:phasefunction}
	{t^{\gamma a}_b \over {\left.\bar{t}_{\bar{b}}^{\bar{\gamma} \bar{a}}\right.}^*} = R^{\gamma \bar{b}}_{\bar{a}} \F{0\bar{a}}{\gamma \bar{b}}{\bar{\gamma}\bar{a}}^*\F{0\gamma}{b \bar{a}}{\bar{b}\gamma} \F{b\bar{a}}{\gamma 0}{a\gamma} 
\end{equation}
 comes to help, which can be derived from the diagrammatic equality 
 \begin{equation}
 	\conddeathL{t} \quad = \quad\conddeathR{t}.
 \end{equation}
This equality is a result of the choice made in Eq.~\eqref{eq:tRchoice}. Choosing the inverse braid would have led to the condensate line passing under the cap.

\subsubsection*{General VLCs}
Finally, general VLCs are found using the eigenvalue equations \eqref{eq:vlcEVdiag1}--\eqref{eq:vlcEVdiag3} and normalized using \eqref{eq.orthogonality}.
 Equation \eqref{eq:vlcEVdiag3}, for example, simplifies in this case to 
\[
	\frac{1}{q}\sum_{\substack{a'\in r\\b'\in s}}\! \left( \F{ab}{a'b'}{0c} \!+\!  {r^{a'\gamma}_{a}}^* s^{\gamma b}_{b'}\F{ab}{a'b'}{\gamma c}\right)\! \vlc{r}{s}{t}{a'}{b'}{c}\! =\! \vlc{r}{s}{t}{a}{b}{c}.
\]
It may happen that  the eigenvector we find is not unique, and also the phase is not fixed by the normalization condition. 
It appears that by cross checking with the other eigenvalue equations \eqref{eq:vlcEVdiag1} and \eqref{eq:vlcEVdiag2} all information can be obtained. 
Currently, we are busy implementing this scheme for the family of $SU(2)_k$ theories for general condensates with a single non-trivial boson \cite{RE2018}.


\paragraph*{Example: $SU(2)_4$}Let us take $SU(2)_4$, with charges $0,1,2,3,4$ as an example. The fusion rules and other topological data are gathered in Table~\ref{tab:su2_4_topdat}. There is one boson $\gamma=4$ in the spectrum, and it turns out that $\varphi = 0 +4$  indeed satisfies all the consistency conditions for a condensate. We have quantum embedding index $q=2$ in this case. The table $\varphi \times a$ reveals the spectrum of the $\mathcal{T}$-theory,
\[
\begin{array}{lllll}
	\varphi &\times &1 &=& (1+3)\\
	\varphi &\times &2 &= &2_+ + 2_-\\
	\varphi &\times &3 &= &(1+3)\\
	\varphi&\times &4 &= &(0+4),
\end{array}
\]
\emph{i.e.} we find four charges $\varphi = (0+4),~(1+3),~2_+$ and $2_-$ (parentheses emphasize that we mean elementary charges in the $\cat{T}$-theory). The notation $2_{\pm}$ signifies that the charge $2$ splits in the condensation into two distinguishable charges of $\mathcal{T}$. These both lift to $2$, but differ in a sign in the VLCs involving $\gamma$, clarifying the mechanism of the splitting of charges. Since $1$ and $3$ have different topological twists, the charge $(1+3)$ gets confined. The charges $\varphi,\,2_+,\,2_-$ correspond to $0,\,3,\,\bar{3}$ of $SU(3)_1$ as has been noted in Ref.~\onlinecite{Bais2009}. 

\begin{table}
\begin{align*}
&\begin{array}{|l | lll | }
  \hline
  \multicolumn{4}{| l |}{SU(2)_{4}}\\
  \hline
  0  &d_0 = 1             &h_0=0							&\phase{0}=1\\
  1  &d_1 = \sqrt{3}   &h_1=\frac{1}{8}			&\phase{1}=-1\\
  2  &d_2 = 2             &h_2 = \frac{1}{3}			&\phase{2}=1\\	
  3  &d_3 = \sqrt{3}   &h_3=\frac{5}{8}			&\phase{3}=-1\\
  4  &d_4 =1              &h_4=1							&\phase{4}=1\\
  \hline
\end{array}\\
&\begin{array}{llll}
	1\times 1 = 0+2&&&\\
	1\times 2 = 1 + 3&2\times 2 = 0+2+4&&\\
	1\times 3 = 2+4	&2\times 3 = 1+3 &3\times 3=0+2&\\
	1\times 4 = 3		&2\times 4 = 2		&4\times 3=1 &4\times 4 = 0
\end{array}
\end{align*}
\caption{\label{tab:su2_4_topdat}Data for $SU(2)_4$. Top: quantum dimensions, topological spins and Frobenius-Schur indicator. Bottom: fusion rules.}
\end{table}

The procedure outlined above can be carried out by hand or implemented algorithmically. To illustrate the outcome, let us list the vertices for $SU(2)_4$ 
\begin{align}
	\sqrt{q} \ \condensatemeetsanyon{(1+3)} &=\quad \anyon{1}+\quad\anyon{3} + e^{i \frac{5\pi }{4}}\vertexket{4}{1}{3}{}  +e^{i \frac{7\pi }{4}} \vertexket{4}{3}{1}{} \\
	\sqrt{q} \ \condensatemeetsanyon{2_+} &= \quad\anyon{2} + \vertexket{4}{2}{2}{} \\
	\sqrt{q}\ \condensatemeetsanyon{2_-} &= \quad\anyon{2}  - \vertexket{4}{3}{1}{} \\
\end{align}
with $q = 2$. Note that the charge $(1+3)$ is confined, but the expansion of the vertex is perfectly well-defined.

\paragraph*{S matrix}
The $S$-matrix  of $SU(2)_4$ is
\[
	S_{ab} = 	
	\begin{pmatrix}
		1		&\sqrt{3}	&2		&\sqrt{3}	&1\\
		\sqrt{3}	&\sqrt{3}	&0		&-\sqrt{3}	&-\sqrt{3}\\
		2		&0		&-2		&0		&2\\
		\sqrt{3}	&-\sqrt{3}	&0		&\sqrt{3}	&-\sqrt{3}\\
		1		&-\sqrt{3}	&2		&-\sqrt{3}	&1\\
		\end{pmatrix}\;\;\;,
\]
while $\tilde{S}_{ab}$ is easily obtained 
\[
\tilde{S}_{ab}=
\begin{pmatrix}
0		&0			&0		&0		&0\\
0		&0			&0		&0		&0\\
0		&0		&2\sqrt{3}i	&0		&0\\
0		&0			&0		&0		&0\\
0		&0			&0		&0		&0\\
\end{pmatrix}\;\;\;.
\]
Note that indeed the non-zero contribution comes from the charge 2, which is the only charge that splits in the condensation $SU(2)_4 \to SU(3)_1$. It is now straightforward to obtain $S_{st}$ from \eqref{eq.smatU}. We get 
\[
	S_{st} = \begin{array}{c | c c c c }
		s,t		&(0+4)		&(1+3)				&2_{+}			&2_{-}		\\
					\hline
		(0+4)		&1		&0					&1				&1					\\
		(1+3)		&0		&0					&0				&0		\\		
		2_{+}		&1		&0					&e^{2\pi i/3}	&e^{-2\pi i/3}			\\
		2_{-}		&1	 	&0					&e^{-2\pi i/3}	&e^{2\pi i/3}			\\
		\end{array},
\]
and recognize $(1+3)$ as the confined charge of $\cat{T}$ --- its row and column are identically zero. The residual matrix
\[
	S_{uv}=
	 \begin{pmatrix}
		&1						&1				&1					\\
		&1						&e^{2\pi i/3}	&e^{-2\pi i/3}			\\
		&1	 					&e^{-2\pi i/3}	&e^{2\pi i/3}			\\
	\end{pmatrix}
\]
is indeed the not yet normalized  $S$-matrix of $SU(3)_{1}$. To obtain the normalized unitary  $S$-matrix we divide by  $\QD_{\cat{U}} = \sqrt{3}$.

\subsection{Single boson with triple-vacuum vertex}
%

For the more interesting case when $\gamma$ is not a simple current we  have 
\begin{equation}
\begin{split}
\varphi &= 0 + \gamma \\
\gamma \times \gamma &= 0+\gamma+\ldots\label{eq:condfus}
\end{split}
\end{equation}
The condensate vertex expands as
 \begin{align}
\condensatevertexket =	\frac{1}{\sqrt{q}} &\left\{ \vacvertexket{}{}{} + \vacLket{}{\gamma}{\gamma} +\right. \nonumber\\
										 &\left.\vacRket{\gamma}{}{\gamma} +	\vacBket{\gamma}{\gamma}{}  +\phi\ \vertexket{\gamma}{\gamma}{\gamma}{} \right\}.
\end{align}
The consistency condition  \eqref{eq.condensatecondition}  implies
\[
	\phi \twototwoSvacL{}{0}{\gamma}{\gamma}{\gamma}{}{}\quad = \quad \phi^* \vertexbra{\gamma}{\gamma}{\gamma}{}\;\;,
\]
which gives the phase of $\phi$ as
\[\label{eq.condensatephase}
	\frac{\phi}{|\phi|} =\pm\F{0\gamma}{\gamma\gamma}{\gamma\gamma}^{\frac{1}{2}*}\;\;.
\]
The choice of sign corresponds precisely to the gauge freedom for the VLCs discussed above. 
We will choose the positive sign by convention.
The modulus can easily be induced by the following instance of the consistency condition
\[
	\twoanyons{\gamma}{\gamma} + |\phi|^2 \twototwoS{\gamma}{\gamma}{\gamma}{\gamma}{\gamma}{}{} \quad= \quad \twototwoMvacM{\gamma}{\gamma}{\gamma}{\gamma} + |\phi|^2 \twototwoM{\gamma}{\gamma}{\gamma}{\gamma}{\gamma}{}{}\;\;.
\]
Rewriting the left hand side using $F$-symbols and solving for $|\phi|$ yields
\[\label{eq.condensatemodulus}
	|\phi| = \left(\frac{\delta_{0c} - \F{\gamma\gamma}{\gamma\gamma}{0c}}{\F{\gamma\gamma}{\gamma\gamma}{\gamma c}-\delta_{\gamma c}}		\right)^{1/2}.
\]
Equation (\ref{eq.condensatemodulus}) has to hold for any charge $c$ from the theory $\cat{A}$, showing that it is a rather stringent condition on the condensate. 
This way, we obtain
\[
	\phi \equiv \F{0\gamma}{\gamma\gamma}{\gamma\gamma}^{*\frac{1}{2}} \left(\frac{\F{\gamma\gamma}{\gamma\gamma}{0\gamma}}{1-\F{\gamma\gamma}{\gamma\gamma}{\gamma\gamma}}		\right)^{1/2}.
\]
Note that this returns  $\phi=0$ if $N_{\gamma\gamma}^{\gamma}=0$, in which case we recover the situation discussed earlier.

The condition for the propagator \eqref{eq:propagatorequations} applied to $t^{\gamma a}_a$ leads to the general solution
\begin{align}
	t_{\pm a}^{\gamma a} = &\pm\frac{1}{2} \F{0a}{\gamma a}{\gamma a}^{*\frac{1}{2}}\left( \pm \F{0a}{\gamma a}{\gamma a}^{*\frac{1}{2}} \phi^* \F{\gamma a}{\gamma a}{\gamma a}\right.\nonumber\\
							&\left.+\sqrt{\F{0a}{\gamma a}{\gamma a}^{*} \phi^{*2} \F{\gamma a}{\gamma a}{\gamma a}^2+4\F{\gamma a}{\gamma a}{0a} } \right)\;\;.
\end{align}
The choice of sign again leads to two distinguishable $\cat{T}$-particles $t_{\pm}$. 

The $t^{\gamma a}_b$ for $a\neq b$ can be solved subsequently. 
 The $t_{a}^{ \gamma a}$ 
appear in the solution for the general $|t^{\gamma a}_b|$\footnote{One might worry choosing the wrong sign in $t^{\gamma a}_{\pm a}$ when calculating the $t^{\gamma a}_{\pm b}$. Choosing non-corresponding signs, however, just produces $t^{\gamma a}_b = 0$.},
\[\label{eq:tab}
	|t^{\gamma a}_b|  = \sqrt{\phi^*t_a^{\gamma a} \F{\gamma a}{\gamma a}{\gamma b} +\F{\gamma a}{\gamma a}{0b} }\;\;.
\]
The phase is harder to compute.  The consistency condition implies (in fact for arbitrary condensates)
\[\label{eq.phasephinotzeropm}
	t^{\gamma b}_a= \F{0a}{\gamma b}{\bar{\gamma} a}^* {t_a^{\bar{\gamma} b}}^*.
\]
Due to the gauge freedom, one can often choose some of the phases for $t^{\gamma a}_b$ freely. The phase of $t^{\gamma b}_{a}$ then follows from \eqref{eq.phasephinotzeropm}. For self-dual particles equation 
\eqref{eq:phasefunction} provides the phase up to a sign, which is the residual gauge freedom in that case. 


\paragraph*{Example: $SU(2)_{10}$}An example of this more intricate case is the $SU(2)_{10}$ theory with charges labeled by integers from 0 to 10. 
The quantum dimensions, spins and Frobenius-Schur indicators are listed in Table~\ref{tab:su2_10_topdat}.
\begin{table}[h]
\[\begin{array}{|l | l ll|}
  \hline
  \multicolumn{4}{| l |}{SU(2)_{10}}\\
  \hline
  0  &d_0 = 1            				 	&h_0=0 		&\phase{0}=1\\
  1  &d_1 = \sqrt{2+\sqrt{3}}  		 	&h_1=\frac{1}{16}&\phase{1}=-1\\
  2  &d_2 =  1+\sqrt{3}          			&h_2 = \frac{1}{6}&\phase{2}=1\\
  3  &d_3 =\sqrt{2}+ \sqrt{2+\sqrt{3}}  		&h_3=\frac{5}{16}&\phase{3}=-1\\
  4  &d_4 =2+\sqrt{3}			              	&h_4=\frac{1}{2}&\phase{4}=1\\
    5  &d_5 = 2\sqrt{2+\sqrt{3}}           		&h_5=\frac{35}{48}&\phase{5}=-1\\
  6  &d_6 =2+ \sqrt{3}   				&h_6=1&\phase{6}=1\\
  7  &d_7 = \sqrt{2}+\sqrt{2+\sqrt{3}}             &h_7 = \frac{21}{16}&\phase{7}=-1\\
  8  &d_8 =1+ \sqrt{3}   				&h_8=\frac{5}{3}&\phase{8}=1\\
  9  &d_9 =\sqrt{2+\sqrt{3}}              		&h_9=\frac{33}{16}&\phase{9}=-1\\
   10  &d_{10} = 1             				&h_{10}=\frac{5}{2}&\phase{10}=1\\
  \hline
\end{array}\]
\caption{\label{tab:su2_10_topdat} The quantum dimensions, topological spins and Frobenius-Schur indicators for $SU(2)_{10}$.}
\end{table}

We will consider $\varphi = 0+6$. This time $6\in 6\times 6 =0+2+4+6+8$ leading to non-zero $\phi$. One can check, using the general topological data for $SU(2)_k$ in the Appendix \ref{ch.su2k}  that Eq. (\ref{eq.condensatephase}) and (\ref{eq.condensatemodulus}) give the result $\phi=2^{1/4} i$. The condensate vertex is therefore
\begin{multline}
	\sqrt{q}\condensatevertexket = \vacvertexket{0}{0}{0} + \vacLket{0}{6}{6} + \vacRket{6}{0}{6}\\\vspace{10pt}
	+ 	\vacBket{6}{6}{0}  +2^{1/4} i\ \vertexket{6}{6}{6}{}	
\end{multline}
 The table $\varphi \times a$ again reveals the spectrum of $\mathcal{T}$  (we use brackets to group the charges accordingly):
\[
\begin{array}{lllrl}
	\varphi &\times &1 &=&(1+5+7)\\
	\varphi &\times &2 &= &(2+4+6+8)\\
	\varphi &\times &3 &= &(3+7) + (3+5+9)\\
	\varphi&\times &4 &= &(4+10) + (2+4+6+8)\\
	\varphi&\times &5 &= &(1+5+7)+(3+5+9)\\
	\varphi&\times &6 &= &(0+4) + (2+4+6+8)\\
	\varphi&\times &7 &= &(1+5+7) + (3+7)\\
	\varphi&\times &8 &= &(2+4+6+8)\\
	\varphi&\times &9 &= &(3 +5 + 9)\\
	\varphi&\times &10 &= &(4+10)\\
\end{array}
\]
The $\mathcal{U}$-theory corresponds to $SO(5)_1$ in this case \cite{Bais2009}, with unconfined charges $(0+6), (3+7), (4+10)$. The embedding index is $q=3+\sqrt{3}$.

The $SU(2)_{10}$ vertices look messier, but the procedure to obtain the coefficients is the same.  We present the $t^{\gamma a}_a$ parts of the expansion, as these are most important in the calculation of observables. The vertices for the unconfined charges are
\begin{align}	 
	 \sqrt{q}\ \condensatemeetsanyon{\hspace{.2cm}(3+7)} =& {i\over 2^{1/4}}\vertexket{6}{3}{3}{}  -{i\over 2^{1/4}} \vertexket{6}{7}{7}{}+\dots \nn\\
	\sqrt{q}\  \condensatemeetsanyon{\hspace{.2cm}(4+10)} =&-i 2^{1/4} \vertexket{6}{4}{4}{}+\dots
\end{align}  
while the vertices for the confined charges are
\begin{align}
	 \sqrt{q}\ \condensatemeetsanyon{\hspace{.5cm}(1+5+7)} &= i(2-\sqrt{3})^{1/4}\vertexket{6}{5}{5}{} \nn\\
	 																		&+i(14-8\sqrt{3})^{1/4}\vertexket{6}{5}{5}{}+\dots\nn\\
	 \sqrt{q}\ \condensatemeetsanyon{ \hspace{.9cm}(2+4+6+8)} &={ i(-1+\sqrt{3}) \over 2^{3/4}}\vertexket{6}{4}{4}{} \\
	 																			&+{ i(-1+\sqrt{3}) \over 2^{3/4}}\vertexket{6}{6}{6}{}+\dots\nn\\
	 \sqrt{q}\ \condensatemeetsanyon{\hspace{.5cm}(3+5+9)} &= - i(14-8\sqrt{3})^{1/4}\vertexket{6}{3}{3}{}\nn\\
	 																		& - i(2-\sqrt{3})\vertexket{6}{5}{5}{}+\dots\nn\ 
\end{align}
The off-diagonal VLCs  obey ${t^{\gamma b}_{a}}^* = \F{0a}{\gamma b}{\gamma a} t^{\gamma a}_{b}$, but we will not print them explicitly here.

Using these, we obtain the $S$-matrix
\begin{equation}
	S_{uv} = \begin{pmatrix}
		1 &\sqrt{2} &1\\
		\sqrt{2} &0 & - \sqrt{2}\\
		1&-\sqrt{2}& 1
	\end{pmatrix}.
\end{equation}
This is the same $S$-matrix as the Ising anyon model. The difference between the Ising model and $SO(5)_1$ resides in the spin of the non-simple current $(3+7)$ which is $5/16$ for $SO(5)_1$ and $1/16$ in the Ising anyon model. (There is a whole family of anyon models with the fusion rules of Ising \cite{Kitaev2006}.)


\subsection{Noncondensable bosons}
To appreciate the fact that the fusion condition on the condensate yields nontrivial constraints, consider a system with five layers of Fibonacci anyons all with the same chirality. This is described by the product of five Fibonacci theories. 

A single Fibonacci theory has a single nontrivial anyonic charge $\tau$ with nontrivial  fusion rule
\begin{equation}
	\tau\times \tau = 0+\tau.
\end{equation}
This fusion rule leads to the Fibonacci sequence when counting the possible ways to fuse an increasing number of $\tau$ anyons back to the vacuum. It implies that the quantum dimension of $\tau$ equals the golden ratio
\begin{equation}
	d_\tau = {1+\sqrt{5}\over 2}.
\end{equation}
It is possible to compute  the $F$- and $R$-symbols by hand in this case by using the constraints set by the pentagon and hexagon relations \cite{Preskill2004}. The result is 
\begin{equation}
		F = 
		\begin{pmatrix}
			d_\tau^{-1}& d_\tau^{-{1\over 2}}\\
			d_\tau^{-{1\over 2}}&-d_\tau^{-1}
		\end{pmatrix}\qquad
		R^{\tau \tau}_0 = e^{-i 4\pi/5},\quad R^{\tau\tau}_{\tau} = e^{i 3\pi/5}
\end{equation}
(where $F_{ef} = \F{\tau \tau\tau}{\tau}{ef}$). It follows that $\theta_\tau = e^{i 4\pi/4}$ and $\phase{\tau}=1$.

Now take five copies of the Fibonacci theory. Charges are now labeled $(a_1,a_2,a_3,a_4,a_5)$ with $a_i = 0, \tau$ and all symbols are multiplicative. The charge $\gamma = (\tau,\tau,\tau,\tau,\tau)$ has $\theta_\gamma =1$. It fuses to the vacuum, so any fusion product of $\gamma$'s contains a charge with trivial spin such that the conditions of the original TSB scheme from Ref.~\onlinecite{Bais2009} and as written in Subsection~\ref{subsec:TSBold}  are satisfied. 

The trouble arises when we try to satisfy our diagrammatic conditions. We have 
\begin{equation}
	\braidket{\gamma}{\gamma}{\gamma}{} = e^{i 3 \pi} \vertexket{\gamma}{\gamma}{\gamma}{}\;,
\end{equation}
which means we can never satisfy Eq.~\eqref{eq:braidconditioncondensate}.
We could try to relax the condition on braiding but it turns out that the condition on fusion also leads to problems. In particular,  for $c_0 = (0,0,0,0,0)$ and $c_1 = (0,0,0,0,1)$ we find
\begin{align}
\left(\frac{\delta_{0c_0} - \F{\gamma\gamma}{\gamma\gamma}{0c_0}}{\F{\gamma\gamma}{\gamma\gamma}{\gamma c_0}-\delta_{\gamma c_0}}		\right) &=\sqrt{1+\sqrt{5}\over 2}\\
\left(\frac{\delta_{0c_1} - \F{\gamma\gamma}{\gamma\gamma}{0c_1}}{\F{\gamma\gamma}{\gamma\gamma}{\gamma c_1}-\delta_{\gamma c_1}}		\right) &=\sqrt{-2+\sqrt{5}}
\end{align}
such that  Eq.~\eqref{eq.condensatemodulus} can not be satisfied. 

One must conclude that, although $\gamma$ is a boson, it cannot condense. It is in fact a mathematical theorem that any number of copies of  Fibonacci theories with the same chirality does not admit algebra objects, or in our language, stable condensates \cite{2011_Davydov_arXiv_1103.3537}.

\section{Conclusion and outlook}
In this work we studied Bose condensation in  anyon models in a
graphical language. Based on the assumption that the condensate acts
as the vacuum for a new theory, constructed as a subset of all the
diagrams of the parent phase, we find consistency conditions on the
condensate (equating it to a Frobenius algebra in the context of
MTCs). The diagrams of the broken phase correspond to superpositions
following the identification of charges due to the condensate, but in
order to solve the consistency requirements on the condensate and
obtain the correct expansion of the diagrams one has to introduce
vertex lifting coefficients (VLCs) governing the expansion of the
elemantary fusion and splitting vertices. The VLCs are the main
novelty of the present paper. They allow for the explicit calculation
of diagrams in the broken phase and give access to the full
topological data straight from the formalism. We showed how one can
obtain the VLCs in many physically relevant cases and worked out
illustrative examples. In particular did we give the calculation in some detail
for the breaking from $SU(2)_4 \rightarrow SU(3)_1$, and  for 
$SU(2)_{10} \rightarrow SO(5)_1$ showing how the $S$-matrix could be
obtained.

Let us conclude with some comments on a number of topics where we expect our findings to be useful.
\paragraph*{Topological data} As mentioned before, the presented
results form an alternative route to the calculation of topological
data (i.e. $F$-symbols and $R$-msymbols) which is in general  a
difficult task.
It complements the route of  explicitly solving  the hexagon and pentagon equations
(which becomes  computationally demanding from typically
five or six particle types\cite{Bonderson:2007zz}) or using quantum group
representation theory which give access to  $SU(2)_k$ and similar
series. The results we obtained in principle allows to obtain all
theories that can be constructed from, say, some $SU(2)_k$ theory by
condensing the bosons (at least for condensates of the form
$\varphi=0+\gamma$).

We should also mention the computer program KAC~\cite{KAC} at this point, which generates the fusion coefficients, conformal weights, the $S$-matrix and some more data for rational CFTs based on affine Lie algebras \emph{and} their simple current extensions.  In Ref.~\onlinecite{1996_Fuchs_NPB_473} it is shown how KAC computes the $S$-matrix for simple current extensions.  It is known that conformal  extensions can be reinterpreted as Bose condensates \cite{Bais2009} opening up a possibility to extend the functionality of KAC. Further development of the technology presented in this paper could in principle allow to include the $F$-symbols and $S$-matrix for general condensates by computing the VLCs, as as long as the $F$-symbols for the parent theory are known. A fully functional algorithm to compute the VLCs, however, needs considerable additional work  especially when $\varphi \neq 0+\gamma$. 
\paragraph*{Bulk-boundary correspondence}Untill recently it was not
quite clear what the precise role of the $\mathcal{T}$-theory, which
appears `halfway' in the symmetry breaking formalism, was. In particular the question
which of the sectors in $\sT$ are gapless and which ones describe massive degrees of freedom has been
the subject of discussion \cite{Grosfeld2008,Bais2009b}.
Moreover, in recent work \cite{PhysRevX.3.021009}, it was found that the existence of {\em protected} edge modes that live
between a topological phase and a vacuum depends on the absence of a so-called `Lagrangian subgroup' in the
fusion algebra of the theory. The definition of this Lagrangian subgroup coincides with our definition of
a Bose condensate for which the $\cat{U}$ theory becomes trivial
(\emph{i.e.} only the new vacuum is unconfined), at least for the
Abelian examples discussed in this paper. In recent work\cite{2014arXiv1407.5790B} we argued that
 the confined charges (those in $\mathcal{T}$ but not $\mathcal{U}$)
 appear on the boundary as massive solitons while the unconfined charges correspond to massless
edge modes. The connection between gapless \emph{vs.} gapped modes and the  
findings in Ref.~\onlinecite{2012CMaPh.313..351K} deserves more attention with
a possible role for the VLCs.

A related issue is the possibility of 1D condensates, as considered by Kong in Ref.~\onlinecite{2012CMaPh.313..351K}. These are condensates that satisfy the fusion condition but not the braiding condition, which can also occur in $\cat{T}$-like theories that have consistent fusion but no consistent braiding and therefore only  planar diagrams.  Extending the formalism developed in this paper one should be able to compute VLCs for this type of condensates as well, leading to fully computational diagrammatics. 

\paragraph*{Explicit models}In earlier work \cite{Bais:2011iu}, it was found that the modular $S$-matrix is a good quantity to characterize the
topological order of a phase of matter. For the models studied in that work, discrete lattice gauge theories, it was possible
to measure the expectation value of the $S$-matrix by inserting a pair of quasiparticle worldline-loops that formed a Hopf link.
One may wonder how general an order parameter the $S$-matrix is, since for many systems such an insertion might not be possible.
However in recent work \cite{PhysRevB.85.235151,2013arXiv1308.3878Z} it was pointed out that one can also extract the $S$-matrix from the overlap between different
ground states on the torus --- an approach which can in principle be applied to any quantum many-body system.
Since the present work enables the calculation of $S$-matrices in the different phases of a topologically ordered system
one may predict the value of this order parameter in phases that are related to one another by Bose condensation.

The relevance of the present work to Levin-Wen (or string-net) models
\cite{Levin:2004mi} can be twofold.
First of all, explicit wave functions for the ground state of these models in terms of tensor product states are known \cite{PhysRevB.79.085118,PhysRevB.79.085119}.
They are expressed in terms of the $F$-symbols of the underlying anyon model. The phase diagram in these models
contains states that are related to one another by Bose condensation. Since the present work allows one to relate the $F$-symbols
of two different theories using VLCs, it may lead to explicit expressions for the ground state wave function in the presence of a
Bose condensate.
Secondly, in this work we find expressions for the $F$-matrices of the $\sT$ theory, in other words, $F$-matrices that satisfy the
pentagon relations but not necessarily the hexagon relations --- the mathematical name for such a structure is a {\em spherical fusion category}. The Hamiltonian of string-net models does not necessarily require consistent braiding and the $F$-matrices we obtain for the $\sT$ theory could therefore serve as input. It is generally believed\cite{Burnell20102550,2012CMaPh.313..351K} that the TQFT describing the topological order of such models is given by the center of the spherical fusion category. In the mathematical literature it is known that the center of our $\sT$ theory equals $\sA \, \times \, \overline{\sU}\,$ \cite{davydov2010}.
Mainly the microscopic origins of the resulting topological phase and
the relation to Bose condensation are very interesting from a physical
point of view.

Finally, a Lagrangian formulation of topological symmetry breaking would be
interesting. This would most naturally use Chern-Simons theory or CFT
as starting point. The use of more traditional field
theoretical language would give  a lot of physical insight concerning  the connection to conventional symmetry breaking. A
difficulty  is however the nonlocality of
the order parameters for topological order and at present it is not
quite clear how, say, a perturbing term driving the transition can be
included consistently. Partial progress was made in
\onlinecite{2014arXiv1407.5790B} where the vertex operator expectation
value on the boundary was identified as an order parameter.

Despite the tremendous progress of the past few years, the level of understanding
of condensate-induced transitions in topological phases is still far-off from the level
of understanding of ordinary symmetry breaking phase transitions, and
the field is wide open for future research.

The authors would like to thank Eric Opdam, Joost Slingerland, Roland van der Veen and Steve Simon for inspiring discussions.
ISE is funded by the Foundation for Fundamental Research on Matter (FOM), which is part of the Netherlands Organisation for Scientific Research (NWO). JCR is funded by the TOPNES grant from the UK EPSRC (EP/I031014/1).

\appendix

\section{\label{sec.rules}Anyon models}

\subsection{Particle spectrum}

\paragraph*{Fusion algebra}An anyon model $\cat{A}$ has a finite collection of topological charges $\{a,b,c,\dots\}$ (or anyons), which obey fusion rules 
\[\label{EQfusionrules}
        a \times b = b \times a = \sum_{c\in \cat{A}}        N_{ab}^c c\;\;.
\]
We assume no fusion multiplicities: $N_{ab}^c$ only to take the values $0$ and $1$. In general they are  non-negative integers which introduces additional indices on the vertices to keep track of fusion / splitting channel.

There is a unique trivial particle, the  vacuum, which we  denote with $0\in\cat{A}$. It has the property that $0\times a =a\times 0=a$ for all $a$.  Fusion is associative such that $(a\times b)\times c = a\times (b\times c)\equiv a\times b\times c$. Each charge $a\in\cat{A}$  has a unique conjugate charge $\bar{a}\in \cat{A}$ such that $a$ and $\bar{a}$ can annihilate:  $a\times \bar{a}=0+\dots$. If some superposition of charges reads $s = a+\dots$  we use the notation $a\in s$ (for example $0\in a\times \bar{a}$).

An important number is the quantum dimension $d_a$ of the charge $a$, which is the largest eigenvalue of the fusion matrix $N_{a}$ with components $(N_{a})_{bc}=N_{ab}^c$. It is the asymptotic growth of the dimension of the topological Hilbert space spanned by anyons of charge $a$: take a fusion product of $N$  anyons of charge $a$ then the dimension of the topological Hilbert for these anyons aproaches  $d_a^N$ in the large $N$ limit. The quantum dimensions form a one-dimensional representation of the fusion algebra
\[
        d_a d_b 
        = \sum_{c} N^c_{ab} d_c\;\;.
\]
The \emph{total quantum dimension} of the theory is defined as $\QD_\cat{A} = \sqrt{\sum_a d_a^2}$. 

\paragraph*{Propagator}To construct diagrams we associate a directed line to each charge label  $a$, representing the anyon propagating in time (which we  take  flowing upward). Reversing the orientation of a line segment is equivalent to conjugating the charge, so
\[
        \anyon{a}\quad =\quad\conjanyon{\bar{a}}\;\;.
\]
The vacuum line is usually left out of the diagrams. When made explicit, it is mostly drawn dotted.

\paragraph*{Topological spin}The topological spin $\theta_a$, also called twist factor, is associated with a $2\pi$ rotation of an anyon of charge $a$ and is diagramatically defined by 
the twisted lines
\[ 
\twistright{a}\quad =\quad \theta_a \longanyon{a}\quad =\quad \twistleft{a}\;\;,
\] 
and
\[ 
\antitwistright{a}\quad =\quad \theta_a^* \longanyon{a}\quad =\quad \antitwistleft{a}\;\;.
\] 
When applicable, $\theta_a$ is related to the (ordinary angular momentum) spin or CFT conformal scaling dimension $h_a$ of $a$, by
\[
  \theta_a = e^{i2\pi h_a}\;\;.
\]
In any case, it is often convenient to give $h_a$ instead of $\theta_a$.

\subsection{Fusion spaces}
Operators on anyons are constructed from elementary splitting spaces $V^{ab}_c$, which are complex vector spaces of dimension $N_{ab}^c $. We pick an element $\ket{a,b;c}$ for all $V^{ab}_c$ and associate to it a vertex
\[ \label{eq.vertexdef}
        \ket{a,b;c} = \left(\frac{d_c}{d_ad_b}\right)^{1/4}     \vertexket{a}{b}{c}{}\;\;.
\]
Note that $\ket{a,b;c}$ as well as the corresponding diagram are necessarily zero when $N_{ab}^c =0$ (and non-zero otherwise).
The normalization factor $(d_c/d_ad_b)^{1/4}$ ensures that bending lines up and down will at worst give a phase, so the evaluation of diagrams is maximally invariant under topological manipulations (see Eq.~\eqref{eq:FSindicator}). We refer to the diagram on the right hand side of Eq.~\eqref{eq.vertexdef} as a {\em splitting vertex}.

Dual to the splitting space we have the fusion space $V_{ab}^c=(V^{ab}_c)^*$, with dual states $\bra{c;a,b}$ and {\em fusion vertex}
\[
        \bra{c;a,b} = \left(\frac{d_c}{d_ad_b}\right)^{1/4}\vertexbra{a}{b}{c}{}\;\;.
\]      
The splitting and fusion vertices are the elementary building blocks for more complicated operators, that can be formed by stacking them such that the charge lines connect.
For example, we can now write down the inner product $\braket{c;a,b}{a,b;c'}$ as an operator $c \to c$ as 
\[\label{EQvertexbraket}
        \vertexbraket{c}{a}{b}{c'}{}{} = \delta_{cc'}\sqrt{\frac{d_a d_b}{d_c}} \quad   \anyon{c}\;\;,
\]
which encodes diagrammatically that anyonic charge is conserved. 

The identity operator on a pair of anyons with charges $a$ and $b$  is 
\[
        \mathbb{I}_{ab} = \sum_{c} \ket{a,b;c}\bra{c;a,b}\;\;,
\]
which can now be written as
\[ \label{EQidentity}
        \anyon{a}\quad\anyon{b} = \sum_{c} \sqrt{\frac{d_c}{d_a d_b}}\quad \twototwoM{a}{b}{a}{b}{c}{}{}\;\;.
\]
All equations of diagrams can be applied locally in bigger, more complicated diagrams to do  calculations. The notation $V_{a_1\dots a_n}^{a'_1\dots a'_m}$ is generally used for operators taking anyons $a_1,\dots,a_n$ to anyons $a'_1,\dots,a'_m$.

\paragraph*{Unitarity} We will consider unitary theories only. The conjugate of a diagrammatically given operator is obtained by mirroring the diagram in the horizontal plane and then reversing all arrows. Coefficients are complex conjugated, \emph{e.g.}
\[
        \left[ A \vertexket{a}{b}{c}{}{}\right]^\dag = A^* \vertexbra{a}{b}{c}{}{}.
\]

\paragraph*{$F$-symbols}Let us take the splitting space $V_d^{abc}$. We can represent the states in this space as superpositions of diagrams of the form
\[
        \onetothreeL{a}{b}{c}{d}{e}{}{} \;\; ,
\]
however we might as well have represented them using diagrams of the form
\[
        \onetothreeR{a}{b}{c}{d}{e}{}{} \;\; .
\]
These are merely two representations in terms of different basis states. For consistency, the two representations have to be related by a unitary transformation. This is  an essential piece of data for the anyon model captured in a set of so called $F$-symbols.\footnote{These are analogues to the    $6j$-symbols from the theory of angular momentum and are sometimes called $q6j$-symbols in the quantum group context.} These
are complex numbers $\F{abc}{d}{ef}$, that implement the $F$-move
\[
        \onetothreeL{a}{b}{c}{d}{e}{}{}  = 
        \sum_{f} 
        \F{abc}{d}{ef} \onetothreeR{a}{b}{c}{d}{f}{}{}\;\;,
\]
(when the diagram on the right evaluates to zero, we take the corresponding $F$-symbol as zero as well). Unitarity amounts to
\[
        \Fvar{abc}{d}{fe}{\dagger} = \F{abc}{d}{ef}^* = \Fvar{abc}{d}{fe}{-1}\;\;.
\]
The quantum dimension is related to the $F$-symbols via
\[
        d_a = |\F{a\bar{a}a}{a}{00}|^{-1}\;\;.
\]
There is an important gauge freedom in the $F$-symbols corresponding to a choice of phase $u^{ab}_c$ for all elementary splitting vertices,
\begin{equation}\label{Fsymgauge}
	\F{abc}{d}{ef}\to \frac{u^{af}_d u^{bc}_f}{u^{ab}_e u^{ec}_d} \F{abc}{d}{ef}.
\end{equation} 
In order for the theory to be consistent, the set of $F$-symbols should satisfy the so called pentagon equations,
\begin{equation}
	\F{fcd}{e}{gl}\F{abl}{e}{fk} = \sum_{h} \F{abc}{g}{fh}\F{ahd}{e}{gk} \F{bcd}{k}{hl},
\end{equation}
(see Refs.~\onlinecite{Kitaev2006,Bonderson:2007zz} for details).

It is extremely convenient to define $F$-symbols for diagrams with two-anyons coming in and two-anyons coming out,
\[
  \twototwoS{a}{b}{c}{d}{e}{}{} = \sum_{f} \F{ab}{cd}{ef}\twototwoM{a}{b}{c}{d}{f}{}{}.
\]
With the use of (\ref{EQidentity}) and (\ref{EQvertexbraket}) one may deduce
\begin{align}
  \F{ab}{ab}{0c} &= \sqrt{\frac{d_c}{d_a d_b}} N_{ab}^c \;\;, \\
  \F{ab}{cd}{ef} &=\sqrt{\frac{d_e d_f}{d_a d_d}}     
  \F{ceb}{f}{ad}^*\;\; .
\end{align}
These alternative $F$-moves can be used to change a splitting vertex with one leg bent down into a fusion vertex, \emph{et cetera}. This gives equalities like
\[
        \twototwoSvacL{0}{c}{a}{b}{\bar{a}}{}{} = \F{0c}{ab}{\bar{a}c} \vertexbra{a}{b}{c}{}\;\;,
\] 
and 
\[
        \twototwoSvacR{a}{b}{c}{0}{\bar{b}}{}{} =\F{ab}{c0}{\bar{b}c} \vertexket{a}{b}{c}{}\;\;,
\]
and so on. The symbols $\F{0c}{ab}{\bar{a}c}$ and $\F{ab}{c0}{\bar{b}c}$ are in fact phases. An important case is when above manipulations are used to straighten a charge line. This gives a factor $\phase{a}\equiv \F{0a}{a0}{\bar{a}a} = d_a \F{a\bar{a}{a}}{a}{00}$:
\[\label{eq:FSindicator}
\twototwoSvacLvacR{a}{\bar{a}} = \phase{a}\quad\anyon{a}\;\;.
\]
For most $a$ these can be set to 1 by a gauge transformation of the $F$-symbols, but for self-conjugate charges it is a gauge invariant quantity known as the Frobenius-Schur indicator. For $a =\bar{a}$, one has $\phase{a}=\pm 1$. 

For the expert reader we note that we do not use additional flags in the cup and cap diagrams, but choose to make explicit use of the Frobenius-Schur indicator to straighten charge lines.

\paragraph*{Topological Hilbert space} Anyon models be understood as conventional quantum mechanics on the topological Hilbert space. For a system with overall neutral  anyonic charge containing  anyons $a_1,\dots, a_n$ this is the space $V_0^{a_1\dots a_n}$. General states are of the form
\begin{equation}
	\ket{\psi}=\sum_{a_1,a_2,a_3,a_4,c } {\psi_{a_1a_2a_3a_4c} \over (d_{a_1} d_{a_2} d_{a_3} d_{a_4} d_{c})^{1/4}} \fouranyonstate{a_1}{a_2}{a_3}{a_4}{c}{}{},
\end{equation}
which may be thought of as the operator creating the state 
 from the vacuum.

\subsection{Braiding}

\begin{figure}[h!]
\includegraphics[width=0.7\columnwidth]{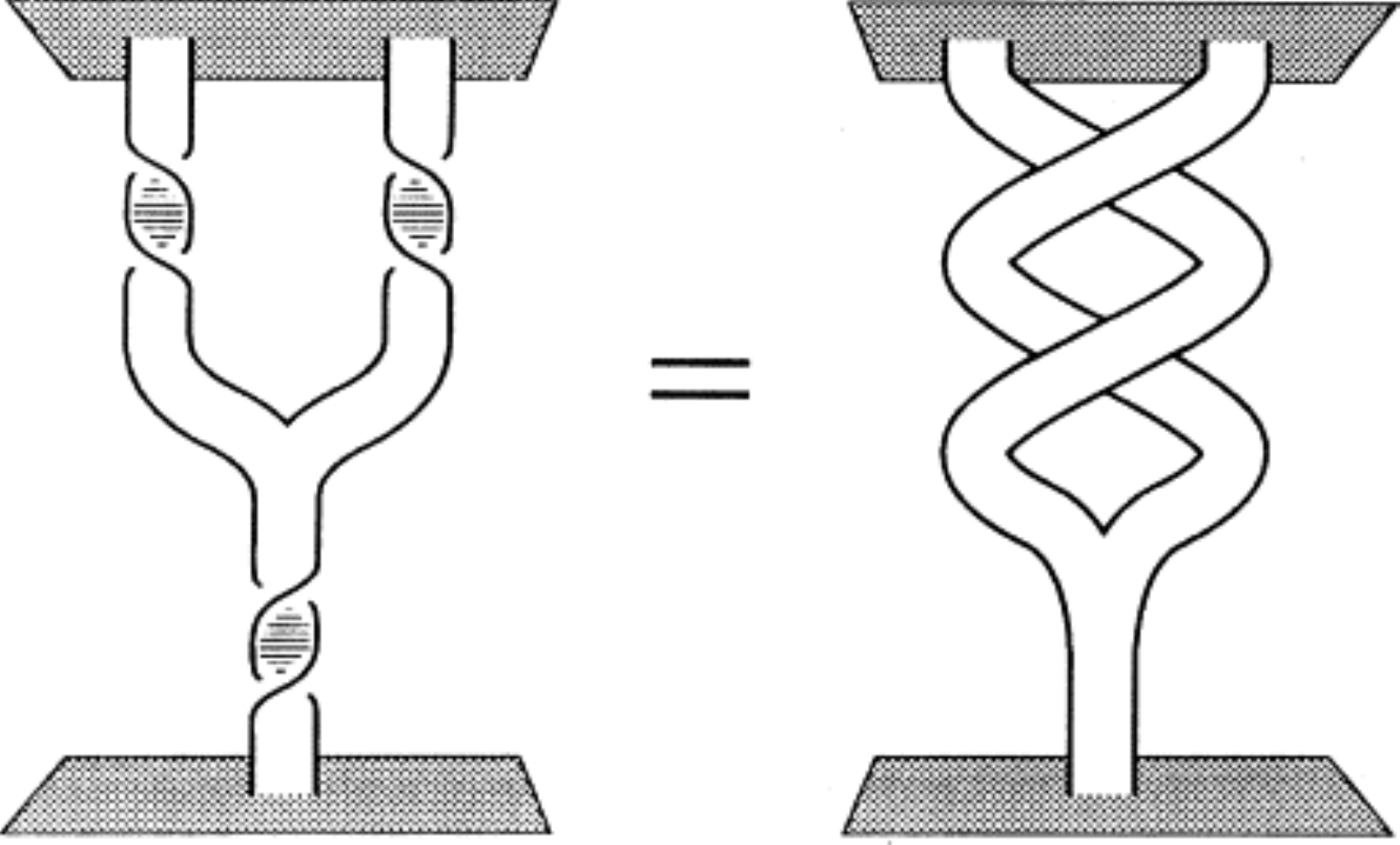}
\caption{\label{fig.suspenders}Suspenders diagram: pictorial
  representation of the monodromy equation~\eqref{eq:monodromy} (taken
  from Ref.~\onlinecite{Bais2009}).}
\end{figure}

\paragraph*{$R$-matrix} A characteristic property of anyons is their non-trivial exchange statistics. The effect of two anyons switching places in the system is taken into account by the braiding operators or $R$-matrices, which are written as
\[
  R_{ab} =\quad \braid{a}{b},\qquad R^{\dagger}_{ab} = R^{-1}_{ab} =\quad \antibraid{a}{b}.
\]
They are defined by their action on basis states of the elementary spaces $V_{c}^{ab}$ captured in a set of $R$-symbols $R_c^{ab}$. These lead to the diagrammatic  $R$-moves
\begin{align}
  \braidket{a}{b}{c}{} &= R^{ab}_{c} \vertexket{b}{a}{c}{}\;,
\end{align} 
and
\begin{align}
  \braidbra{a}{b}{c}{} &= (R^{ab}_{c})^* \vertexbra{b}{a}{c}{}.
\end{align} 
The full braiding operator is then
\[
  \braid{a}{b}\quad = \sum_{c} \sqrt{\frac{d_c}{d_a d_b}} R^{ab}_{c} \twototwoM{b}{a}{a}{b}{c}{}{}.
\]
A similar equation holds for the inverse operation. Note that  unitarity implies that $(R^{ab}_c)^{-1} = (R^{ab}_c)^{*}$.
Also the following relation has to hold:
\[
        \theta_a = \phase{a}(R^{\bar{a}a}_0)^*.
\]
The effect of a double braiding -- or monodromy -- of two anyons is governed by the monodromy equation
\[\label{eq:monodromy}
        \doublebraidket{a}{b}{c}{} = \frac{\theta_c}{\theta_a\theta_b}\vertexket{a}{b}{c}{}\;\;,
\]
which gives the  operator identity
\[
        \doublebraid{a}{b} = \sum_{c} \frac{\theta_c}{\theta_a\theta_b}\sqrt{\frac{d_c}{d_a d_b}}\ \twototwoM{a}{b}{a}{b}{c}{}{}\;\;.
\]
When thinking of the charge lines as ribbons, it is a matter of topological manipulation to see that the monodromy equation holds, see the so-called ``suspenders diagram'' in Figure~\ref{fig.suspenders}.

\subsection{Evaluation of diagrams}
\paragraph*{Tensor product and entanglement} 
The tensor product of two operators $X$ and $Y$ is given by the diagram
\[
	\tensoroperator{X\otimes Y}\quad =\quad \operator{X}{}{}{}{}\quad \operator{Y}{}{}{}{}\;\; .
\]
The notion of entanglement therefore has an appealing graphical visualization: states or operators can be written as a tensor product when they are equivalent to disjoint diagrams without  nontrivial charge lines connecting the parts.
The quantum trace of an  operator $X$, denoted $\qTr X$, is constructed diagrammatically by closing the diagram with loops that match the outgoing lines on top with the incoming lines on the bottom  at the same position
\[
 \qTr\left[     \operator{X}{a_1}{a_n}{a'_1}{a'_n} \right] = \delta_{a_1a'_1}\dots\delta_{a_na'_n} \quantumtrace{X}{a_1}{a_n}\;\;.
\]
\paragraph*{Topological $S$-matrix}
The topological $S$-matrix  is defined as
\[
S_{ab} = \frac{1}{\QD_\cat{A}} \ \smatrix{a}{b}\;\;.
\]
It encodes a wealth of information about the theory $\cat{A}$.  By applying the monodromy equation, we find
\[
  S_{ab} = \frac{1}{\QD_\cat{A}} \sum_c N_{ab}^c \frac{\theta_c}{\theta_a \theta_b} d_c\;\;.
\]
The theory is called modular when $S_{ab}$ is non-degenerate (and in that case unitary). Together with the  $T$-matrix with coefficients
\[
  T_{ab} = e^{2 \pi i c/24} \theta_a \delta_{ab}
\]
it forms a representation of the modular group $SL(2,\mathbb{Z})$ with defining relations $(ST)^3 =S^2=C$ and $S^4 = 1$, where $C_{ab} =  \delta_{a\bar{b}}$ is  the charge conjugation matrix. Here $c$ is the topological central charge, which is equal to the central charge of the corresponding CFT mod 24 when applicable. It can be determined mod 8 from the twist factors and quantum dimensions by the relation
\[
        \exp\left(\frac{2\pi i\;  c}{8}\right) =  \frac{1}{\QD_\cat{A}}\sum_a d_a^2 \theta_a.
\]
The $S$-matrix gives direct access to the quantum dimensions of the charges and the total quantum  dimension. The fusion rules can be derived via the Verlinde formula
\[\label{eq.verlinde}
  N_{ab}^c = \sum_{x} \frac{ S_{ax} S_{bx} S_{\bar{c} x} }{ S_{0x} }\;\;.
\]
The $S$-matrix elements can in principle be measured by certain  interferometry measurements \cite{Bonderson:2007zz,Bonderson2009787,Bonderson2007}. It can also be used as an order parameter for topological order \cite{Bais:2011iu,PhysRevB.85.235151}. Recently it was used to determine the non-abelian order in a model of interacting lattice bosons\cite{2014_Zhu_PRL_112}.

\section{Topological data for $SU(2)_k$ theories}\label{ch.su2k}
The $SU(2)_{k}$-theories have a particle spectrum that can be understood as truncated versions of the representation theory of $SU(2)$.
These theories are realized as TQFTs by a Chern-Simons theory \cite{Witten:1988hf} with gauge group $SU(2)$ and coupling constant $k$. Mathematically,
the information presented below can all be cast in the form of the representation theory of the $q$-deformation of $SU(2)$ with $q=e^{i\frac{2\pi}{k+2}}$.

The charges are labelled by integers $a=0,1,\dots,k$. The fusion rules are given by 
\begin{align}
        a \x b  = &|a-b| + (|a-b|+2)+\dots\\
                  &+\min\{a+b,\, 2k-a-b\},
\end{align}
i.e. $N_{ab}^c = 1$ when $|a-b|\leq c\leq \min\{a+b, 2k-a-b\}$ and $a+b+c = 0\mod{2}$, and zero otherwise.

For the $F$-symbols, one has the general formula
\[\label{eq.FsymbolsSUk}
  \F{abc}{d}{ef} = i^{a+b+c+d }\sqrt{\q{e+1} \q{f+1}}\qsixj{a}{b}{c}{d}{e}{f}
\]
where
\begin{multline}
  \qsixj{a}{b}{c}{d}{e}{f}
      =\Delta(a,b,e) \Delta(e,c,d) \Delta(b,c,f)\Delta(a,f,d)\\
      \times\sum_z \left\{ 
        \frac{
          (-1)^{z}\q{z+1}! }{
          \q{z-\frac{a+b+e}{2}} !\q{z-\frac{e+c+d}{2}}!\q{z-\frac{b+c+f}{2}}!
        }\right.\\
    \left.\x\frac{1}{
          \q{z-\frac{a+f+d}{2}}!\q{\frac{a+b+c+d}{2}-z}!
          }\right.\\
    \left.\x\frac{1}{
        \q{\frac{a+e+c+f}{2}-z}!\q{\frac{b+e+d+f}{2}-z}!
          }\right\}       
\end{multline}
with
\begin{align}
  \Delta(a,b,c)&=\sqrt{ \frac{
          \q{\frac{-a+b+c}{2}}! \q{\frac{a-b+c}{2}}!\q{\frac{a+b-c}{2}}! 
          }{
          \q{\frac{a+b+c}{2}+1}!} } \\
          \q{n}!&=\prod_{m=1}^n \q{m},\qquad
          \q{n}={q^{n/2}-q^{-n/2}\over q^{1/2} - q^{-1/2}}   
\end{align}
The sum over $z$ should run over all integers for which the $q$-factorials are well-defined, i.e. such that none of the arguments become less than zero. This condition depends on the level $k$. The expression for  $\Delta$ is only well-defined for admissible triples $(a,b,c)$, by which we mean that $a+b+c=0 \mod{2}$ and $|a-b|\leq c\leq a+b$ (we will take it to be zero for other triples, implementing consistency with the fusion rules). Note that $\Delta$ is invariant under permutations of its arguments.

The $R$-symbols are given by the general equation
\[
        R^{ab}_{c} = i^{c-a-b} q^{\frac{1}{8}(c(c+2)-a(a+2)-b(b+2))},
\]
which in turn gives a simple expression for the topological spins
\[
        \theta_a = e^{2\pi i \frac{a(a+2)}{4(k+2)}}.
\]

The quantum dimensions of the theory are
\[
        d_a = \frac{\sin\left(\frac{(a+1)\pi}{k+2} \right)}{\sin\left(\frac{\pi}{k+2}  \right)}
\]
The topological central charge is $c=3k/(k+2)$.

%


%

\begin{widetext}
\section{VLC properties and relations\label{ch.vlcprops}}

\hspace{-.4cm}\begin{minipage}{\linewidth}
\vspace{-1cm}
\begin{table}[H]
\caption{One-to-one consistency conditions}
\hrule\vspace{-3.5mm}
\begin{align}
\span\text{\sc{Diagrammatic equation}} & \span\text{\sc{Algebraic equation}} \vspace{1mm}\nonumber \\
\hline\vspace{1mm} \nonumber\\
\anyon{t} \; &= \; \sum_{a \in t} \;\; \anyon{a} &    \\
\condensatebubblemeetsanyon{t} \; &=\halfbraidcond{t}= \;\; \anyon{t} & \frac{1}{q}\sum_{b \in t} \sum_{\gamma \in\varphi}  |t^{\gamma a}_{b}|^2 \sqrt{ \frac{d_\gamma d_b}{d_a} }&= 1\\
\projector{t} \; &=  \; \delta_{t\in \sU} \;\;\; \anyon{t}&   \frac{1}{q} \sum_{b \in t} \sum_{\gamma \in\varphi}|t^{\gamma a}_b|^2 \sqrt{\frac{d_\gamma d_b}{d_a}} \frac{\theta_a}{\theta_b}   &=
\begin{cases} 1\;\;&\text{if $t\in \cat{U}$}\\ 0 \;\; &\text{if $t\notin \cat{U}$}\end{cases}
\\
\vertexbraket{t'}{r}{s}{t}{}{}\; &=\; \delta_{tt'} \sqrt{\frac{d_r d_s}{d_t}} \quad\anyon{t} 
& \sum_{a,b}\left( { d_a d_b d_t\over d_r d_s d_c}\right)^{\frac{1}{2}} \vlc{r}{s}{t}{a}{b}{c} \vlc{r}{s}{t'}{a}{b}{c}^* &= \delta_{tt'}\quad \forall \; c \label{eq:ortho1}\\
\vertexbraketbubbleL{s'}{r}{t}{s}\; &=\; \delta_{ss'} \sqrt{\frac{d_r d_t}{d_s}} \quad \anyon{s}
& \sum_{a,c}\left( { d_a d_c d_s\over d_r d_t d_b}\right)^{\frac{1}{2}} \vlc{r}{s}{t}{a}{b}{c} \vlc{r}{s'}{t}{a}{b}{c}^* &= \delta_{ss'}\quad \forall \; c\label{eq:ortho2}\\
\vertexbraketbubbleR{r'}{t}{s}{r}\; &=\; \delta_{rr'} \sqrt{\frac{d_s d_t}{d_r}} \quad\anyon{r}
& \sum_{b,c}\left( { d_b d_c d_r\over d_s d_t d_a}\right)^{\frac{1}{2}} \vlc{r}{s}{t}{a}{b}{c} \vlc{r'}{s}{t}{a}{b}{c}^* &= \delta_{rr'}\quad \forall \; c\label{eq:ortho3}
\end{align}
\hrule
\end{table}

\vspace{-.6cm}
\begin{table}[H]
\caption{One-to-two  consistency conditions}
\hrule\vspace{-3.5mm}
\begin{align}
\span\text{\sc{Diagrammatic equation}} & \span\text{\sc{Algebraic equation}} \vspace{1mm}\nonumber \\
\hline\vspace{1mm} \nonumber\\
\newvacRket{}{}{t} &\equiv \condensateRbraidket{}{}{t}{} & t^{a\gamma}_b &\equiv t^{\gamma a}_b R^{\gamma a }_b \;\;\; \forall \;a,b \in t\\
	\conddeathL{t} \; &= \ \conddeathR{t} & 	 {t^{\gamma a}_b \over {\left.\bar{t}_{\bar{b}}^{\bar{\gamma} \bar{a}}\right.}^*}&=R^{\gamma \bar{b}}_{\bar{a}} \F{0\bar{a}}{\gamma \bar{b}}{\bar{\gamma}\bar{a}}^*\F{0\gamma}{b \bar{a}}{\bar{b}\gamma} \F{b\bar{a}}{\gamma 0}{a\gamma}  \\
\condensatetwototwoSanyonRight{t} \; &= \; \condensatevertexbraanyon{t} & \F{0a}{\gamma b} {\gamma a} t^{\gamma a}_b &= t^{\bar{\gamma} b *}_a \\
\condensatetwototwoSanyonLvacR{t} \; &= \; \newvacRket{}{}{t} & \F{a \bar{\gamma}}{b 0} {\gamma b} t^{b \gamma *}_{a} &= t^{a \bar{\gamma}}_{b} \\
\condensatetwotooneRSanyondeath{\bar{t}} \; &= \; \condensatetwotooneManyondeath{\bar{t}} & \F{0 \gamma}{\bar{a}b}{a\gamma} t^{a\gamma}_{b} &= {\left.\bar{t}^{\bar{a}b}_\gamma\right.}^* \\
\condensateLonetotwoSanyonbirth{t} \; &= \; \newvacBket{t}{}{} & \F{b\bar{a}}{\gamma 0}{a \gamma} t^{\gamma a *}_b &= t^{b\bar{a}}_\gamma \\
\twotooneR{\bar{r}}{r}{s}{t} &= \F{\varphi s}{\bar{r} t}{r s}  \vertexbra{\bar{r}}{t}{s}{} & \F{0b}{\bar{a}c}{ab} \vlc{r}{s}{t}{a}{b}{c} &= \F{\varphi s}{\bar{r}{t}}{rs}
\vlc{\bar{r}}{t}{s}{\bar{a}}{c}{b}^* \\
\anyonLonetotwoS{t}{r}{s}{\bar{s}} &=  \F{rs}{t\varphi}{\bar{s}t}\vertexket{r}{s}{t}{} & 
\F{ab}{c0}{\bar{b}c} \vlc{t}{s}{r}{c}{\bar{b}}{a}^* &= \F{rs}{t\varphi}{\bar{s}{t}} \vlc{r}{s}{t}{a}{b}{c} 
\end{align}
\hrule
\end{table}

\end{minipage}

\begin{table}[H]
\caption{Two-to-two consistency conditions}
\vspace{1mm}
\hrule\vspace{-3.5mm}
\begin{align}
\span\text{\sc{Diagrammatic equation}} & \span\text{\sc{Algebraic equation}} \vspace{1mm}\nonumber \\
\hline\vspace{1mm} \nonumber\\
\condensatetwototwoSanyonR{t}{t} &= \condensatetwototwoManyonR{t}{t}{t} & 
\F{\gamma a}{\gamma b}{0c} + \phi^* t^{\gamma a}_b \F{\gamma a}{\gamma b}{\gamma c} &= t^{\gamma a}_c t^{b\gamma *}_c 
\;\; (\text{for $a,b\in t$ and $\varphi = 0+\gamma$})\\
\condensatetwototwoSanyonL{t} &= \condensatetwototwoManyonL{t}{t}{t} & \F{a\gamma}{b \gamma}{0c} + \phi t^{b\gamma *}_a \F{a\gamma}{b\gamma}{\gamma c} &= t^{a\gamma}_c t^{b\gamma *}_c 
\;\; (\text{for $a,b\in t$ and $\varphi = 0+\gamma$})\\
\condensatetwototwoSanyonstraight{t} &=	\condensatetwototwoManyonstraight{t} & \sum_c t^{\gamma c *}_a t^{c\gamma}_b \F{a\gamma}{\gamma b}{cd} &= t^{a\gamma}_d t^{\gamma b *}_d \\
\condensatetwototwoSanyonbend{t} &= \phase{t} \condensatetwototwoManyonstraightMirror{t} &  \sum_c t^{b \bar{a}}_\gamma \bar{t}^{\bar{c}a}_\gamma \F{\gamma a}{b\gamma}{\bar{c}d} &= \phase{t} t^{\gamma a}_{d} t^{b\gamma *}_d\;\; \left( \Rightarrow \phase{t} \!=\! \phase{a}\;\; \forall \; a\in t \right)\\
\condensatetwototwoSanyondeath{t} &= \condensatetwototwoManyondeath{t} & \sum_{\bar{c}} t^{a\bar{c}*}_\gamma \bar{t}^{\bar{c}\gamma}_{\bar{b}} \F{\gamma \gamma}{a\bar{b}}{\bar{c}d} &= \delta_{d0} + \delta_{d\gamma} \phi {t^{a\bar{b}}_\gamma}^*\;\;\; (\text{for $\varphi = 0+ \gamma$}) \\
\condensatetwototwoSanyonbirth{t} &= \condensatetwototwoManyonbirth{t} & \sum_c {t^{\gamma c}_a}^* t^{c\bar{b}}_\gamma \F{a\bar{b}}{\gamma \gamma}{cd} &= \delta_{d0} + \delta_{d\gamma} \phi^* t^{a\bar{b}}_\gamma \;\;\; (\text{for $\varphi = 0+ \gamma$}) \\
\condensatetwototwoScondensateLdown{r}{s}{t}{} &= \condensatetwototwoMcondensateLdown{r}{s}{t}{} & \sum_{a'} {t^{\gamma a'}_a}^* \vlc{r}{s}{t}{a'}{b}{c} &= \sum_{c'} t^{\gamma c *}_{c'} \vlc{r}{s}{t}{a}{b}{c'} \\
 \condensatetwototwoScondensateRup{t}{r}{s}{} &= \condensatetwototwoMcondensateRup{t}{r}{s}{} & \sum_{b'} s^{b'\gamma}_b \vlc{r}{s}{t}{a}{b'}{c}^* &= \sum_{c'} t^{c\gamma}_{c'} \vlc{r}{s}{t}{a}{b}{c'}^* \\
\condensatetwototwoScondensateMiddle{}{}{r}{s} &= \F{rs}{rs}{\varphi t} \twototwoM{r}{s}{r}{s}{t}{}{} &  \F{rs}{rs}{\phi t} \vlc{r}{s}{t}{a}{b}{c} \vlc{r}{s}{t}{a'}{b'}{c}^* 
&=\frac{1}{q}\sum_{\gamma\in\varphi} {r_{a}^{a'\gamma}}^* s_{b'}^{\gamma b}\F{ab}{a'b'}{\gamma c}
\\
&&&=\frac{1}{q}\F{ab}{a'b'}{0c} + \frac{1}{q}\F{ab}{a'b'}{\gamma c} r^{a'\gamma *}_a s^{\gamma b}_{b'} \;\;\; (\text{for $\varphi = 0+\gamma$}) \nn
\end{align}
\hrule
\end{table}

\end{widetext}



\end{document}